\documentclass[aps,superscriptaddress,floatfix,twocolumn]{revtex4-1}
\usepackage[latin9]{inputenc}
\usepackage{amsmath}
\usepackage{amssymb}
\usepackage{graphicx}
\usepackage{esint}
\usepackage{epstopdf}
\usepackage{braket}
\usepackage{color}
\usepackage{leftindex}
\usepackage[colorlinks=true,citecolor=blue,linkcolor=blue,urlcolor=blue]{hyperref}
\usepackage{graphicx}
\usepackage{stackengine}
\usepackage{nicematrix}

\makeatletter
\@ifundefined{textcolor}{}
{
 \definecolor{BLACK}{gray}{0}
 \definecolor{WHITE}{gray}{1}
 \definecolor{RED}{rgb}{1,0,0}
 \definecolor{GREEN}{rgb}{0,1,0}
 \definecolor{BLUE}{rgb}{0,0,1}
 \definecolor{CYAN}{cmyk}{1,0,0,0}
 \definecolor{MAGENTA}{cmyk}{0,1,0,0}
 \definecolor{YELLOW}{cmyk}{0,0,1,0}
}

\makeatother

\begin{document}

\title{Resonant analogue configurations in atomic condensates}

\author{Juan Ram\'on Mu\~noz de Nova}
\email{jrmnova@fis.ucm.es}
\affiliation{Departamento de F\'isica de Materiales, Universidad Complutense de Madrid, 28040 Madrid, Spain}

\author{Pablo Fern\'andez Palacios}
\affiliation{Instituto de Energ\'ia Solar, Universidad Polit\'ecnica de Madrid, 28040 Madrid, Spain}

\author{Pedro Alc\'azar Guerrero}
\affiliation{Catalan Institute of Nanoscience and Nanotechnology (ICN2), CSIC and BIST, Campus UAB, Bellaterra, 08193 Barcelona, Spain}

\affiliation{Department of Physics, Campus UAB, Bellaterra, 08193 Barcelona, Spain}

\author{Ivar Zapata}
\affiliation{mrHouston Tech Solutions, 28002 Madrid, Spain}

\author{Fernando Sols}
\affiliation{Departamento de F\'isica de Materiales, Universidad Complutense de
Madrid, 28040 Madrid, Spain}

%JR: Robustness and universality of the Hawking effect due to the microscopic description; trans-Planckian problem. Static instability. Update ATLAS and all CMS citations. GP/GL analogy.

\begin{abstract}
As a contribution to a memorial volume, we provide a comprehensive discussion of resonant configurations in analogue gravity, focusing on its implementation in atomic condensates and combining review features with original insights and calculations. In particular, we jointly analyze the analogues of the Andreev and Hawking effects using a microscopic description based on the Bogoliubov approximation. We perform a detailed study of the thermality of the Andreev and Hawking spectra for canonical black-hole solutions, finding that both can be described by a gray-body distribution to a very good approximation. We contemplate several resonant scenarios whose efficiency to enhance anomalous scattering processes is compared to that of non-resonant setups. The presence of quantum signatures in analogue configurations, such as the violation of Cauchy-Schwarz inequalities or entanglement, is analyzed, observing that resonant configurations highly increase the entanglement signal, especially for the Andreev effect. We also discuss how these results have served as inspiration for the rapidly expanding field of quantum information in high-energy colliders. Finally, we study the physics of black-hole lasers as further examples of resonant analogue structures, distinguishing three stages in its time evolution. For short times, we compute the linear and non-linear spectrum for different models. For intermediate times, we generalize the current analysis of the BHL-BCL crossover. For long times, we discuss the emerging concept of spontaneous Floquet state and its potential implications.
\end{abstract}
\date{\today}

\maketitle

\section{Introduction}\label{sec:Intro}

In the 70s, Stephen Hawking \cite{Hawking1974} predicted, on the basis of a semiclassical calculation (in which fields are quantized but the background spacetime is treated as classical), that black holes spontaneously emit radiation with a Planckian spectrum. This became known as Hawking radiation, one of the most celebrated predictions of modern theoretical physics. As it involves thermodynamics, quantum mechanics, and general relativity, understanding Hawking radiation is regarded as the first step towards a quantum theory of gravity. However, its detection in an astrophysical scenario is quite unlikely because its effective temperature of emission, the Hawking temperature, is of the order of $T_H\sim 10^{-8}\,\textrm{K}$ for a black hole of several solar masses, much lower for instance than the temperature of the cosmic microwave background, $T_{\rm{CMB}}\simeq 2.7\,\textrm{K}$.

It was later noted by Unruh \cite{Unruh1981} that the equations of motion describing the fluctuations of an irrotational flow are formally analogue to those of a massless scalar field in a curved spacetime described by the so-called acoustic metric. This connection established the rich field of analogue gravity, where tabletop experiments are used to replicate gravitational effects in a controlled setup. A number of vastly different systems have been proposed for implementing analogue experiments, including atomic Bose-Einstein condensates \cite{Garay2000,Lahav2010}, water waves \cite{Schutzhold2002,Weinfurtner2011}, non-linear optical fibers \cite{Philbin2008,Drori2019}, ion rings \cite{Horstmann2010,Wittemer2019}, quantum fluids of light \cite{Carusotto2013,Nguyen2015}, and even superconducting transmon qubits \cite{Shi2023}. 

In the specific case of Hawking radiation, the subsonic/supersonic regions of a flowing fluid are akin to the exterior/interior of a black hole, since sound cannot travel upstream in a supersonic flow, in the same way as light is trapped inside a black hole. As a result, the analogue of the event horizon is provided by the subsonic/supersonic interface, and Hawking radiation is associated with the spontaneous emission of correlated phonons into the subsonic and supersonic regions. A classical stimulated version of this emission has been observed in hydraulic \cite{Weinfurtner2011,Euve2016} and optical \cite{Drori2019} analogues. Due to their intrinsic quantumness, well-controlled behavior, and low temperature, Bose-Einstein condensates are the most promising candidates to exhibit the genuine Hawking effect, of a quantum nature. Indeed, the study of quantum field theory in curved spacetimes using atomic condensates can be regarded as another paradigm of quantum simulation \cite{Viermann2022}. 

From the theoretical point of view, the initial proposal of Ref. \cite{Garay2000} was subsequently expanded in a number of works \cite{Leonhardt2003a,Balbinot2008,Carusotto2008,Macher2009a,Recati2009,Zapata2011,Larre2012,deNova2014,Finazzi2014,Busch2014,deNova2015,Michel2016}. From the experimental point of view, the first analogue black hole in an atomic condensate was achieved by the Technion group in 2010 \cite{Lahav2010}. After progressively improving the setup \cite{Shammass2012,Schley2013,Steinhauer2014}, the Technion experiment provided the first claimed detection of the Hawking effect in 2016 \cite{Steinhauer2016}, which was later confirmed by a precise agreement with the theoretical predictions \cite{deNova2019,Kolobov2021}, including the measurement of the Hawking temperature and the observation of the stationarity of the spontaneous emission of Hawking radiation. The detection of the Hawking effect is far from being the end of the road for the field, and analogues of the dynamical Casimir effect \cite{Jaskula2012}, Sakharov oscillations \cite{Hung2013}, superradiance \cite{Torres2017}, inflation \cite{Eckel2018}, Unruh effect \cite{Hu2019}, quasi-normal ringdown \cite{Torres2020}, backreaction \cite{Patrick2021} or cosmological particle creation \cite{Steinhauer2022a}, have also been observed in the laboratory.  

Interestingly, subsonic/supersonic interfaces in condensates provide yet another insightful conceptual connection, since it was shown \cite{Zapata2009a} that there also takes place the analogue of the Andreev reflection in superconductors \cite{Andreev1964}, where an electron (hole) incident on a normal/superconductor interface from the normal side is reflected as a hole (electron). The Andreev reflection has already been observed in superconductors \cite{Bozhko1982,Benistant1983}, as well as in superfluid $^3$He \cite{Enrico1993,Okuda1998}, but there is not yet experimental evidence of this behavior in condensates. 

During the quest for the detection of the Hawking effect, it was proposed \cite{Zapata2011} that setups involving multiple internal reflections, similar to those occurring in a Fabri-Perot interferometer, might be advantageous due to the non-thermal character of the resulting spectrum. The precedent existed of the black-hole laser (BHL) \cite{Corley1999}, where multiple internal reflections on a pair of horizons result in an increasing unstable output of Hawking radiation. Moreover, it was later shown that resonant configurations may enhance the quantum contribution of the Hawking effect \cite{deNova2014,deNova2015}.

%When the emphasis is on the achievement of stationary Hawking radiation with a r, the attention turns towards resonant structures hosting multiple  reflection processes 

In this work, we provide a comprehensive discussion of resonant analogue configurations in atomic condensates and their most important features, including an original unified analysis of the Andreev and Hawking effects. We will devote special attention to key contributions from Renaud Parentani, highlighted throughout the article.

%, which here are understood in their spontaneous versions

We begin by providing a general introduction to gravitational analogues in atomic condensates in Section \ref{sec:HawkingCondensates}. In this respect, the work by Macher and Parentani \cite{Macher2009,Macher2009a}, along with that of Recati, Pavloff, and Carusotto \cite{Recati2009}, represented a milestone, since it established the microscopic analogy of the Hawking effect without resorting to any effective metric. This microscopic approach is the framework upon which we will develop our results. In particular, we draw an interesting analogy between the Bogoliubov-de Gennes (BdG) equations and the relativistic Klein-Gordon equation, which provides an insightful way to address the quantization of the problem. We discuss how the bosonic version of the Andreev reflection emerges in the interior of an analogue black hole \cite{Zapata2009a} and gives rise to a spontaneous emission of quasiparticles, to which we refer as the Andreev effect, in analogy with the Hawking effect. We assess the thermality of canonical analogue configurations, finding that their Andreev spectrum is also Planckian to a very good approximation due to its universal scaling at low frequencies \cite{deNova2014}, an unnoticed result in the literature.

Section \ref{sec:ResonantHawking} is devoted to the study of Andreev-Hawking processes in resonant analogue structures. Resonant Andreev scenarios have been also considered in normal/superconductor junctions \cite{Prada2005}. Here, we examine a double barrier structure with alternating supersonic and subsonic segments in the region between the two barriers, in whose original proposal Parentani directly contributed \cite{Zapata2011}. Another setup analyzed in this section is a resonant flat-profile structure, involving a stationary homogeneous flowing condensate with a piecewise modulated speed of sound, first discussed in Ref. \cite{deNova2014}.  Our results show that the Andreev signal is also highly enhanced by resonant peaks in the spectrum, and it may even become larger than the Hawking signal, in contrast with typical analogue configurations.

It was noticed that an optical lattice, with its characteristic multiple barrier structure, might enhance some of the properties of a resonant structure \cite{deNova2014a,deNova2017b}. The second part of Sec. \ref{sec:ResonantHawking} addresses the long-lasting quasi-stationary outcoupling of a condensate through an optical lattice, which may have an ideal uniform or a Gaussian shape, with a characteristic time scale much longer than that of conventional black-hole configurations.

Section \ref{sec:QuantumHR} analyzes the  quantumness of the Andreev and Hawking effects by borrowing concepts and techniques from quantum optics \cite{Walls2008}. This approach has also been recently advocated in the original gravitational scenario \cite{Scully2022}. With the help of these tools, we jointly characterize the Andreev and Hawking effects as the spontaneous production of hybrid Andreev-Hawking modes from the non-degenerate parametric amplification of the vacuum. The quantumness of the Andreev-Hawking effect is evaluated through different quantum correlations such as the violation of Cauchy-Schwarz (CS) inequalities or entanglement. Originally, the violation of CS inequalities by quantum Hawking radiation was first discussed in Ref. \cite{deNova2014}. In an almost parallel effort, Busch and Parentani \cite{Busch2014}, as well as Finazzi and Carusotto \cite{Finazzi2014}, analyzed the entanglement of Hawking radiation in condensates using the generalized Peres-Horodecki criterion \cite{Simon2000}. The inspiring visit of Renaud Parentani to our group in Madrid in November 2013 motivated us to unify both approaches within a common framework in Ref. \cite{deNova2015}. All these techniques were later employed in the detection of the entanglement of Hawking radiation in the 2016 experiment \cite{Steinhauer2016}. 

In the last part of Sec. \ref{sec:QuantumHR}, we briefly discuss how the study of quantum correlations in the Andreev-Hawking effect has motivated the research on quantum information in high-energy colliders \cite{Afik2021}. This has rapidly become an emergent field of research by itself, which has already achieved its first milestone with the pioneering observation of entanglement in quarks by the ATLAS and CMS collaborations at the Large Hadron Collider (LHC) \cite{ATLAS2024,CMS2024}, representing also the highest-energy detection of entanglement ever. A pedagogical introduction to this fascinating topic for a readership outside the high-energy field is presented in Ref. \cite{Afik2022}. 

%In this context, another remarkable phenomenon is the black-hole laser (BHL) effect \cite{Corley1999}, i.e., the self-amplification of Hawking radiation due to successive reflections between a pair of horizons, leading to the emergence of dynamical instabilities in the excitation spectrum. 

%can emerge from a resonant configuration, devoting special attention to its non-linear regime and the resulting CES state.

Section \ref{sec:BHL} addresses the emergence of a black-hole laser in resonant configurations. In an atomic condensate, the BHL effect arises because of its superluminal dispersion relation, which allows the radiation reflected at the inner horizon to travel back to the outer one, further stimulating the production of Andreev-Hawking radiation \cite{Leonhardt2003,Barcelo2006,Jain2007,Coutant2010,Finazzi2010,Bermudez2018,Burkle2018}. Other analogue setups have been proposed to observe the BHL effect \cite{Faccio_2012,Peloquin2016,RinconEstrada2021,Katayama2021}. The work by Parentani and collaborators \cite{Coutant2010,Finazzi2010} was instrumental in determining the properties of a BHL, including the first full microscopic BdG computation of the spectrum of dynamical instabilities, in analogy with the microscopic derivation of the Hawking effect  \cite{Recati2009,Macher2009a}. Parentani and Michel also pioneered the study of the non-linear regime of a BHL \cite{Michel2013}, achieved once the initial instability has grown up to saturation, and the numerical study of its dynamics \cite{Michel2015}, in parallel to the work by Mu\~noz de Nova, Finazzi, and Carusotto \cite{deNova2016}.

Our discussion of the black-hole laser further extends the original work of Parentani in all the stages of its time evolution. At short times, using the protocol to construct BHL solutions of Ref. \cite{deNova2017a}, we compute the linear and non-linear spectrum for different BHL models, including that of a double barrier structure, original of the present work. Our results confirm all the trends anticipated in Ref. \cite{Michel2013}.

At intermediate times in the evolution of a BHL, one has to take into account the Bogoliubov-Cherenkov-Landau (BCL) mode present in the supersonic region \cite{Carusotto2006}, which is analogous to the undulation in hydraulic setups \cite{Coutant2012}. Because of its zero-frequency, the BCL mode is resonantly excited by any obstacle in the flow and overshadows the BHL effect in real experiments \cite{Steinhauer2014,Kolobov2021,Steinhauer2022}. The BHL-BCL problem has attracted a number of studies in the theoretical literature \cite{Tettamanti2016,Steinhauer2017,Wang2016,Wang2017,Llorente2019,Tettamanti2021,Kolobov2021,Steinhauer2022,deNova2023}, and the observation of the BHL effect still remains a major challenge in the analogue field. Here, we generalize the discussion in Ref. \cite{deNova2023} of the BHL-BCL crossover, originally based on a flat-profile model, underlining the crucial role played by the $\mathbb{Z}_2$ symmetry of a quantum BHL, first predicted by Michel and Parentani \cite{Michel2013}.

For sufficiently long times, the BHL displays a dynamical phase diagram where it can only reach two states \cite{deNova2016,deNova2021}: the true non-linear ground state or the so-called Continuous Emission of Solitons (CES) state, which represents a realization of a spontaneous Floquet state \cite{deNova2022}. The original conception of spontaneous Floquet state was heavily influenced by richful discussions with Renaud Parentani during the visit of one of us (JRMdN) to Orsay in 2015. Here we analyze in detail the CES state arising from a flat-profile BHL solution, and discuss interesting implications of spontaneous Floquet states, including a specific and tangible realization of time operator in quantum mechanics.

The inspiration of Renaud Parentani, and in some cases his direct involvement, is a common thread of the work discussed in this article.

% A specific mechanism of BCL-stimulated Hawking radiation was identified in 2021 \cite{Kolobov2021}, whose role in the 2014 experiment was recently confirmed \cite{Steinhauer2022}.
% Originally, due to its stimulated character, it was thought that the observation of the BHL effect would be a first step towards the dreamed observation of the Hawking effect. In actuality, it has been rather the opposite. This is because in a condensate the BHL effect arises in a finite supersonic region, which is energetically unstable according to the Landau criterion so any static perturbation will resonantly produce Bogoliubov-Cherenkov-Landau (BCL) radiation \cite{Carusotto2006}, the analogue of the undulation in hydraulic setups \cite{Coutant2012}. Furthermore, the BHL modes are expected to contain similar wavevectors and frequencies to those of the BCL wave, as the latter is also stimulated by the scattering of Hawking radiation at the inner horizon. Therefore, experimental attempts to isolate the BHL effect will be hindered by a background BCL signal. 

%Section \ref{sec:ResonantHawking} discusses resonant Hawking radiation in analogue configurations, examining a realistic experimental implementation using an optical lattice. 

\section{Andreev and Hawking effects in atomic condensates}\label{sec:HawkingCondensates}

\subsection{Gross-Pitaevskii and Bogoliubov-de Gennes equations}\label{subsec:GPBdG}

We begin by reviewing the basic concepts and techniques for the study of atomic condensates. We consider the following general second-quantization Hamiltonian for interacting bosons \cite{Fetter2003}:
\begin{equation}\label{eq:HamiltonianManyBody}
\hat{H}=\int\mathrm{d}\mathbf{x}~\hat{\Psi}^{\dagger}(\mathbf{x})\left[-\frac{\hbar^2}{2m}\nabla^2+V(\mathbf{x},t)+\frac{g}{2}\hat{\Psi}^{\dagger}(\mathbf{x})\hat{\Psi}(\mathbf{x})\right]\hat{\Psi}(\mathbf{x}),
\end{equation}
where $m$ is the mass of the atoms, $V(\mathbf{x},t)$ is some external potential, and the bosons interact through the contact pseudopotential $W(\mathbf{x}-\mathbf{x}')=g\delta(\mathbf{x}-\mathbf{x}')$ \cite{Pitaevskii2003}. The field operator $\hat{\Psi}(\mathbf{x})$ satisfies the canonical commutation relation $[\Psi(\mathbf{x}),\Psi^\dagger (\mathbf{x}')]=\delta(\mathbf{x}-\mathbf{x}')$, which leads to the Heisenberg equation of motion
\begin{equation}\label{eq:HeisenbergEquationOfMotion}
    i\hbar\partial_t\hat{\Psi}(\mathbf{x},t)=\left[-\frac{\hbar^2}{2m}\nabla^2+V(\mathbf{x},t)+g\hat{\Psi}^{\dagger}(\mathbf{x},t)\hat{\Psi}(\mathbf{x},t)\right]\hat{\Psi}(\mathbf{x},t).
\end{equation}
Close to $T=0$, the condensate can be described by a coherent state, characterized by a macroscopic wavefunction $\Psi(\mathbf{x},t)$ that is normalized to the total particle number,
\begin{equation}\label{eq:Normalization}\int\mathrm{d}\mathbf{x}~|\Psi(\mathbf{x},t)|^2=N.
\end{equation}
Quantum fluctuations around the condensate are accounted by expanding the field operator around its coherent expectation value (see Sec. \ref{subsec:QuantumOptics} for a thorough discussion on coherent states) as 
\begin{equation}
   \hat{\Psi}(\mathbf{x},t)=\Psi(\mathbf{x},t)+\hat{\varphi}(\mathbf{x},t).
\end{equation}
Plugging this expansion into Eq. (\ref{eq:HeisenbergEquationOfMotion}) yields,  at leading order, the \textit{time-dependent} Gross-Pitaevskii (GP) equation,
\begin{equation}\label{eq:TDGP}
    i\hbar\partial_t \Psi(\mathbf{x},t)=\left[-\frac{\hbar^2}{2m}\nabla^2+V(\mathbf{x},t)+g|\Psi(\mathbf{x},t)|^2\right]\Psi(\mathbf{x},t),
\end{equation}
and, at linear order in the quantum fluctuations, the \textit{time-dependent} Bogoliubov-de Gennes (BdG) equations
\begin{equation}\label{eq:TDBdG} i\hbar\partial_t\hat{\Phi}=M(t)\hat{\Phi},\,\hat{\Phi}=\left[\begin{array}{c}\hat{\varphi}\\ \hat{\varphi}^\dagger\end{array}\right],\, M(t)=\left[\begin{array}{cc} N(t) & A(t)\\
-A^*(t) &-N(t)\end{array}\right],
\end{equation}
where
\begin{equation}\label{eq:TDBdGMatrix}
    N(t)=-\dfrac{\hbar^2}{2m}\nabla^2+V(\mathbf{x},t)+2g|\Psi(\mathbf{x},t)|^2,~A(t)=\Psi^2(\mathbf{x},t).
\end{equation}
The GP equation is thus a non-linear Schr\"odinger equation that describes the condensate dynamics, where the non-linearity stems from the interactions between the condensate atoms, while the linear dynamics of the quantum fluctuations is governed by the BdG equations. Notice that the BdG equations also describe the linear dynamics of  the fluctuations of the GP wavefunction $\Psi'(\mathbf{x},t)$ around a reference solution $\Psi(\mathbf{x},t)$, $\Psi'(\mathbf{x},t)=\Psi(\mathbf{x},t)+\varphi(\mathbf{x},t)$, resulting in the substitution $\hat{\varphi}\to \varphi$ in Eq. (\ref{eq:TDBdG}).

%$\Psi(\mathbf{x},t)\to \Psi_0(\mathbf{x},t)$ and 
%in Eqs. (\ref{eq:TDBdG}), (\ref{eq:TDBdGMatrix}).
We focus on time-independent configurations, $V(\mathbf{x},t)=V(\mathbf{x})$, and consider stationary condensates, which are accounted by 
\begin{equation}\label{eq:StationaryFieldExpansion}
   \hat{\Psi}(\mathbf{x},t)=\left[\Psi_0(\mathbf{x})+\hat{\varphi}(\mathbf{x},t)\right]e^{-i\mu t/\hbar},
\end{equation}
$\mu$ being the chemical potential. This results in the \textit{time-independent} GP equation 
\begin{equation}\label{eq:TIGP}
    \mu \Psi_0(\mathbf{x})=\left[-\frac{\hbar^2}{2m}\nabla^2+V(\mathbf{x})+g|\Psi_0(\mathbf{x})|^2\right]\Psi_0(\mathbf{x}),
\end{equation}
and the \textit{stationary} BdG equations
\begin{equation}\label{eq:TIBdG} i\hbar\partial_t\hat{\Phi}=M_0\hat{\Phi},\,\hat{\Phi}=\left[\begin{array}{c}\hat{\varphi}\\ \hat{\varphi}^\dagger\end{array}\right],~M_0=\left[\begin{array}{cc} N_0 & A_0\\
-A_0^* &-N_0\end{array}\right],
\end{equation}
where now
\begin{equation}\label{eq:TIBdGMatrix}
    N_0=-\dfrac{\hbar^2}{2m}\nabla^2+V(\mathbf{x})+2g|\Psi_0(\mathbf{x})|^2-\mu,~A_0=\Psi^2_0(\mathbf{x}).
\end{equation}
Since $M_0$ is time-independent, any solution to the stationary BdG equations can be expanded in terms of a complete set of eigenmodes
\begin{equation}\label{eq:BdGEigenmode}
    M_0z_n=\epsilon_n z_n,~z_n\equiv \left[\begin{array}{c}u_n\\ v_n\end{array}\right].
\end{equation}
The matrix operator $M_0$ is non-Hermitian and it can possess complex eigenvalues, representing dynamical instabilities which grow exponentially in time. For the present moment, we assume that the system is dynamically stable and ignore the presence of Nambu-Goldstone modes; we will come back later to this issue in Sec. \ref{sec:BHL}. 

Even though $M_0$ is non-Hermitian, the BdG eigenmodes do form an orthonormal basis under the inner product
\begin{equation}\label{eq:BdGProduct}
    (z_n|z_m)\equiv\braket{z_n|\sigma_z| z_m} =\int\mathrm{d}\mathbf{x}~[u^*_nu_m-v^*_nv_m], 
\end{equation}
with $\braket{z_n|z_m}$ the standard scalar product for two spinors and $\sigma_i$ the usual Pauli matrices. This is because $\Lambda\equiv \sigma_z M_0$ is indeed Hermitian, and thus $M_0$ is pseudo-Hermitian, i.e.,
\begin{equation}\label{eq:PseudoHermitian}
    (z_n|M_0z_m)=\braket{z_n|\Lambda z_m}=\braket{\Lambda z_n|z_m}=(M_0z_n|z_m),
\end{equation}
which implies the conservation of the inner product between solutions of the BdG equations and the orthogonality between eigenmodes,
\begin{equation}\label{eq:EigenOrto}
    (\epsilon_m-\epsilon^*_n)(z_n|z_m)=0.
\end{equation}
Actually, the conservation of the norm for any solution $z$ of the BdG equations, $i\hbar\partial_t z=M_0z$, can be derived from a continuity equation, in analogy with the Schr\"odinger equation [see Eq. (\ref{eq:SchrodingerCurrent})],
\begin{align}\label{eq:QuasiparticleCurrent}
    &\partial_t(z^\dagger\sigma_z z)+\boldsymbol{\nabla}\cdot \mathbf{j}=0,\\
    \nonumber &\mathbf{j}=-\frac{i\hbar}{2m}\left[u^*\boldsymbol{\nabla} u-u\boldsymbol{\nabla}u^*+v^*\boldsymbol{\nabla} v-v\boldsymbol{\nabla}v^*\right],
\end{align}
where $u,v$ are the components of the spinor $z$ and $\mathbf{j}$ is the quasiparticle current. 

However, in contrast to the Schr\"odinger case, both $M_0$ and the inner product are not positive definite. Indeed, by noticing that $\sigma_x M^*_0 \sigma_x=-M_0$ and $\sigma_x \sigma_z \sigma_x=-\sigma_z$, we can define a conjugate mode as $\bar{z}_n\equiv \sigma_x z^*_n$, which has opposite eigenvalue and norm  
\begin{equation}\label{eq:ConjugateRelations}
    M_0\bar{z}_n=-\epsilon^*_n \bar{z}_n,~(z_n|z_m)=-(\bar{z}_n|\bar{z}_m)^*.
\end{equation}
This symmetry stems from that of the field spinor $\hat{\Phi}$, which is self-conjugate, $\hat{\bar{\Phi}}=\hat{\Phi}$. Unless otherwise stated, the modes $z_n$ are chosen with positive norm, $(z_n|z_n)=1$.

The above properties of the inner product suggest that the correct analogy for the BdG equations should rather be established with the Klein-Gordon (KG) equation for an Hermitian scalar field,
\begin{equation}\label{eq:KleinGordon}
    \left[\square-\frac{m^2c^2}{\hbar^2}\right]\hat{\phi}=0,~\square\equiv \partial_\mu \partial^\mu=\nabla^2-\frac{1}{c^2}\partial^2_t, 
\end{equation}
where we take the Minkowski metric as $\eta_{\mu\nu}=\textrm{diag}[-1,1,1,1]$. By invoking its canonical momentum $\hat{\Pi}(\mathbf{x})$, which satisfies the equation of motion $\hat{\Pi}=\hbar \partial_t\hat{\phi}$ and the commutation relation $[\hat{\phi}(\mathbf{x}),\hat{\Pi}(\mathbf{x}')]=i\delta(\mathbf{x}-\mathbf{x}')$, the KG equation can be recasted as the BdG equations (\ref{eq:TIBdG}),
\begin{equation}\label{eq:TIKG} i\hbar\partial_t\hat{\Phi}=M_{0}\hat{\Phi},\,\hat{\Phi}=\left[\begin{array}{c}\hat{\phi}\\ i\hat{\Pi}\end{array}\right],~M_{0}=\left[\begin{array}{cc} 0 & 1\\
H_0 &0 \end{array}\right], 
\end{equation}
with $H_0\equiv -(\hbar c \nabla)^2+m^2c^4$. The KG modes are derived from the eigenvalue problem 
\begin{equation}
    M_0z_n=\epsilon_n z_n,~z_n\equiv \left[\begin{array}{c}\phi_n\\ \chi_n\end{array}\right],
\end{equation}
equivalent to the more usual equation $\epsilon^2_n\phi_n=H_0\phi_n$. Notice that $M_0$ is again non-Hermitian, while $\Lambda=\sigma_x M_0$ is, which implies the conservation of the KG inner product 
\begin{equation}\label{eq:KGProduct}
    (z_n|z_m)\equiv\braket{z_n|\sigma_x| z_m} =\int\mathrm{d}\mathbf{x}~[\phi^*_n\chi_m+\chi^*_n\phi_m].
\end{equation}
Conjugate solutions are defined now as $\bar{z}_n\equiv \sigma_z z^*_n$, since $\sigma_z M^*_0 \sigma_z=-M_0$ and $\sigma_z \sigma_x \sigma_z=-\sigma_x$, so Eq. (\ref{eq:ConjugateRelations}) is satisfied. Moreover, the field spinor is also self-conjugate; this property can be directly traced back here to the Hermitian character of the field $\hat{\phi}$. 

The field spinor can be expanded in terms of the complete set of eigenmodes $\{z_n,\bar{z}_n\}$. In both BdG and Hermitian KG cases, the self-conjugate character of $\hat{\Phi}$ implies that this expansion is of the form
\begin{equation}\label{eq:QuantumFieldFluctuations}
   \hat{\Phi}(\mathbf{x},t)=\sum_{n}\hat{a}_{n}(t)z_{n}(\mathbf{x})+\hat{a}^{\dagger}_{n}(t)\bar{z}_{n}(\mathbf{x}),
\end{equation}
where $\hat{a}_n(t)\equiv (z_n|\hat{\Phi}(t))$ is the quantum amplitude of the mode $z_n$. The canonical commutation rules for the field spinor can be expressed in matrix form as 
\begin{equation}
    [\hat{\Phi}(\mathbf{x}),\hat{\Phi}^\dagger(\mathbf{x}')]=\sigma_i\delta(\mathbf{x}-\mathbf{x}'),
\end{equation}
with $\sigma_i$ the Pauli matrix characterizing the corresponding inner product, Eqs. (\ref{eq:BdGProduct}), (\ref{eq:KGProduct}). Using this relation, it is straightforward to prove that the quantum amplitudes $\hat{a}_{n}$ behave as bosonic annhihilation operators,
\begin{equation}\label{eq:Aniquilacion}
    [\hat{a}_n,\hat{a}^\dagger_m]=[(z_n|\hat{\Phi}),(\hat{\Phi}|z_m)]=(z_n|z_m)=\delta_{nm},
\end{equation}
whose equation of motion is simply
\begin{equation}\label{eq:BogoliubovQuantum}
    i\hbar\partial_t \hat{a}_n=(z_n|M_0\hat{\Phi})=\hbar\omega_n\hat{a}_n\Longrightarrow \hat{a}_n(t)=\hat{a}_ne^{-i\omega_nt},
\end{equation}
$\hbar \omega_n=\epsilon_n$ being the frequency of the mode. Hence, we arrive at the usual result
\begin{equation}\label{eq:QuantumFieldFluctuationsExplicit}
   \hat{\Phi}(\mathbf{x},t)=\sum_{n}\hat{a}_{n}z_{n}(\mathbf{x})e^{-i\omega_n t}+\hat{a}^{\dagger}_{n} \bar{z}_{n}(\mathbf{x})e^{i\omega_n t}.
\end{equation}
Remarkably, this expansion allows to diagonalize the KG Hamiltonian in an elegant and compact way:
\begin{eqnarray}
    \hat{H}_{\rm{KG}}&=&\frac{1}{2}\int\mathrm{d}\mathbf{x}~\hat{\Pi}^2+(\hbar c)^2|\boldsymbol{\nabla}\hat{\phi}|^2+m^2c^4\hat{\phi}^2\\
    \nonumber &=&\frac{1}{2}(\hat{\Phi}|M_0\hat{\Phi})=\sum_{n} \epsilon_n\left(\hat{a}^{\dagger}_{n}\hat{a}_{n}+\frac{1}{2}\right).
\end{eqnarray}
In the BdG case, the field expansion diagonalizes the grand-canonical Hamiltonian $\hat{K}=\hat{H}-\mu \hat{N}$, with
\begin{equation}\label{eq:ParticleNumberOperator}
\hat{N}=\int\mathrm{d}\mathbf{x}~\hat{\Psi}^{\dagger}(\mathbf{x})\hat{\Psi}(\mathbf{x})
\end{equation}
the particle-number operator. After expanding up to quadratic order, in the spirit of the Bogoliubov approximation, one obtains:
\begin{equation}\label{eq:GrandCanonicalEnergy}
    \hat{K}\simeq K_0+K'_V+\frac{1}{2}(\hat{\Phi}|M_0\hat{\Phi})=K_0+K_V+\sum_{n} \epsilon_n \hat{a}^{\dagger}_{n}\hat{a}_{n},
\end{equation}
where $K_0\equiv K[\Psi_0]$ is the mean-field energy of the condensate, obtained by replacing $\hat{\Psi}$ by $\Psi_0$, and
\begin{eqnarray}
     \nonumber K'_V&=&\frac{1}{2}\int\mathrm{d}\mathbf{x}~[\hat{\varphi}^\dagger,N_0\hat{\varphi}+A_0\hat{\varphi}^\dagger]=-\frac{1}{2}\sum_{n}\epsilon_n\braket{z_n|z_n},\\
     K_V&=&K'_V+\frac{1}{2}\sum_{n}\epsilon_n=-\sum_{n}\int\mathrm{d}\mathbf{x}~\epsilon_n|v_n|^2 %\int\mathrm{d}\mathbf{x}~\epsilon_n(|u_n(\mathbf{x})|^2+|v_n(\mathbf{x})|^2)
\end{eqnarray}
are $c$-number contributions arising from the zero-point motion of the quasiparticles. Notice that the grand-canonical Hamiltonian $\hat{K}$ is the one governing the dynamics instead of $\hat{H}$, since we have extracted the global phase $e^{-i\mu t/\hbar}$ in Eq. (\ref{eq:StationaryFieldExpansion}). In fact, the time-independent GP equation (\ref{eq:TIGP}) is precisely the condition for $\Psi_0$ to be an extreme of the grand-canonical energy $K$, which leads to the identification of the non-linear GP eigenvalue as the chemical potential $\mu$, and to the absence of linear terms in the field fluctuations in Eq. (\ref{eq:GrandCanonicalEnergy}). The precise nature of the extreme is obtained by considering small fluctuations of the stationary GP wavefunction: 
\begin{eqnarray}\label{eq:energeticstability}
\nonumber    \delta K&\equiv& K[\Psi'_0]-K[\Psi_0]\simeq \frac{1}{2}(\Phi|M_0\Phi)=\frac{1}{2}\braket{\Phi|\Lambda|\Phi},\\
\Psi'_0(\mathbf{x})&=&\Psi_0(\mathbf{x})+\varphi(\mathbf{x}),~~ \Phi=\left[\begin{array}{c}\varphi\\ \varphi^*\end{array}\right].
\end{eqnarray} %$\Psi_0(\mathbf{x})\to \Psi'_0(\mathbf{x})=\Psi_0(\mathbf{x})+\varphi(\mathbf{x})$, which results in
If $\Lambda$ is a positive-definite operator, then $\Psi_0$ is a minimum and the system is energetically stable. In that case,
\begin{equation}\label{eq:energeticdynamical}
    \braket{z_n|\Lambda|z_n}=(z_n|M_0z_n)=\epsilon_n(z_n|z_n)>0,
\end{equation}
so all energies are positive, $\epsilon_n>0$, and the ground state is the quasiparticle vacuum  $\hat{a}_{n}\ket{0}=0$. If $\Lambda$ is not positive definite, we can have negative-energy modes, denoted as anomalous, while positive-energy modes are denoted as normal. As a result, the system is energetically unstable.

In general, the quantum state $\hat{\rho}$ of an ensemble of bosons at thermal equilibrium at a temperature $T$ is 
\begin{equation}
    \hat{\rho}=\frac{e^{-\beta \hat{K}}}{Z},
\end{equation}
with $\beta=1/k_B T$ and $Z=\textrm{Tr}(e^{-\beta \hat{K}})$ the partition function. Within the Bogoliubov approximation, this leads in an energetically stable condensate to a thermal Planckian distribution for the quasiparticle occupation number:
\begin{equation}\label{eq:PlanckianQuasiparticle}
\braket{\hat{a}^{\dagger}_{n}\hat{a}_{m}}=\textrm{Tr}(\hat{a}^{\dagger}_{n}\hat{a}_{m}\hat{\rho})=\frac{\delta_{nm}}{e^{\beta\epsilon_n}-1}.
\end{equation}

\subsection{Gravitational analogy}

We now review how the original gravitational analogy \cite{Unruh1981} was established using a fluid flow. Specifically, we consider the Euler equations for an ideal irrotational barotropic flow,
\begin{eqnarray}\label{eq:EulerVelocity}
    0&=&\partial_t \rho+\boldsymbol{\nabla}\cdot\mathbf{J},~\mathbf{J}= \rho \mathbf{v},\\
    \nonumber  [\partial_t +\mathbf{v}\cdot\boldsymbol{\nabla}]\mathbf{v} &\equiv& D_t\mathbf{v}=-\frac{1}{m}\boldsymbol{\nabla}V-\frac{1}{\rho}\boldsymbol{\nabla}P.
\end{eqnarray}
where $\rho$ is the mass density of the fluid, $\mathbf{J}$ is the current, $V$ is some external potential (e.g., gravity), $D_t$ is the total derivative, and $P$ is the local pressure. The irrotationality condition $\boldsymbol{\nabla}\times\mathbf{v}=0$ implies that the flow is potential, i.e., $\mathbf{v}=\boldsymbol{\nabla}\phi$. By invoking the barotropic condition, $P=P(\rho)$, we can arrive at a simplified equation for the flow potential $\phi$,
\begin{eqnarray}\label{eq:EulerPotential}
    0&=&\partial_t \rho+\boldsymbol{\nabla}\cdot(\rho \boldsymbol{\nabla}\phi),\\
    \nonumber  \partial_t\phi &=& -\frac{|\boldsymbol{\nabla}\phi|^2}{2}-\frac{V}{m}-h(\rho),~\frac{dh}{d\rho}=\frac{1}{\rho}\frac{dP}{d\rho}.
\end{eqnarray}
In the usual case of an isentropic flow, $h(\rho)$ is the specific enthalpy. The gravitational analogy emerges when considering small fluctuations $\delta\rho,\,\delta \phi$ of the density and the flow potential around a certain background solution characterized by $\rho,\,\phi$. Specifically, after expanding up to linear order, we get
\begin{eqnarray}\label{eq:EulerPotentialLinear}
    D_t\frac{\delta\rho}{\rho} &=&-\frac{1}{\rho}\boldsymbol{\nabla}\cdot(\rho\boldsymbol{\nabla}\delta\phi),\\
    \nonumber  D_t\delta \phi &=& -c^2\frac{\delta \rho}{\rho},~c^2\equiv \frac{dP}{d\rho},
\end{eqnarray}
$c$ being the local speed of sound. By combining both equations, one arrives at a single equation for the flow potential fluctuations that can be rewritten as
\begin{equation}
    \square \delta \phi=\nabla_\mu \nabla^\mu  \delta \phi=\frac{1}{\sqrt{-g}}\partial_\mu(\sqrt{-g}g^{\mu\nu}\partial_\nu\delta \phi)=0,
\end{equation}
which is precisely the covariant form of the KG equation (\ref{eq:KleinGordon}) for a massless scalar field in a curved spacetime described by the metric
\begin{equation}\label{eq:RelativisticMetric}
g_{\mu\nu}(x)=\frac{\rho(x)}{c(x)}\left[\begin{array}{cc}-[c^2(x)-v^2(x)]& -\mathbf{v}^T(x) \\
-\mathbf{v}(x)& \delta_{ij}\\
\end{array}\right],~x\equiv(t,\mathbf{x}),
\end{equation}
whose line element simply reads
\begin{equation}\label{eq:LineElement}
ds^2=\frac{\rho(x)}{c(x)}\left[-c^2(x)dt^2+|d\mathbf{x}-\mathbf{v}(x)dt|^2\right].
\end{equation}
The metric $g_{\mu\nu}$ is known as the hydrodynamic (or acoustic) metric, and parametrizes a whole class of metrics. Thus, we can use fluid flows, accessible to us in the laboratory, to study gravitational phenomena that can be mimicked by an acoustic metric. Specifically, we can address the physics of black holes since the acoustic metric presents horizons at the subsonic/supersonic interfaces, where $v(x)=c(x)$, denoted as acoustic horizons. In fact, the Schwarzschild metric can be rewritten using the Gullstrand-Painlev\'e coordinates as a stationary acoustic metric with
\begin{equation}
    c(\mathbf{x})=c,~\mathbf{v}(\mathbf{x})=c \sqrt{\frac{r_S}{r}}\frac{\mathbf{x}}{r},~r_S=\frac{2GM}{c^2},
\end{equation}
where $r_S$ is the Schwarzschild radius. Thus, the exterior/interior of a black hole is analogous to the subsonic/supersonic regions of a flowing fluid. A more thorough discussion about ergoregions and horizons in acoustic metrics is presented in Ref. \cite{Visser1998}.

In condensates, the gravitational analogy emerges within the Bogoliubov formalism in the so-called hydrodynamic approximation. For that purpose, we invoke the Madelung decomposition of the GP wavefunction,
\begin{equation}
\Psi(\mathbf{x},t)=\sqrt{n(\mathbf{x},t)}e^{i\theta(\mathbf{x},t)},
\end{equation}
which leads to a pair of hydrodynamic equations after rewriting the time-dependent GP equation (\ref{eq:TDGP}) in terms of the condensate phase and density:
\begin{eqnarray}\label{eq:EulerPhaseCondensate}
    0&=&\partial_t n+\boldsymbol{\nabla}\cdot(n \mathbf{v}),~\mathbf{v}=\frac{\hbar\boldsymbol{\nabla}\theta}{m},\\
    \nonumber\frac{\hbar \partial_t \theta}{m} &=&\frac{\hbar^2}{2m^2\sqrt{n}}\nabla^2\sqrt{n}-\frac{v^2}{2}-\frac{V}{m}-\frac{gn}{m}.
\end{eqnarray}
The first line is a continuity equation, from where we identify the particle current  
\begin{equation}\label{eq:SchrodingerCurrent}
    \mathbf{J}=n\mathbf{v}=-\frac{i\hbar}{2m}\left[\Psi^*\boldsymbol{\nabla} \Psi-\Psi\boldsymbol{\nabla}\Psi^*\right]
\end{equation}
and the flow velocity $\mathbf{v}$; notice that $n=|\Psi|^2$ is the particle density of the condensate, related to the mass density as $\rho=n\cdot m$. Since $\mathbf{v}$ is proportional to the gradient of the phase, the resulting flow is irrotational, with a flow potential $\phi=\hbar \theta/m$. The second line provides the dynamics for the potential flow, from where we can identify the local pressure as
\begin{equation}
    h=\frac{gn}{m}\Longrightarrow P=\frac{gn^2}{2}.
\end{equation}
The only genuine quantum-mechanical term involving $\hbar$ in Eq. (\ref{eq:EulerPhaseCondensate}) is the so-called quantum potential
\begin{equation}
    Q\equiv -\frac{\hbar^2}{2m\sqrt{n}}\nabla^2\sqrt{n}.
\end{equation}
Hence, in the hydrodynamic approximation, where $Q$ is negligible, the GP equation reduces to the Euler equation for an ideal irrotational barotropic flow, from where the gravitational analogy is retrieved, as originally shown in Ref. \cite{Garay2000}. 

Further insight on the quantum aspects of the gravitational analogy can be obtained from the time-dependent BdG equations (\ref{eq:TDBdG}). By using the relative quantum fluctuations, $\hat{\varphi}(\mathbf{x},t)\equiv\Psi(\mathbf{x},t)\hat{\chi}(\mathbf{x},t)$, we arrive at 
\begin{equation}\label{eq:BdGfieldequationTIHydrodynamic}
i\hbar D_t\hat{\chi}=\left[T_n+mc^2\right]\hat{\chi}+mc^2\hat{\chi}^{\dagger},~T_n\equiv -\frac{\hbar^2}{2mn }\boldsymbol{\nabla}\cdot(n\boldsymbol{\nabla}),
\end{equation}% \\T_n&\equiv &-\frac{\hbar^2}{2mn }\nabla (n\nabla)
where the speed of sound is simply found to be $c^2=h=gn/m$. The relative quantum fluctuations can be in turn expressed in terms of the more physical density and phase fluctuations from 
\begin{eqnarray}
   \nonumber \hat{\Psi}(\mathbf{x},t)&=&\Psi(\mathbf{x},t)+\hat{\varphi}(\mathbf{x},t)=\Psi(\mathbf{x},t)[1+\hat{\chi}(\mathbf{x},t)]\\
   &=&\sqrt{n(\mathbf{x},t)+\delta\hat{n}(\mathbf{x},t)}e^{i[\theta(\mathbf{x},t)+\delta\hat{\theta}(\mathbf{x},t)]}.
\end{eqnarray}
Expanding up to linear order in the density and phase fluctuations yields
\begin{eqnarray}\label{eq:BdGfieldequationTIHydrodynamic2}
\delta \hat{n}(\mathbf{x},t)&=&n(\mathbf{x},t)\left[\hat{\chi}(\mathbf{x},t)+\hat{\chi}^{\dagger}(\mathbf{x},t)\right],\\
\nonumber \delta \hat{\theta}(\mathbf{x},t)&=&-\frac{i}{2}\left[\hat{\chi}(\mathbf{x},t)-\hat{\chi}^{\dagger}(\mathbf{x},t)\right].
\end{eqnarray}
We can then rewrite the BdG equations (\ref{eq:BdGfieldequationTIHydrodynamic}) as
\begin{eqnarray}\label{eq:BdGfieldequationTIHydrodynamicPhase}
D_t\frac{\delta \hat{n}}{n}&=&-\frac{1}{n}\boldsymbol{\nabla}\cdot\left[n\boldsymbol{\nabla}\frac{\hbar\delta \hat{\theta}}{m}\right],\\
\nonumber D_t\frac{\hbar \delta \hat{\theta}}{m}&=&-\left[\frac{T_n}{2m}+c^2\right]\frac{\delta \hat{n}}{n}.
\end{eqnarray}
So far, within the Bogoliubov approximation, these equations are \textit{exact}, resulting from a change of variables $\{\hat{\varphi},\hat{\varphi}^\dagger\}\to\{\delta\hat{n},\delta\hat{\theta}\}$. Now, if we assume that we are in the Thomas-Fermi regime, where the background condensate density smoothly varies on a sufficiently large length scale, in the long-wavelength limit we can neglect the contribution of $T_n$ at the r.h.s. of the second line. This precisely amounts to work in the hydrodynamic approximation, where all the contributions from the quantum potential are neglected, and we retrieve the quantum version of Eq. (\ref{eq:EulerPotentialLinear}), from where we find that $\square\delta\hat{\theta}=0$.

Therefore, in condensates, the gravitational analogy emerges in the hydrodynamic approximation (equivalent to work in the long-wavelength limit of the BdG equations above a condensate within the Thomas-Fermi regime) as an equation of motion for the phase fluctuations which mimics that of a massless scalar field in a curved spacetime described by an acoustic metric (\ref{eq:RelativisticMetric}).

\subsection{Microscopic Hawking effect}

\begin{figure*}[t]
\begin{tabular}{@{}ccc@{}}\bigskip\bigskip \stackinset{l}{0pt}{t}{0pt}{(a)}{\includegraphics[width=0.33\textwidth]{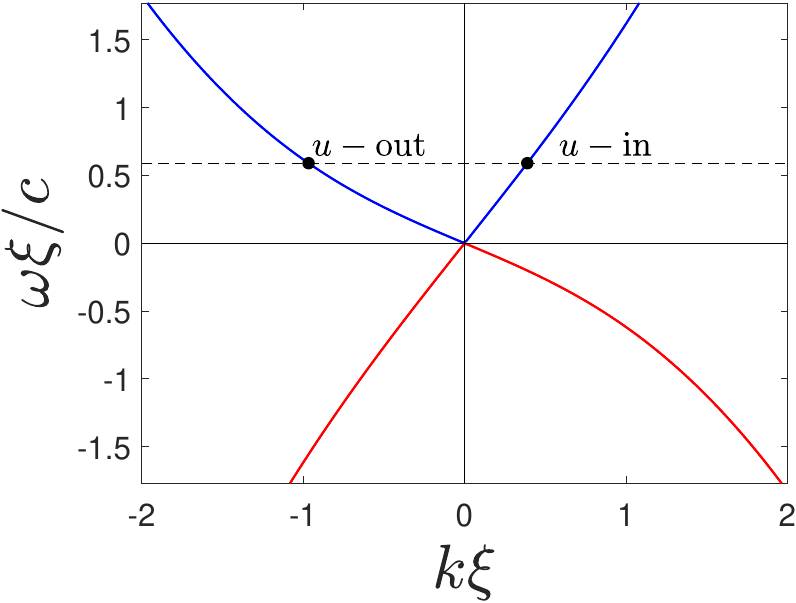}}&
    \stackinset{l}{0pt}{t}{0pt}{(b)}
{\includegraphics[width=0.33\textwidth]{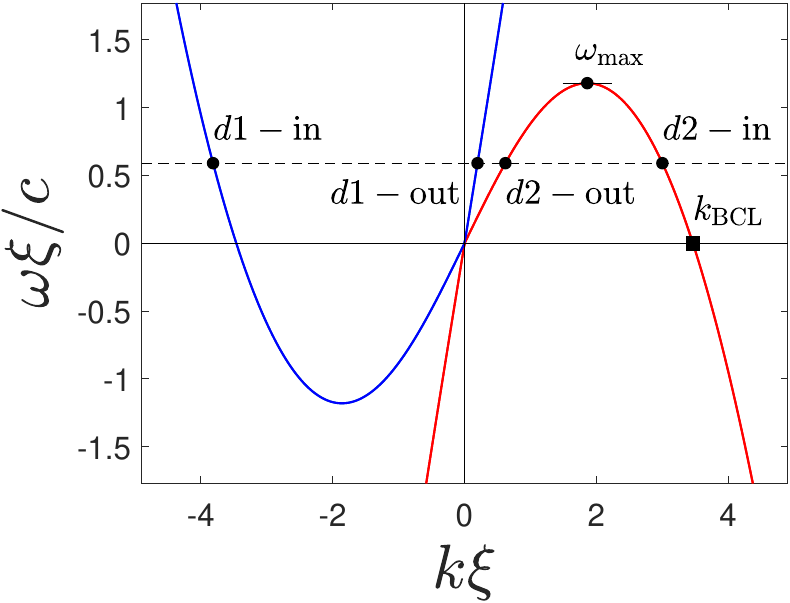}} & 
    \stackinset{l}{0pt}{t}{0pt}{(c)}
{\includegraphics[width=0.32\textwidth]{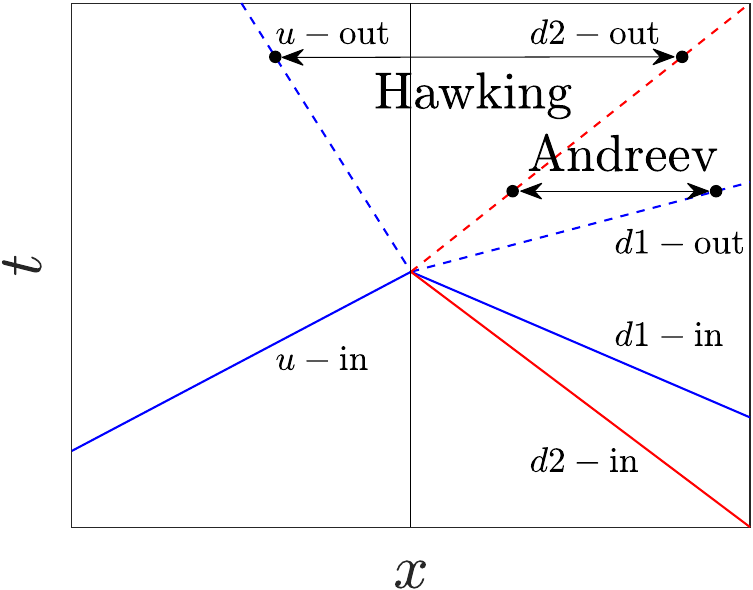}} 
\end{tabular}
\caption{Dispersion relation of a homogeneous flowing condensate. The blue/red lines signal the $\pm$ branches of Eq. (\ref{eq:Dispersion}). a) Subsonic regime with Mach number $M=v/c=0.5$. b) Supersonic regime with Mach number $M=v/c=2$. For a certain frequency below the cutoff frequency $\omega_{\rm{max}}$, all wavevectors are purely real (horizontal dashed line). The BCL mode has zero frequency and finite wavevector $k_{\rm{BCL}}$ (black square). c) Schematic spacetime diagram of the modes in a)-b) when placed in an analogue black hole, where the upstream/downstream region is subsonic/supersonic. The position of the acoustic horizon is signaled by a vertical solid black line. Incoming/outgoing modes are depicted as solid/dashed lines. Horizontal arrows indicate the characteristic correlations of both the Hawking and Andreev effects.}
\label{fig:DispersionRelation}
\end{figure*}

Due to the low temperature and genuine quantumness of Bose-Einstein condensates, the gravitational analogy allows to study there the Hawking effect, which is translated into the spontaneous emission of phonon radiation by an acoustic horizon. In a pair of seminal works, it was shown by Macher and Parentani \cite{Macher2009a}, and by Recati, Pavloff and Carusotto \cite{Recati2009}, that the Hawking effect can be studied within the full microscopic Bogoliubov framework without the need of invoking the hydrodynamic approximation or even any metric at all.

%black (white) holes are described by flows undergoing a subsonic/supersonic (supersonic/subsonic) transition at the 

We now derive the Hawking effect from a microscopic approach along the lines of Refs. \cite{Recati2009,Macher2009a}. For simplicity, hereafter we focus on one-dimensional (1D) condensates, where $x$ will label the 1D spatial coordinate. Acoustic horizons are then reduced to discrete points where the flow undergoes a subsonic/supersonic transition. 

We start by considering a stationary GP plane-wave solution
\begin{equation}
\Psi_0(x)=\sqrt{n}e^{i(qx+\theta_0)}.
\end{equation}
The associated BdG spectrum, resulting from Eq. (\ref{eq:BdGEigenmode}), is also described by plane waves with wavevector $k$ and energy $\epsilon=\hbar\omega$, as given by the dispersion relation
\begin{equation}\label{eq:dispersionrelation}
\left(\omega-vk\right)^{2}=\Omega^2(k)=c^{2}k^{2}+\frac{\hbar^2k^{4}}{4m^2}=c^{2}k^{2}\left[1+\frac{(k\xi)^2}{4}\right],
\end{equation}
where 
\begin{equation}
    c=\sqrt{\frac{gn}{m}},~v=\frac{\hbar q}{m}
\end{equation}
are the homogeneous sound and flow speeds. This is nothing else than the usual Bogoliubov dispersion relation $\Omega(k)$ for a homogeneous condensate at equilibrium plus a Doppler shift $\omega\to\omega-vk$, resulting from the background condensate flow, which tilts the sound cones. Remarkably, the Bogoliubov dispersion relation is superluminal, with the healing length $\xi\equiv \hbar/mc$ playing the role of a Planck length scale that controls the UV physics.

For given $\omega$, the dispersion relation (\ref{eq:dispersionrelation}) provides $4$ wavevectors, labeled as $k_a(\omega)$ and given by the roots of the fourth order polynomial, which can be either real (describing propagating solutions) or complex (describing exponentially growing/decaying solutions). The corresponding BdG spinor for each wavevector $k_a$ reads
\begin{eqnarray} \label{eq:PlaneWaveSpinors}
s_{a,\omega}(x) & = & \frac{e^{ik_{a}\left(\omega\right)x}}{\sqrt{2\pi|w_{a}\left(\omega\right)|}}\left[\begin{array}{c}
e^{i(qx+\theta_0)}u_{a}(\omega)\\
e^{-i(qx+\theta_0)}v_{a}(\omega)
\end{array}\right] \nonumber\\
\left[\begin{array}{c}
\nonumber u_{a}(\omega)\\
v_{a}(\omega)
\end{array}\right]&=&N_a(\omega)\left[\begin{array}{c}
\frac{\hbar k_{a}^{2}\left(\omega\right)}{2m}+[\omega-vk_{a}\left(\omega\right)]\\
\frac{\hbar k_{a}^{2}\left(\omega\right)}{2m}-[\omega-vk_{a}\left(\omega\right)]
\end{array}\right]\\
N_a(\omega)&=&\left(\frac{m}
{2\hbar k_{a}^{2}\left(\omega\right)\left|\omega-vk_{a}\left(\omega\right)\right|}\right)^{\frac{1}{2}},
\end{eqnarray}
with $u_a(\omega),v_a(\omega)$ the usual Bogoliubov components for a homogeneous condensate, satisfying $|u_{a}(\omega)|^2-|v_{a}(\omega)|^2=\pm 1$, and $w_{a}\left(\omega\right)\equiv\left[dk_{a}\left(\omega\right)/d\omega\right]^{-1}$ the group velocity, included here in order to normalize the propagating modes in frequency domain, $(s_{a,\omega}|s_{a,\omega'})=\pm\delta\left(\omega-\omega'\right)$; all normalization factors can be removed for complex wavevector solutions, where they do not play any role. It is easy to check that the dispersion relation (\ref{eq:dispersionrelation}) possess the symmetry $k_{a}(\omega)=-k_a(-\omega)$, which implies $\bar{s}_{a,\omega}=s_{a,-\omega}$. Hence, the $\pm$ branches of 
\begin{equation}\label{eq:Dispersion}
    \omega(k)=vk\pm \Omega(k), 
\end{equation}
depicted in blue (red) in Fig. \ref{fig:DispersionRelation}, respectively, are conjugate of each other, with the $\pm$ sign also corresponding to the norm of the modes.

% corresponds to the 
%$\pm$ norm of the propagating modes

The dispersion relation displays two qualitatively different regimes, depending on whether the flow is subsonic ($v<c$) or supersonic ($v>c$). In the subsonic regime, Fig. \ref{fig:DispersionRelation}a, there are $2$ real wavevectors and $2$ complex ones. The propagating solutions are labeled as $u-\rm{in}$ and $u-\rm{out}$, where ``in'' (``out'') indicates if the group velocity is positive (negative); the motivation behind this notation will become clearer later. The flow is energetically stable since all modes have positive energy. In the supersonic regime, Fig. \ref{fig:DispersionRelation}b, for any frequency $-\omega_{\rm{max}}<\omega<\omega_{\rm{max}}$, there are $4$ different propagating solutions, where the cutoff frequency $\omega_{\rm{max}}$ is given by  
\begin{eqnarray}\label{eq:CutoffFrequency}
    \omega_{\rm{max}}&=&vk_{\rm{max}}-\Omega(k_{\rm{max}}),\\
    \nonumber \left(k_{\rm{max}}\xi\right)^2&=&\frac{M^2-4+M\sqrt{M^2+8}}{2},
\end{eqnarray}
$M=v/c$ being the Mach number, with $M>1$ for a supersonic flow. The ``in'' (``out'') label is reverted here and now indicates if the group velocity is negative (positive). The $d1$ modes are those normal, with positive energy, while the $d2$ modes are those anomalous, with negative energy. The presence of the anomalous $d2$ modes reveals the energetic instability of a supersonic flow. This is a consequence of the Landau criterion for superfluidity, which predicts the appearance of a zero-frequency mode in a supersonic flow, the celebrated Bogoliubov-Cherenkov-Landau (BCL) mode, with a finite wavevector $\hbar k_{\rm{BCL}}=2m\sqrt{v^2-c^2}$ computed by $\Omega(k_{\rm{BCL}})=vk_{\rm{BCL}}$ (black square in Fig. \ref{fig:DispersionRelation}b). This results in the coherent excitation of the BCL mode by the presence of any obstacle in a supersonic flow, spoiling its superfluidity \cite{Carusotto2006}; see Eq. (\ref{eq:BCLStimulation}) and ensuing discussion for more details. Nevertheless, supersonic flows are dynamically stable \cite{Mayoral2011}, since energetic instability is only a necessary condition for dynamical instability, but not sufficient \cite{Wu2003}. This is directly seen from Eq. (\ref{eq:energeticdynamical}): the presence of complex modes (dynamical instability), which have zero norm by virtue of Eq. (\ref{eq:EigenOrto}), necessarily implies that $\Lambda$ cannot be a positive-definite operator (energetic instability).

% arise from the $+$ branch in Eq. (\ref{eq:Dispersion}), while the $d2$ modes arise from the $-$ branch, being the conjugates of the negative energy modes in the $+$ branch.

We are now in a position to study a black-hole (BH) solution, defined here as a stationary 1D GP solution with two asymptotic homogeneous regions, one subsonic and one supersonic, flowing from subsonic to supersonic. When the flow travels from supersonic to subsonic, we have a white hole (WH) solution, the time reversal of a BH (obtained simply by conjugation of the wavefunction). Continuity of the GP wavefunction implies that both BH and WH solutions always possess, at least, one acoustic horizon, where $v(x)=c(x)$. We denote the region between the two asymptotic regions, in which the acoustic horizon is located, as the scattering region. The emergence of BH configurations is expected on quite general grounds, since it was shown by Michel, Parentani and Zegers \cite{Michel2016} that stationary asymptotically uniform flows are attractor solutions, providing an analogue version of the celebrated no-hair theorem.

By convention, for BH solutions, we take the flow velocity always positive, so the upstream subsonic region (labeled as ``u") is located  at $x\rightarrow-\infty$, while the downstream supersonic region (labeled as ``d") is located at $x\rightarrow\infty$. This convention hence matches the notation previously introduced in Fig. \ref{fig:DispersionRelation}, since ``in" modes are incoming (traveling towards the horizon), and ``out" modes are outgoing (traveling outwards). The asymptotic flow velocity, sound speed, and healing length, are labeled as $v_{u,d},c_{u,d},\xi_{u,d}$. 
% , so the asymptotic spinors  %  which accordingly has negative norm.

The BdG modes of a BH solution are asymptotically given in terms of linear combinations of the plane-wave spinors $s_{i-\rm{in/out},\omega}$ of Eq. (\ref{eq:PlaneWaveSpinors}), $i=u,d1,d2$, representing the different incoming and outgoing scattering channels. Throughout this work, we operate just with positive frequencies $\omega>0$, and the remaining part of the spectrum is obtained by conjugation. In particular, the scattering problem for any positive frequency $0<\omega<\omega_{{\rm max}}$ involves the normal $u,d1$ channels, and the conjugate of the anomalous $d2$ channel (horizontal dashed line in Fig. \ref{fig:DispersionRelation}), so $s_{d2-\rm{in/out},\omega}$ has negative norm. The retarded (``in") scattering states $z^{(+)}_{i,\omega}$ are global eigenmodes of the stationary BdG equations (\ref{eq:BdGEigenmode}) with positive frequency, presenting unit amplitude in the asymptotic incoming channel $i$ and zero in the other incoming channels. The  amplitude of the asymptotic ``out" scattering channels is determined by the $S$-matrix, as usual in scattering theory. For example, the scattering state $z^{(+)}_{d2,\omega}$ asymptotically reads
\begin{align}\label{eq:scatteringchannelstate}
&z^{(+)}_{d2,\omega}\left(x\rightarrow-\infty\right)=S_{ud2}\left(\omega\right)s_{u-\rm{out},\omega}(x),\\
\nonumber &z^{(+)}_{d2,\omega}\left(x\rightarrow\infty\right)=s_{d2-\rm{in},\omega}(x)+S_{d1d2}\left(\omega\right)s_{d1-\rm{out},\omega}(x)\\
\nonumber &+S_{d2d2}\left(\omega\right)s_{d2-\rm{out},\omega}(x).
\end{align}
Similar expressions can be provided for the remaining ``in" scattering states. The advanced (``out") scattering states $z^{(-)}_{i,\omega}(x)$ are the outgoing analogues of the ``in" states, having unit amplitude in the outgoing channel $i$ and zero in the other outgoing channels. They are characterized by the inverse of the scattering matrix $S(\omega)$,
\begin{equation}
    z^{(-)}_{i,\omega}(x)=\sum_{j=u,d1,d2} S^{-1}_{ji}(\omega)z^{(+)}_{j,\omega}(x).
\end{equation}
A schematic spacetime diagram of these modes is presented in Fig. \ref{fig:DispersionRelation}c. Retarded scattering states are characterized by one incoming channel (solid lines), which is eventually scattered into the outgoing channels (dashed lines). In contrast, advanced scattering states are labeled by just one outgoing channel, whose time-reversed trajectory scatters at the horizon into all incoming channels.

By invoking the conservation of the quasiparticle current (\ref{eq:QuasiparticleCurrent}) for an arbitrary linear combination of ``in" scattering states, it is shown that the $S$-matrix is pseudo-unitary, i.e.,
\begin{equation}\label{eq:pseudounitarity}
S^{\dagger}\eta S=\eta\equiv{\rm diag}(1,1,-1).
\end{equation}
Thus, $S\in U(2,1)$, which implies $S^{-1}=\eta S^\dagger \eta$. A direct consequence of the pseudounitarity of $S$ is that the scattering states are orthonormal,
\begin{equation}
(z^{(+)}_{i,\omega}|z^{(+)}_{j,\omega'})=(z^{(-)}_{i,\omega}|z^{(-)}_{j,\omega'})=\eta_{ij}\delta(\omega-\omega').
\end{equation}
% \begin{eqnarray}
% \nonumber (z^{(+)}_{i,\omega}|z^{(+)}_{j,\omega'})&=&(z^{(-)}_{i,\omega}|z^{(-)}_{j,\omega'})=\eta_{ij}\delta(\omega-\omega'),\\ %(z^{(-)}_{i,\omega}|z^{(+)}_{j,\omega'})&=&S_{ij}(\omega)\delta(\omega-\omega').
% \end{eqnarray}

Since they form a complete orthonormal basis, the quantum fluctuations of the field operator can be expanded in terms of the scattering states as
\begin{align}\label{eq:BHFieldOperator}
&\nonumber \hat{\Phi}(x) = \displaystyle\sum_{I=u,d1}\int_{0}^{\infty}d\omega\,[z^{(+)}_{I,\omega}(x)\hat{a}_{I}(\omega)+\bar{z}^{(+)}_{I,\omega}(x)\hat{a}_{I}^{\dag}(\omega)]\\
&+\int_{0}^{\omega_{{\rm max}}}d\omega\,[z^{(+)}_{d2,\omega}(x)\hat{a}_{d2}^{\dagger}(\omega)+\bar{z}^{(+)}_{d2,\omega}(x)\hat{a}_{d2}(\omega)].
\end{align}
A similar expression can be written using the ``out'' scattering states after replacing $z^{(+)}_{i,\omega}(x)$ by $z^{(-)}_{i,\omega}(x)$, and the ``in'' quantum amplitudes $\hat{a}_i(\omega)$ by the ``out'' ones $\hat{b}_i(\omega)$, which are related through the scattering matrix as
\begin{equation}\label{eq:inoutmodesrelation}
\left[\begin{array}{c}
\hat{b}_{u}\\
\hat{b}_{d1}\\
\hat{b}_{d2}^{\dagger}
\end{array}\right] = \left[\begin{array}{ccc}S_{uu}&S_{ud1}&S_{ud2}\\
S_{d1u}&S_{d1d1}&S_{d1d2}\\
S_{d2u}&S_{d2d1}&S_{d2d2}\end{array}\right]\left[\begin{array}{c}
\hat{a}_{u}\\
\hat{a}_{d1}\\
\hat{a}_{d2}^{\dagger}
\end{array}\right]\, .
\end{equation}
This is a Bogoliubov relation, mixing annihilation with creation operators. It stems from the anomalous character of the $z^{(\pm)}_{d2,\omega}$ scattering states, which have a negative norm inherited from the corresponding anomalous scattering channels, and hence their amplitudes behave as creation instead of annihilation operators [see Eq. (\ref{eq:Aniquilacion}) and ensuing discussion]. In the following, we reserve lowercase Latin indices $i,j$ to label all channels, $i=u,d1,d2$, while uppercase Latin indices $I,J$ just label normal channels, $I=u,d1$. Lowercase Latin indices $a,b$ will label general BdG modes, either outgoing or incoming, either propagating or not. In addition, the normal-normal and anomalous-anomalous scattering processes (characterized by the $S$-matrix elements $S_{IJ},S_{d2d2}$) are denoted as normal, while the normal-anomalous and anomalous-normal scattering processes (characterized by the $S$-matrix elements $S_{d2I},S_{Id2}$) are denoted as anomalous.

The origin of the Hawking effect is the degeneracy of the vacuum of the Bogoliubov theory, revealed by the anomalous sector of $\hat{K}$, 
\begin{eqnarray}
   \nonumber \hat{K}_{H}&\equiv & \sum_{i,j}\int^{\omega_{\rm{max}}}_0\mathrm{d}\omega~\hbar\omega \hat{a}_{i}^{\dagger}(\omega)\eta_{ij} \hat{a}_{j}(\omega)\\&=&\sum_{i,j}\int^{\omega_{\rm{max}}}_0\mathrm{d}\omega~\hbar\omega \hat{b}_{i}^{\dagger}(\omega)\eta_{ij} \hat{b}_{j}(\omega).
\end{eqnarray}
Both the incoming vacuum $\hat{a}_{i}(\omega)\ket{0_{\rm{in}}}=0$ and the outgoing vacuum $\hat{b}_{i}(\omega)\ket{0_{\rm{out}}}=0$ satisfy $\hat{K}_{H}\ket{0_{\rm{in}}}=\hat{K}_{H}\ket{0_{\rm{out}}}=0$. However, they do not represent the same quantum state, as can be seen from the non-vanishing population of normal outgoing modes in the incoming vacuum,
\begin{equation}\label{eq:hawkingef}
\bra{0_{\rm{in}}}\hat{b}_{I}^{\dagger}(\omega)\hat{b}_{I}(\omega')\ket{0_{\rm{in}}}=\delta(\omega-\omega')|S_{Id2}(\omega)|^2\neq 0.
\end{equation}
The Hawking effect is recovered for $I=u$, representing a spontaneous outgoing flux of particles in the subsonic region (i.e., the exterior of the black hole) in the absence of incoming radiation. This emission is correlated with that of anomalous outgoing $d2$ modes into the supersonic region (horizontal arrow in Fig. \ref{fig:DispersionRelation}c), which are referred to as the partner modes of the Hawking effect. 

% where an electron (hole) incident on a normal/superconductor interface from the normal side is reflected as a hole (electron).

The case $I=d1$ is characterized by the anomalous reflection coefficient $S_{d1d2}$, representing the bosonic analogue of an incident hole that is reflected as a particle in the normal side of a normal/superconductor junction, the celebrated Andreev reflection. In analogy with the Hawking effect, throughout this work we will refer to the spontaneous emission of normal outgoing $d1$ modes into the supersonic region as the Andreev effect. Once again, this emission is correlated with that of anomalous outgoing $d2$ modes (horizontal arrow in Fig. \ref{fig:DispersionRelation}c). A discussion of the equivalence between the Andreev scattering picture and the spontaneous emission of quasiparticle pairs in the context of normal/superconductor interfaces was given in Refs. \cite{Samuelsson2003,Prada2004}. In general, by invoking the Ginzburg-Landau (GL) order parameter, which obeys a non-linear Schr\"odinger equation equivalent to the time-independent GP equation (\ref{eq:TIGP}), we can extend the analogy with superconductivity and identify normal metals with supersonic regions, and superconductors with subsonic regions. In fact, the physics of real black holes and the physics of superconductors are in close relationship \cite{Manikadan2017,Manikadan2020}.

%The case $I=d1$ characterizes the spontaneous version of the Andreev effect in superconductors \cite{Zapata2009a}, to which we will simply refer as the Andreev effect for brevity, consisting in the spontaneous emission of normal outgoing $d1$ modes into the supersonic region, also correlated with partner outgoing $d2$ modes (horizontal arrow in Fig. \ref{fig:DispersionRelation}c). Actually, by using the Ginzburg-Landau (GL) order parameter, which obeys a non-linear Schr\"odinger equation equivalent to the time-independent GP equation (\ref{eq:TIGP}), we can extend the analogy with superconductivity and identify normal metals with supersonic regions, and superconductors with subsonic regions. In fact, the physics of real black holes and the physics of superconductors are in close relationship \cite{Manikadan2017,Manikadan2020}.

\subsection{Analytical BH solutions}

We present here canonical analytical BH solutions in condensates. In the following, we set units and rescale the GP wavefunction as
\begin{equation}
    \hbar=m=c_u=k_B=1,~\Psi_0\to \sqrt{n_u}\Psi_0.
\end{equation}

The simplest BH solution is provided by the flat-profile model, originally introduced in Ref. \cite{Carusotto2008}, where the plane wave $\Psi_0(x)=e^{iqx}$ is a solution of the GP equation (\ref{eq:TDGP}) at all times since the coupling constant $g(x,t)$ and the external potential $V(x,t)$ are tuned in such a way that
\begin{equation}\label{eq:FlatProfileCondition}
    g(x,t)n_u+V(x,t)=E_b,
\end{equation}
with $E_b$ some constant energy that can be subtracted from the Hamiltonian. Nevertheless, even though the flow velocity is constant and homogeneous, $v(x,t)=q$, the BdG modes do experience non-trivial dynamics as the sound speed is $c^2(x,t)=g(x,t)n_u$; see Eq. (\ref{eq:BdGfieldequationTIHydrodynamicPhase}). 

\begin{figure*}[t]
\begin{tabular}{@{}ccc@{}}\stackinset{l}{0pt}{t}{0pt}{(a)}{\includegraphics[width=0.66\columnwidth]{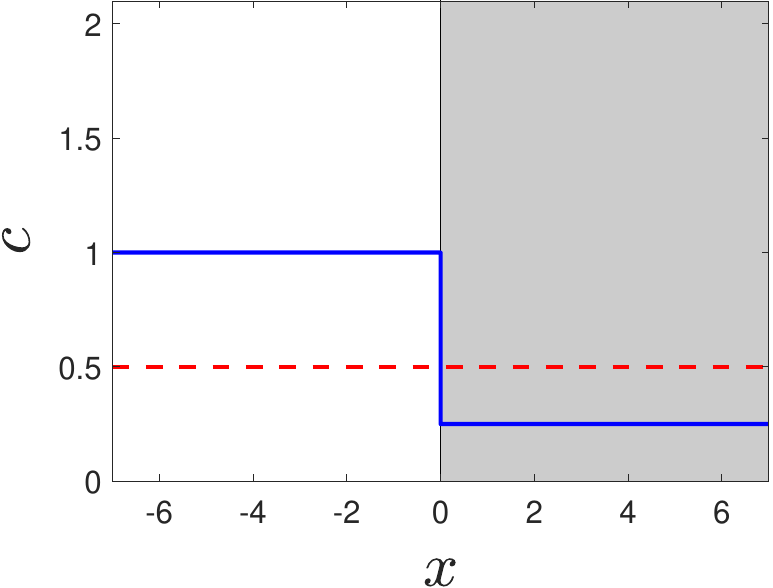}}~ &~
    \stackinset{l}{0pt}{t}{0pt}{(b)}
    {\includegraphics[width=0.66\columnwidth]{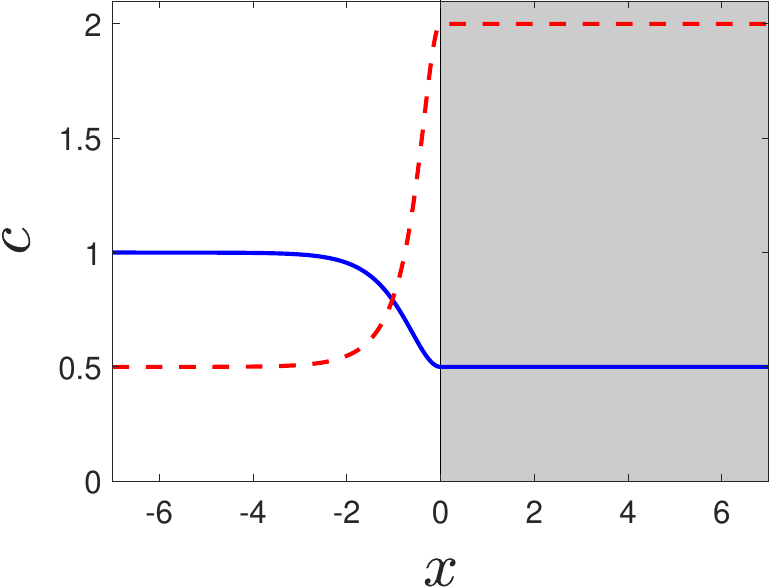}}~ &~ 
    \stackinset{l}{0pt}{t}{0pt}{(c)}{\includegraphics[width=0.66\columnwidth]{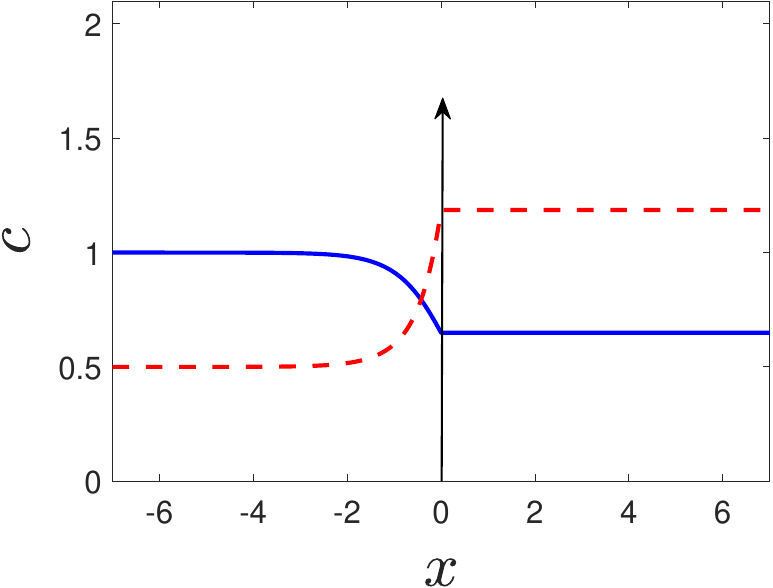}}  \\
    \stackinset{l}{0pt}{t}{0pt}{(d)}{\includegraphics[width=0.66\columnwidth]{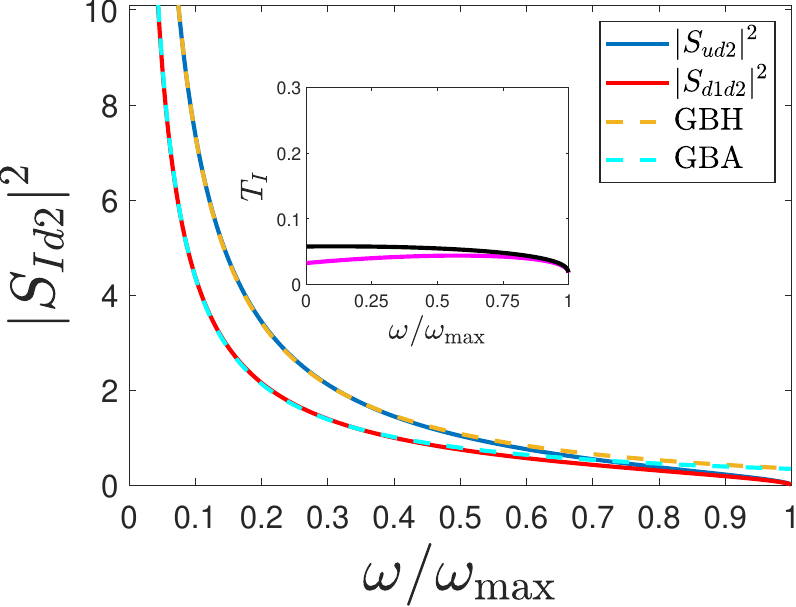}}~ &~
    \stackinset{l}{0pt}{t}{0pt}{(e)}
    {\includegraphics[width=0.66\columnwidth]{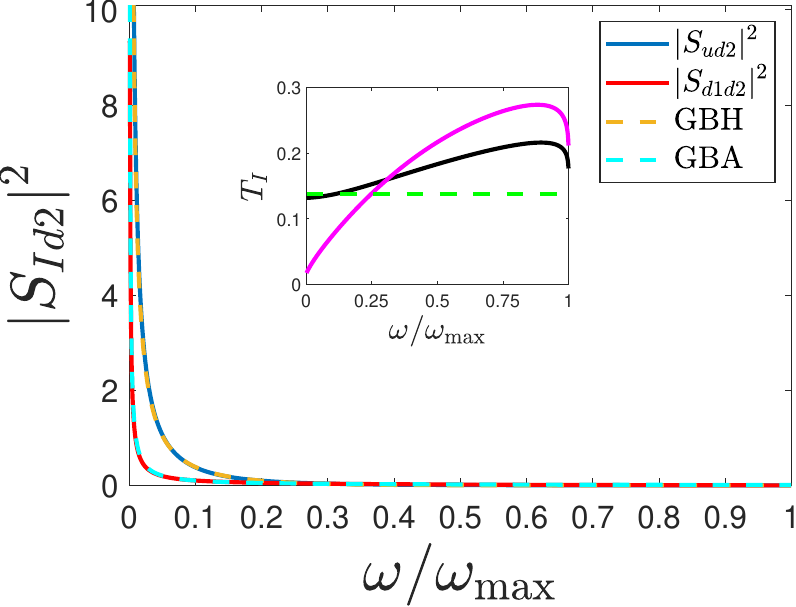}}~ &~ 
    \stackinset{l}{0pt}{t}{0pt}{(f)}{\includegraphics[width=0.66\columnwidth]{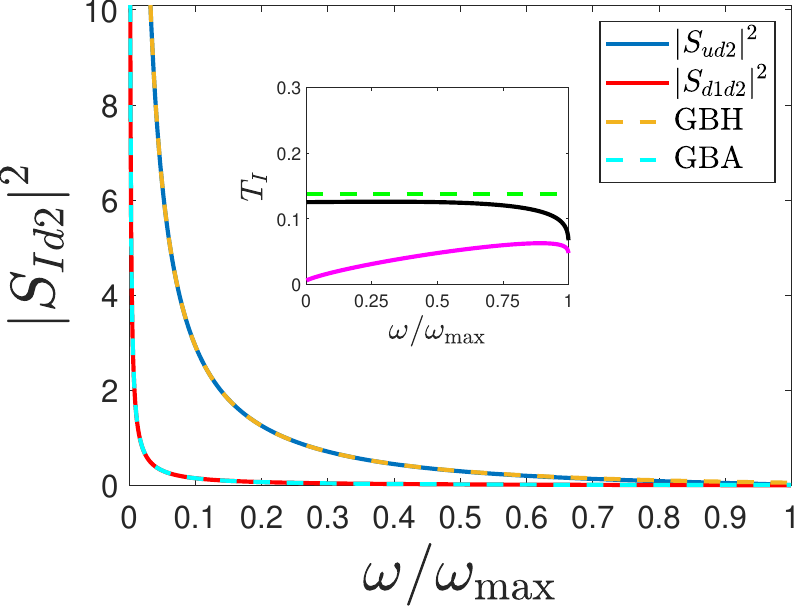}} 
\end{tabular}
\caption{Upper row: Sound (solid blue) and flow (dashed red) velocity profiles for different BH solutions with subsonic Mach number $q=0.5$. a) Flat profile. The shaded area indicates the supersonic region, where $c_2=0.25$. b) Waterfall potential. The shaded area indicates the region where the step potential $V(x)=-V_0\Theta(x)$ is present. c) Delta barrier. The arrow indicates the position of the delta potential $V(x)=Z\delta(x)$. Lower row: d)-f) Hawking (solid blue) and Andreev (solid red) spectra of the BH solutions above. Dashed gold (cyan) line is a gray-body fit of the Hawking (Andreev) spectrum, Eq. (\ref{eq:GrayBody}). Inset: Frequency-dependent Hawking (solid black) and Andreev (solid magenta) temperatures $T_{u,d1}(\omega)$, Eq. (\ref{eq:OmegaTemperature}). Horizontal dashed green line marks the predicted Hawking temperature for a soliton, Eq. (\ref{eq:SolitonHawking}).}
\label{fig:BHModels}
\end{figure*}

In particular, we can choose a time-independent piecewise homogeneous dependence for the coupling constant, $g(x,t)=g(x)$, where the BdG solutions within each homogeneous region are spanned by the plane-wave spinors (\ref{eq:PlaneWaveSpinors}), thus allowing for simple analytical calculations. Specifically, for the BH solution, we take
\begin{equation}\label{eq:FlatProfile}
g(x)n_u=\left\{ \begin{array}{cc}
1, & x< 0,\\
c_2^2, &  x \geq 0,
\end{array}\right.
\end{equation}
with $c_2<q<1$, $c_2$ being the supersonic speed of sound and $q$ representing the subsonic Mach number. Hence, we reach a BH solution whose acoustic horizon is placed at $x=0$, as depicted in Fig. \ref{fig:BHModels}a.

However, in practice, the flat-profile condition (\ref{eq:FlatProfileCondition}) is extremely challenging to implement experimentally. More realistic models of BH configurations only involve external potentials $V(x)$, which can be easily manipulated in the laboratory, leaving the interaction strength $g$ constant. Consequently, since the current $J$ is uniform for a stationary 1D solution as dictated by the continuity equation [first line of Eq. (\ref{eq:EulerPhaseCondensate})], we simply have that
\begin{equation}
    c(x)=\sqrt{n(x)},~v(x)=\frac{J}{n(x)}=\frac{J}{c^2(x)}.
\end{equation}
Typically, these BH solutions are described by a gray soliton in the upstream region
\begin{eqnarray}\label{eq:GraySoliton}
    \nonumber \Psi_{0}(x)&=&e^{i(qx+\theta_0)}\left[q+i\gamma_q(x-x_0)\right],\\
    \gamma_q(x)&\equiv & \sqrt{1-q^2}\tanh(\sqrt{1-q^2}x).
\end{eqnarray}
A gray soliton exponentially approaches a subsonic plane wave $\Psi_0(x)\xrightarrow[x\to\pm\infty]{} e^{iqx}$ since $\gamma_q(\pm \infty)=\pm \sqrt{1-q^2}$, with $q< 1$ the asympotic Mach number $M_u=q$. In our units, $q$ is also the minimum soliton amplitude as well as the value of the conserved current, $J=q$. In the downstream region, the BH solutions are given by the supersonic plane wave
\begin{equation}
    \Psi_0(x)=c_de^{iv_d x},~v_d=\frac{q}{c^2_d},
\end{equation}
with $c_d<v_d$ the corresponding supersonic sound and flow velocities; notice that $v_d$ is fixed by current conservation once $c_d$ is given.

For example, the waterfall configuration \cite{Larre2012} accelerates the atoms to supersonic speeds by means of an attractive step potential of the form
\begin{equation}\label{eq:waterfall}
V(x)=-V_0\Theta(x),~V_0=\frac{1}{2}\left(q^2+\frac{1}{q^2}\right)-1,
\end{equation}
where $\Theta(x)$ is the Heaviside function. This gives rise to a stationary GP wavefunction
\begin{equation}\label{eq:CompactBHWF}
\Psi_0(x)=\left\{ \begin{array}{cc}
e^{iqx}\left[q+i\gamma_q(x)\right], & x< 0,\\
qe^{i\frac{x}{q}}, &  x \geq 0,
\end{array}\right.
\end{equation}
which corresponds to half a gray soliton for $x<0$, and a homogeneous supersonic flow with $c_d=q<v_d=1/q$ for $x>0$, where the attractive potential is present. The resulting BH solution is represented in Fig. \ref{fig:BHModels}b. The waterfall configuration is quite relevant because it provides a simple theoretical model of the Technion experiment \cite{Lahav2010,Steinhauer2014,Steinhauer2016,deNova2019,Kolobov2021}.

Another possibility is to use a repulsive localized potential, which can be modeled by a delta barrier of the form 
\begin{equation}\label{eq:DeltaPotential}
    V(x)=Z\delta(x),~Z=\frac{(1-c^2_d)\sqrt{c^2_d-q^2}}{2c^2_d}.
\end{equation}
This delta potential introduces a discontinuity in the derivative of the GP wavefunction, $\Psi'_0(0^+)-\Psi'_0(0^-)=2Z\Psi_0(0)$. The resulting BH solution is similar to that of the waterfall model:
\begin{equation}\label{eq:CompactDelta}
\Psi_0(x)=\left\{ \begin{array}{cc}
e^{i(qx+\theta_0)}\left[q+i\gamma_q\left(x-x_0\right)\right] & x< 0,\\
c_de^{iv_dx}, &  x \geq 0,
\end{array}\right.
\end{equation}
where $x_0,\theta_0$ are such that the wavefunction is continuous and
\begin{equation}\label{eq:DeltaConstraints}
    c_d=\frac{\sqrt{q^2+\sqrt{q^4+8q^2}}}{2},~v_d=\frac{q}{c^2_d}.
\end{equation}
This BH solution is represented in Fig. \ref{fig:BHModels}c.

%The conservation of the 1D current from the continuity equation and of the chemical potential determines

Remarkably, the scattering states associated to these BH solutions can be also computed analytically because the stationary BdG solutions for a gray soliton (\ref{eq:GraySoliton}) are known (see Ref. \cite{Zapata2011} for the technical details). They take a similar form to the homogeneous solutions (\ref{eq:PlaneWaveSpinors}), namely
\begin{widetext}
\begin{eqnarray}\label{eq:solitonspinors}
\nonumber \zeta_{a,\omega}(x)&=&\frac{e^{ik_{a}(\omega)x}}{\sqrt{2\pi|w_{a}(\omega)|}}\left[\begin{array}{c}
e^{i(qx+\theta_0)}u_{a,\omega}(x)\\
e^{-i(qx+\theta_0)}v_{a,\omega}(x)
\end{array}\right], 
\\ \nonumber \left[\begin{array}{c}
u_{a,\omega}(x)\\
v_{a,\omega}(x)
\end{array}\right]&=&N_a(\omega)\left[\begin{array}{r}
\left(1+\dfrac{k_a(\omega)}{\omega}\left[\dfrac{k_a(\omega)}{2}+i\gamma_q(x-x_0)\right]\right)^2\\
-\left(1-\dfrac{k_a(\omega)}{\omega}\left[\dfrac{k_a(\omega)}{2}+i\gamma_q(x-x_0)\right]\right)^2
\end{array}\right],\\
N_a(\omega)&=&\frac{\omega}{\sqrt{8k^2_a(\omega)|\omega-vk_a(\omega)|}}.
\end{eqnarray}
\end{widetext}
In fact, $\zeta_{a,\omega}(x)\xrightarrow[x\to\pm\infty]{}s_{a,\omega}(x)$ as the soliton asymptotically approaches a subsonic plane wave $e^{iqx}$, whose dispersion relation yields the wavevectors $k_a(\omega)$. Thus, the scattering states $z^{(\pm)}_{i,\omega}$ for the waterfall (\ref{eq:CompactBHWF}) and delta (\ref{eq:CompactDelta}) BH solutions are obtained by matching at $x=0$ the upstream soliton spinors $\zeta_{a,\omega}$ with the corresponding downstream supersonic plane-wave spinors $s_{a,\omega}$; notice that in the subsonic region one also needs to include the evanescent solution $\zeta_{\rm{ev},\omega}$ with complex wavevector $k_{\rm{ev}}(\omega)$ which exponentially decays at $x\to-\infty$, $\textrm{Im}\,k_{\rm{ev}}(\omega)<0$. The flat-profile BH solution does not even involve the soliton spinors $\zeta_{a,\omega}(x)$ because the upstream region is also homogeneous, and the subsonic plane-wave spinors $s_{a,\omega}(x)$ are used instead.

The resulting anomalous scattering coefficients $|S_{ud2}|^2,|S_{d1d2}|^2$ characterizing the Hawking and Andreev effects for each BH solution are depicted in Figs. \ref{fig:BHModels}d-f as solid blue and red lines, respectively. We can compare these results, fully derived within the BdG microscopic framework, with the predictions from the gravitational analogy in the hydrodynamic limit \cite{Leonhardt2003a}
\begin{equation}\label{eq:Hydrodynamic}
|S_{ud2}(\omega)|^2=\frac{1}{e^{\frac{\omega}{T_H}}-1},~|S_{d1d2}(\omega)|^2=0,
\end{equation}
where $T_H$ is the Hawking temperature, given here by \cite{Visser1998}
\begin{equation}
    T_H=\frac{1}{2\pi}\left|c'(x_H)-v'(x_H)\right|,
\end{equation}
with $x_H$ the position of the acoustic horizon. Notice that, for the flat-profile configuration, this temperature is formally infinite due to the discontinuity of the sound speed. On the other hand, for the waterfall and delta configurations, it is easy to see that the acoustic horizon of a gray soliton (\ref{eq:GraySoliton}) is placed where $c(x)=v(x)=q^{\frac{1}{3}}$, yielding a predicted Hawking temperature
\begin{equation}\label{eq:SolitonHawking}
    T_H(q)=\frac{3}{2\pi}\left(1-q^{\frac{2}{3}}\right)\sqrt{1-q^{\frac{4}{3}}}\leq T_H(0)=\frac{3}{2\pi}<\frac{1}{2}.
\end{equation}

A necessary condition for the validity of the hydrodynamic approximation is that $T_H\ll \omega_{\rm{max}}$. It is easily seen from Eq. (\ref{eq:CutoffFrequency}) that $k_{\rm{max}}\leq v_d$ and that $\omega_{\rm{max}}\leq v^2_d/2$; these inequalities are saturated in the infinite supersonic Mach number limit, $c_d=0$. For the flat-profile configuration, this implies a relatively small cutoff frequency as $\omega_{\rm{max}}\leq q^2/2 < 1/2$. For the delta and waterfall configurations, we respectively have that 
\begin{equation}\label{eq:CutoffHawking}
    \omega_{\rm{max}}\leq \frac{8}{\left(q+\sqrt{q^2+8}\right)^2}<1,~\omega_{\rm{max}}\leq \frac{1}{2q^2}.
\end{equation}
% while for the waterfall configuration it is found that $\omega_{\rm{max}}\leq 1/2q^2$. 

As suggested in Ref. \cite{Larre2012}, we can check the agreement with the gravitational analogy by fitting the Hawking spectrum to a gray-body distribution. We also extend this fit to the Andreev spectrum as
\begin{equation}\label{eq:GrayBody}
|S_{Id2}(\omega)|^2=\frac{\Gamma_I}{e^{\frac{\omega}{T_I}}-1},
\end{equation}
% \begin{equation}\label{eq:GrayBody}
% |S_{ud2}(\omega)|^2=\frac{\Gamma_H}{e^{\frac{\omega}{T_H}}-1},~|S_{d1d2}(\omega)|^2=\frac{\Gamma_A}{e^{\frac{\omega}{T_A}}-1},
% \end{equation}
where $\Gamma_{I}$ is a gray-body factor and $T_{I}$ is the effective temperature of the spectrum. In the hydrodynamic limit, we can expect $\Gamma_u\simeq 1$ and $T_u\simeq T_H$. The result of the fit for the Hawking (Andreev) spectrum is depicted as a dashed gold (cyan) line in lower row of Fig. \ref{fig:BHModels}. Another way to quantify the Planckianity, proposed by Macher and Parentani \cite{Macher2009a}, is an $\omega$-dependent temperature, defined through
\begin{equation}\label{eq:OmegaTemperature}
|S_{Id2}(\omega)|^2\equiv \frac{1}{e^{\frac{\omega}{T_I(\omega)}}-1}.
\end{equation}
% \begin{equation}\label{eq:OmegaTemperature}
% |S_{ud2}(\omega)|^2\equiv \frac{1}{e^{\frac{\omega}{T_H(\omega)}}-1},~|S_{d1d2}(\omega)|^2\equiv \frac{1}{e^{\frac{\omega}{T_A(\omega)}}-1}.
% \end{equation}
$T_{u,d1}(\omega)$ is shown in the inset of lower Fig. \ref{fig:BHModels} as a solid black (magenta) line. In the Hawking case, we can compare the result with the predicted Hawking temperature (\ref{eq:SolitonHawking}), indicated as a horizontal dashed green line.

%between the Andreev and Hawking spectrum, and 

A gray-body distribution displays an excellent agreement with the Hawking and Andreev spectra in all cases, even for the flat-profile model (Fig. \ref{fig:BHModels}d), where the system is far from the hydrodynamic regime and the predicted Hawking temperature is formally infinite. The agreement is particularly good at low frequencies because, in general, the scattering coefficients $|S_{id1}(\omega)|^2,|S_{id2}(\omega)|^2$ display a universal scaling $\sim 1/\omega$ at low frequencies \cite{deNova2014} (the remaining column $|S_{iu}(\omega)|^2$ approaches a finite value in this limit). A gray-body distribution can reproduce exactly this behavior up to corrections $O(\omega)$,
\begin{equation}
    \frac{\Gamma}{e^{\frac{\omega}{T}}-1}\simeq \frac{\Gamma T}{\omega}-\frac{\Gamma}{2}+O(\omega).
\end{equation}

Regarding the effective temperatures $T_{I}(\omega)$, we observe that $T_u(\omega)$ depends very mildly on the frequency, approaching the predicted Hawking temperature in the low-frequency limit, while $T_{d1}(\omega)$ depends more strongly on the frequency. This is quite natural, since there is no analytical prediction of a Planckian distribution for the Andreev spectrum. The effective temperature $T_{d1}(\omega)$ also measures the strength of the Andreev effect, which satisfies $|S_{d1d2}(\omega)|^2\ll|S_{ud2}(\omega)|^2$ for both the waterfall and delta models, as expected from the gravitational prediction (\ref{eq:Hydrodynamic}), and $|S_{d1d2}(\omega)|^2\sim |S_{ud2}(\omega)|^2$ for the flat profile. This difference is due to the sharpness of the flat-profile horizon, far from the hydrodynamic limit. 

In the waterfall case, low subsonic Mach numbers imply large supersonic Mach numbers, so $\omega_{\rm{max}}\simeq 1/2q^2$ [see Eq. (\ref{eq:CutoffHawking})]. This large cutoff frequency makes dispersive effects important at intermediate frequencies $\omega \lesssim \omega_{\rm{max}}$, explaining the strong deviations of $T_u(\omega)$ from the predicted Hawking temperature as well as the dominance of the Andreev effect, $T_{d1}(\omega)>T_u(\omega)$ (inset of Fig. \ref{fig:BHModels}e). Nevertheless, the relevant part of the Hawking spectrum is still Planckian (solid blue and dashed gold lines in main Fig. \ref{fig:BHModels}e), with $T_u(\omega)\simeq T_H$, because it is restricted to the low-frequency dispersionless regime due to the smallness of the Hawking temperature, $T_H\ll \omega_{\rm{max}}$. 

Finally, close to $\omega_{\rm{max}}$, the Planckianity is necessarily spoiled for both the Hawking and Andreev spectra since there the anomalous coefficients $|S_{Id2}(\omega)|^2,|S_{d2I}(\omega)|^2$ vanish as a power law $\sim \sqrt{\omega_{\rm{max}}-\omega}$ \cite{deNova2014}. This is revealed by the departure from the gray-body fit, especially in the flat-profile case, and by the sudden drop of $T_{I}(\omega)$ in all configurations.

%On the other limit of the spectrum

\begin{figure*}[t]
\begin{tabular}{@{}cc@{}}\stackinset{l}{0pt}{t}{0pt}{\large{(a)}}{\includegraphics[width=0.98\columnwidth]{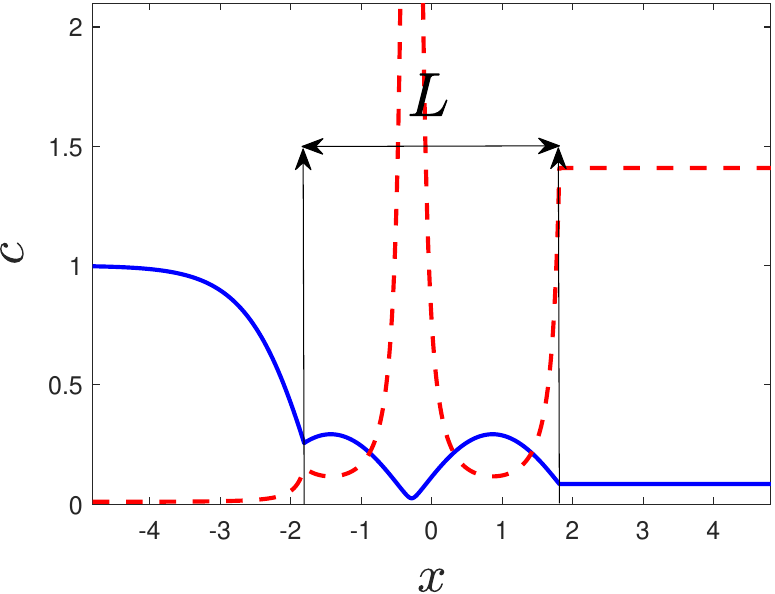}}&
    \stackinset{l}{0pt}{t}{0pt}{\large{(b)}}
    {\includegraphics[width=\columnwidth]{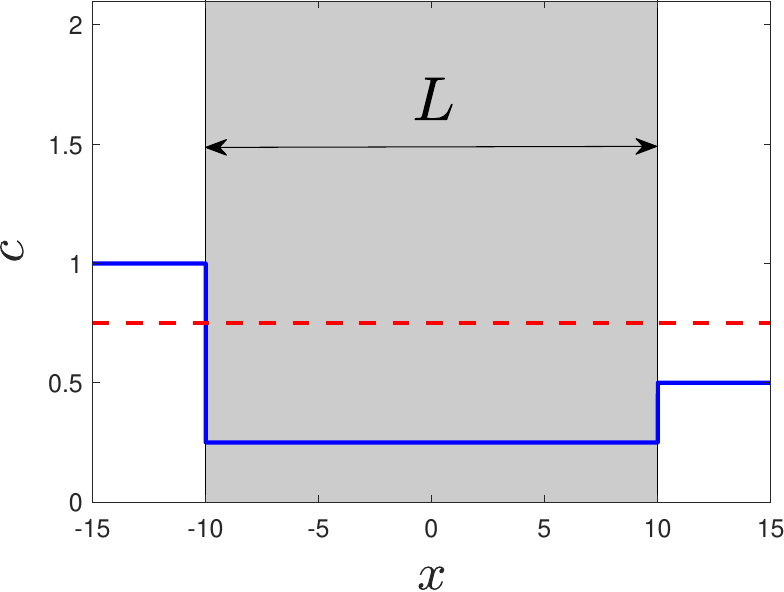}} \\
    \stackinset{l}{0pt}{t}{0pt}{\large{(c)}}{\includegraphics[width=0.98\columnwidth]{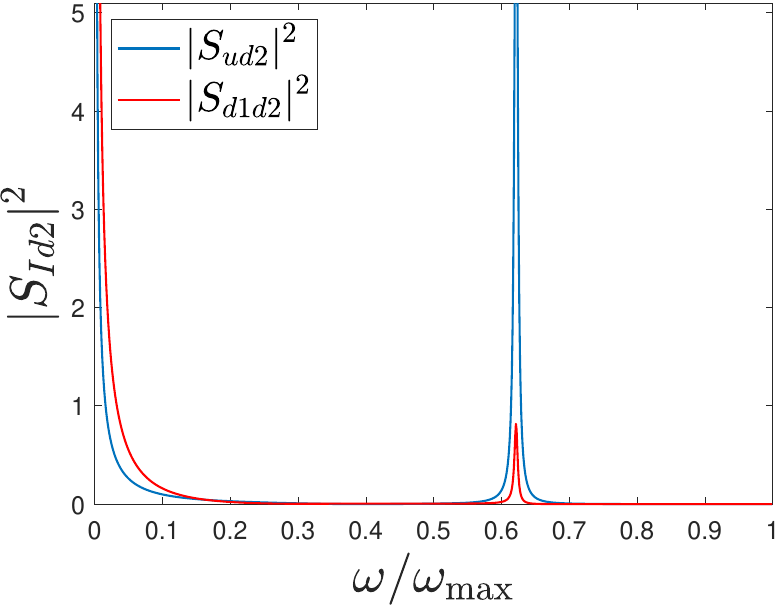}} &
    \stackinset{l}{0pt}{t}{0pt}{\large{(d)}}
    {\includegraphics[width=\columnwidth]{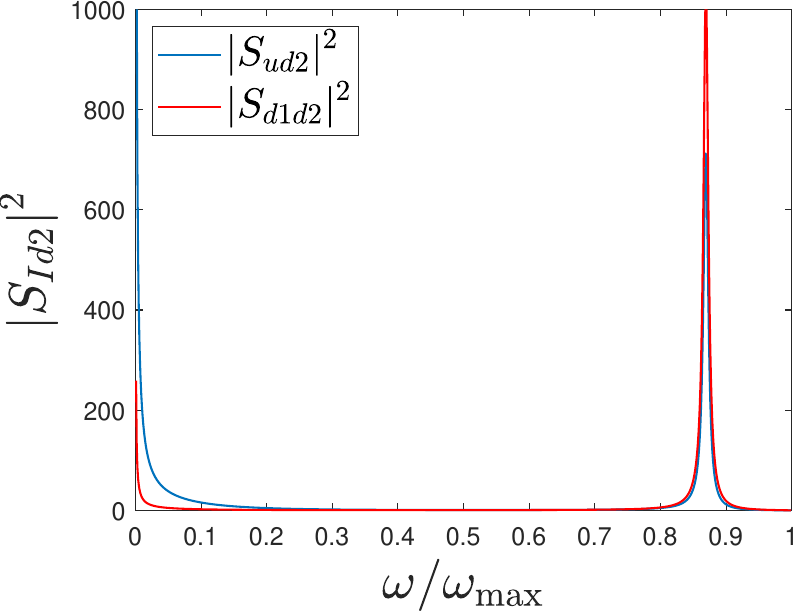}} 
\end{tabular}
\caption{Upper row: Sound (solid blue) and flow (dashed red) velocity profiles for different resonant BH solutions. a) Double delta. The asymptotic subsonic flow velocity is $q=0.01$. The arrows indicate the position of the delta barriers, whose amplitude is $Z=2.2$. The length of the resonant cavity is $L\approx 3.62$. b) Resonant flat-profile. The global flow velocity is $q=0.75$. The shaded area indicates the supersonic resonant cavity of length $L=20$, where the speed of sound is $c_2=0.25$. The downstream supersonic sound speed is $c_3=0.5$. Lower row: c)-d) Hawking (solid blue) and Andreev (solid red) spectra of the BH solutions above.}
\label{fig:Resonant}
\end{figure*}

\section{Resonant Andreev-Hawking radiation}\label{sec:ResonantHawking}

% Here we study a new method that specifically aims at detecting the spontaneous contribution to Hawking radiation. This approach relies on the strong frequency dependence of resonant tunneling through a double-barrier structure. Such a sonic BH analogue behaves as a Fabry–Perot resonator for quasi-particles, with the peculiar feature that quasi-particles propagate linearly against a condensate background, which is itself governed by a nonlinear equation.

% We have studied the flow of an atomic condensate through a double-barrier interface separating regions of subsonic and supersonic flow. Such a setup provides a scenario where Hawking radiation into the subsonic side is enhanced at some frequencies due to multiple scattering of quasi-particles by the two barriers and the modulations of the condensate. The resulting highly non-thermal Hawking radiation presents peaks at frequencies that may lie well above the working temperature and thus can be unambiguously interpreted as stemming from quantum fluctuations of the quasi-particle vacuum. The non-thermal Hawking spectrum emitted by the double-barrier interface represents an important advantage over the cases of single or zero barrier, where the low-frequency zero-point radiation has a thermal character which makes it more difficult to distinguish it from genuinely thermal radiation.

The thermal character of the Andreev and Hawking spectra discussed above makes quite difficult to isolate their signal in a real experiment, as it can be quite easily misidentified or overshadowed by another background thermal component. It was suggested by Zapata, Albert, Parentani and Sols \cite{Zapata2011} that resonant BH configurations could provide a strategic advantage due to their highly non-thermal frequency dependence. We discuss in this section how resonant BH configurations emerge in gravitational analogues and propose possible experimental realizations based on the use of optical lattices.

\subsection{Resonant BH configurations}

The first proposed model of resonant BH configuration \cite{Zapata2011} consisted of a condensate flowing through a double-delta barrier
\begin{equation}\label{eq:DoubleDeltaPotential}
    V(x)=Z\left[\delta\left(x+\frac{L}{2}\right)+\delta\left(x-\frac{L}{2}\right)\right].
\end{equation}
The resulting BH solution, represented in Fig. \ref{fig:Resonant}a, corresponds to a gray soliton for $x<-L/2$ and a supersonic homogeneous plane wave for $x>L/2$, with the same relation between the parameters $q,c_d,v_d$ as for the delta BH solution, Eq. (\ref{eq:DeltaConstraints}). The difference is that now the value of $Z$ is not fine-tuned as in Eq. (\ref{eq:DeltaPotential}), and the stationary GP solution is described by a cnoidal wave for $|x|<L/2$, given in terms of elliptic functions (see Ref. \cite{Zapata2011} for the technical details). Specifically, for a certain value of the asymptotic subsonic flow velocity $q$, there are only stationary GP solutions for $Z\in [Z_{\rm{min}}(q),Z_{\rm{max}}(q)]$, and the number of available solutions grows with the interbarrier distance $L$, displaying an increasing number of cnoidal periods. These BH solutions may include several local acoustic horizons, always an odd number of them to ensure that there is a global transition from an asymptotic subsonic flow to a supersonic one.

The above configuration can be easily generalized to provide analytical BH solutions by combining the three models of Fig. \ref{fig:BHModels}, which yield piecewise homogeneous GP equations, whose solutions are known. As a result, an analytical resonant BH solution is typically given by either a gray soliton or a homogeneous subsonic wave in the upstream region, by a homogeneous supersonic plane wave in the downstream region, and by a cnoidal wave in between; the GP wavefunction in each region is characterized by the same global current $J$ and chemical potential $\mu$. 

A particularly simple model of resonant BH solution, represented in Fig. \ref{fig:Resonant}b, was provided in Ref. \cite{deNova2014} by using a flat-profile piecewise configuration 
\begin{equation}\label{eq:DoubleFlatProfile}
    g(x)n_u=\left\{ \begin{array}{cc}
1, & x< 0,\\
c_2^2, & 0\leq  x \leq L,\\
c_3^2, &  x > L,
\end{array}\right.
\end{equation}
with $c_2,c_3<q$.

Regarding the Andreev and Hawking spectra, they are computed by solving the scattering problem between the asymptotic upstream and downstream channels, similar to that which was solved for the non-resonant BH solutions. However, now there are two matchings, one with the upstream region and one with the downstream region. Due to the periodic character of a cnoidal wave, the corresponding BdG solutions take the form of Bloch waves; analytical solutions can be obtained with the help of mathematical tables (see for instance Ref. \cite{Martone2021}). In practice, a numerical integration of the time-independent BdG equations for given frequency $\omega$ is quite efficient as these can be recast as a simple $4\times 4$ linear system of first-order ordinary differential equations. In the case of the flat-profile resonant configuration, this is not needed, as all the modes involved in each region are the plane-wave spinors of Eq. (\ref{eq:PlaneWaveSpinors}).

The Andreev and Hawking spectra for the BH solutions of Figs. \ref{fig:Resonant}a,b are presented in Figs. \ref{fig:Resonant}c,d, respectively. In the case of the double delta barrier, Fig. \ref{fig:Resonant}c, we observe that apart from the universal thermal $1/\omega$ peak of $|S_{Id2}(\omega)|^2$ at low frequencies, there is a strongly non-thermal peak close to $\simeq 0.6\omega_{\rm{max}}$. This is because the two delta barriers behave as a Fabry-Perot resonator for the anomalous scattering processes of the Andreev and Hawking effects. We note that, as for the single delta barrier, $|S_{d1d2}(\omega)|^2\ll|S_{ud2}(\omega)|^2$.

In the resonant flat-profile case, Fig. \ref{fig:Resonant}d, we observe a similar trend, i.e., apart from the universal thermal peak at low frequencies, there is a highly non-thermal peak at large frequencies. However, unlike for the double delta barrier case, here the Andreev signal is larger than the Hawking signal, $|S_{d1d2}(\omega)|^2 > |S_{ud2}(\omega)|^2$. This highlights that resonant structures may also be useful to enhance the Andreev effect, which is suppressed with respect to the Hawking effect in typical non-resonant configurations. It must be noted that, even though the cavity is much longer than for the double delta barrier, the spectrum only displays one peak. This is due to the smallness of the cutoff frequency $\omega_{\rm{max}}$ for the flat-profile configuration, resulting in low frequencies for the spectrum that are translated into small wavevectors, whose inverse provides the typical length scale for the occurrence of resonances.

%Enlarging the resonant cavity increases the number of resonant peaks, each one with a different strength \cite{Zapata2011,deNova2014}. Remarkably, for the resonant flat-profile BH solution in Fig. \ref{fig:Resonant}d, even though the cavity is much longer than for the double delta barrier in Fig. \ref{fig:Resonant}c, the spectrum only displays one peak. 

\begin{figure*}[t]
\begin{tabular}{@{}ccc@{}}\stackinset{l}{0pt}{t}{0pt}{(a)}{\includegraphics[width=0.62\columnwidth]{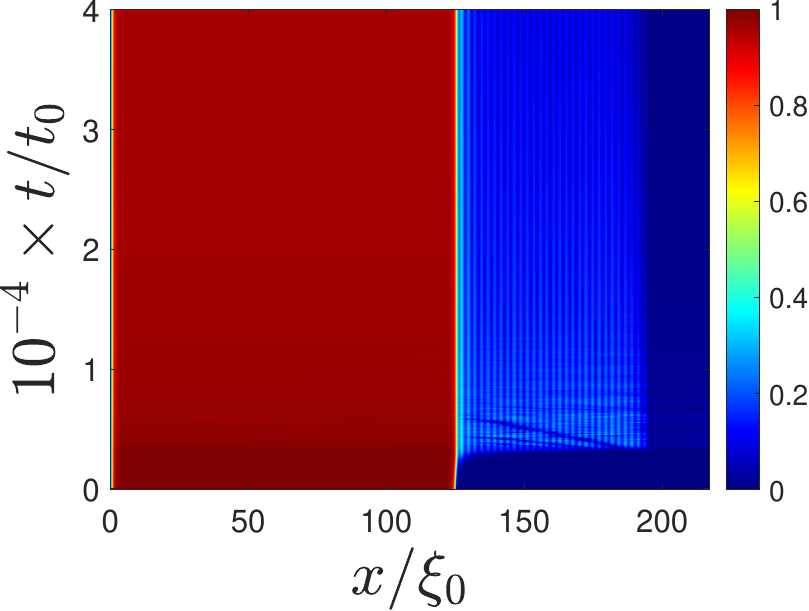}}~ &~
    \stackinset{l}{0pt}{t}{0pt}{(b)}
    {\includegraphics[width=0.61\columnwidth]{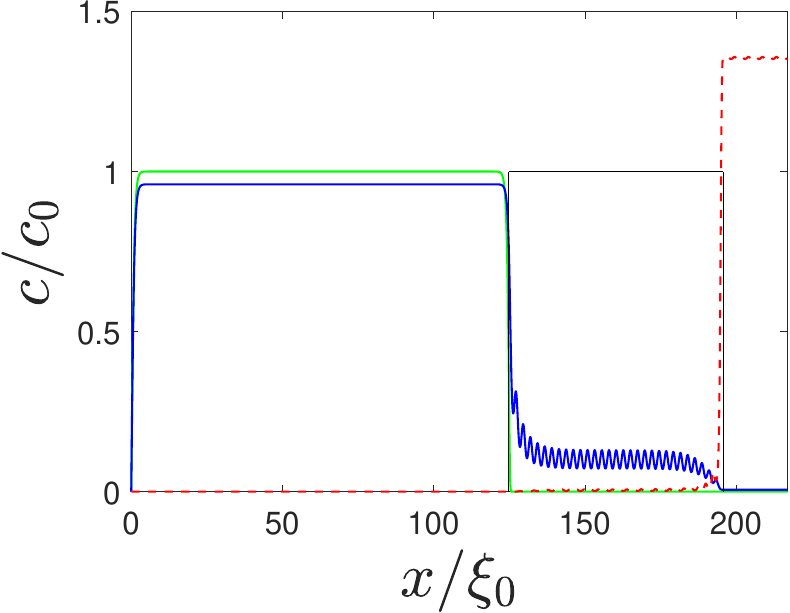}}~ &~ 
    \stackinset{l}{0pt}{t}{0pt}{(c)}{\includegraphics[width=0.66\columnwidth]{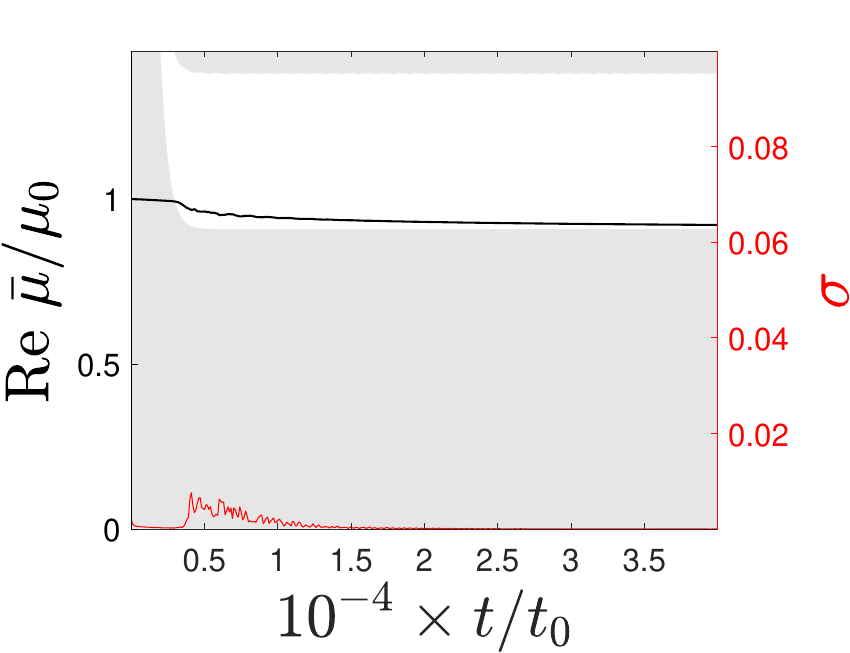}}  \\
    \stackinset{l}{0pt}{t}{0pt}{(d)}{\includegraphics[width=0.62\columnwidth]{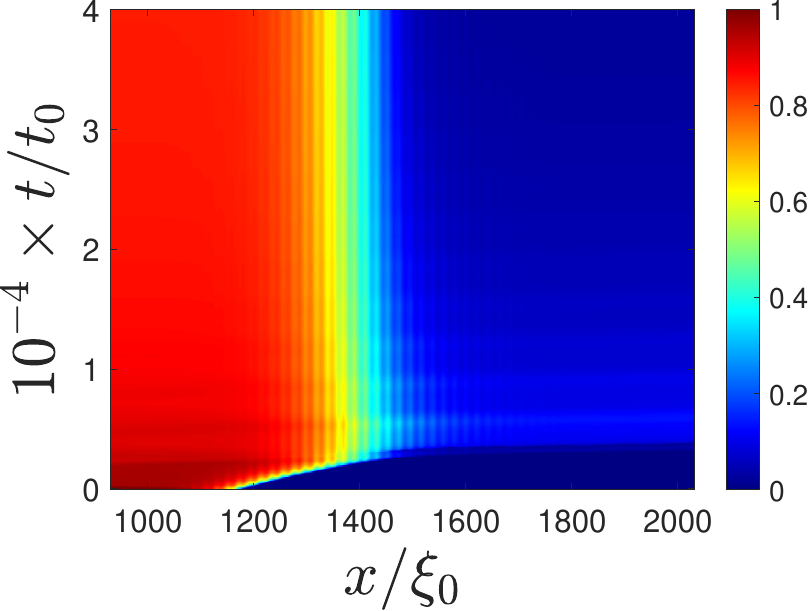}}~ &~
    \stackinset{l}{0pt}{t}{0pt}{(e)}
    {\includegraphics[width=0.61\columnwidth]{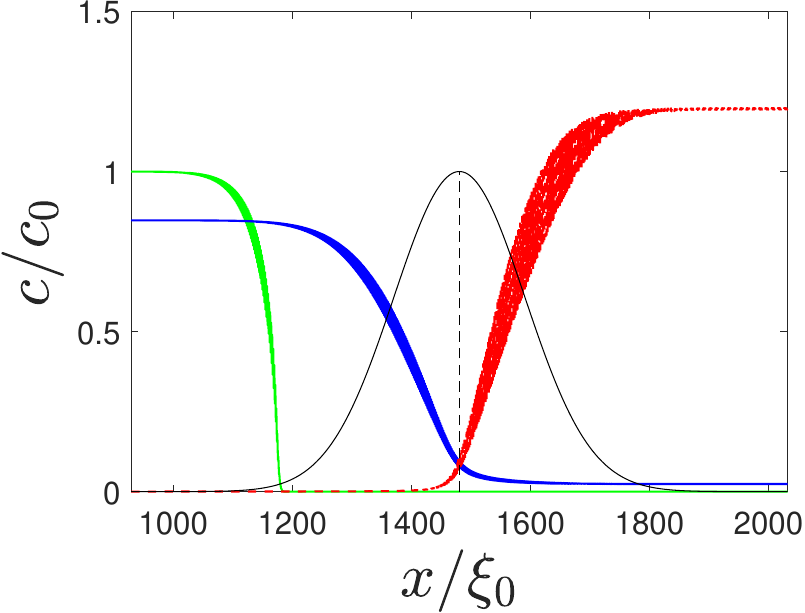}}~ &~ 
    \stackinset{l}{0pt}{t}{0pt}{(f)}{\includegraphics[width=0.66\columnwidth]{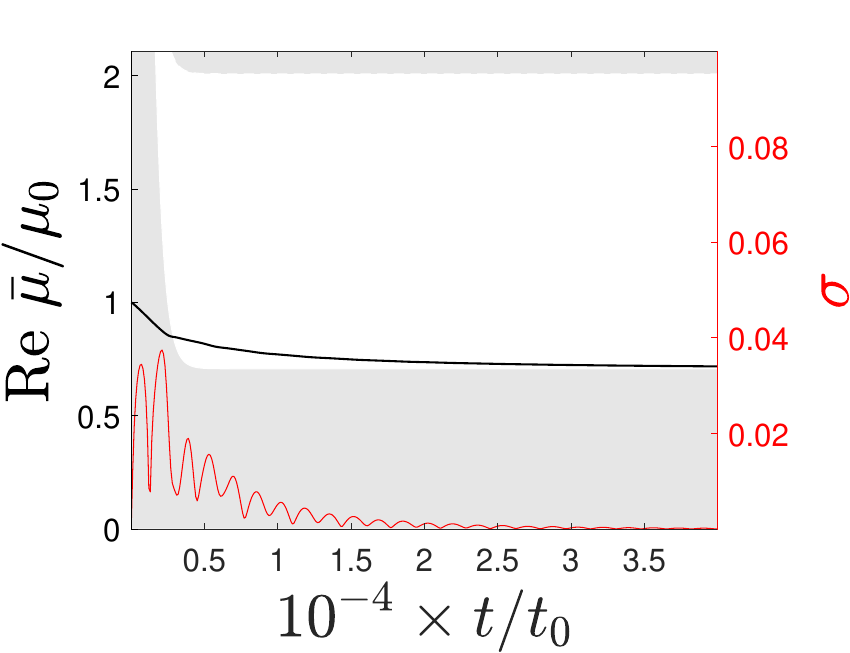}} 
\end{tabular}
\caption{Time evolution of an initially confined condensate which is outcoupled through an optical lattice. Upper row: Ideal optical lattice (\ref{eq:idealOL}) with $L\approx 125 \xi_0$, $d\approx 2.36\xi_0$, and $n_{\rm osc}=30$. The lowering time is $\tau=500 t_0$. a) Time-dependent profile of the sound speed, $c(x,t)/c_0$, with $c(x,t)=|\Psi(x,t)|$. b) Sound (solid blue) and flow (dashed red) velocity profiles of the quasi-stationary regime, evaluated at the last snapshot of a). The initial sound speed profile is depicted in solid green (the initial flow velocity is identically zero), while the optical lattice envelope is shown as a black line. c) Time evolution of the average chemical potential $\bar{\mu}(t)$ (solid black) and its relative fluctuations $\sigma(t)$ (solid red). The instantaneous conduction bands (energy gaps) of the optical lattice are depicted as white (gray) bands. Lower row: d)-f) Same as a)-c) but for a Gaussian optical lattice (\ref{eq:actualpotential}) with $L\approx 1480 \xi_0$, $d\approx 1.73\xi_0$, and $\tilde{w}\approx 220.5\xi_0$.}
\label{fig:OLMF}
\end{figure*}
%The lowering time is $\tau=500 t_0$.

\subsection{Black hole from an outcoupled condensate through an optical lattice}

% \begin{itemize}

%     \item \fs{Discuss motivation NJP 2014 \cite{deNova2014a} as a continuation of NJP 2011 (simpler to use an optical lattice in experiments, according to David Gu\'ery-Odelin).}

% \end{itemize}

%As the resonant cavity enlarges, more and more peaks are expected in the Andreev and Hawking spectra. 

The resonant configurations discussed above displayed a single resonant peak in the spectrum. Interestingly, the opposite limit of a long cavity with many resonant peaks can be experimentally reproduced with the help of an optical lattice, a major paradigm in AMO physics \cite{Orzel2001,Greiner2002,Bakr2009}. In particular, Ref. \cite{deNova2014a} provided a thorough numerical study of the quasi-stationary BH resulting from the outcoupling of a condensate through an optical lattice, which we proceed to discuss. 

A 1D optical lattice can be created from the interference of two fixed-phase lasers of wavelength $\lambda$ whose wavevectors form an angle $\theta$ \cite{Fabre2011}. The resulting potential can be written as
\begin{equation}\label{eq:OLPotential}
    V(x,t)=V(t)f(x-L)\cos^{2}\left[k_L(x-L)\right]
\end{equation}
with $k_L=\pi/d$ and $d=\lambda/\left[2\sin(\theta/2)\right]$ the lattice period. In the above equation, $f(x)$ is a dimensionless function that characterizes the global shape of the optical lattice, accounting for its finite size in real experiments, while $V(t)$ represents its (possibly time-dependent) amplitude. 

In our specific configuration, we assume that our condensate is confined by a high-amplitude barrier placed at $x=0$, modeled by a hard-wall boundary condition for the GP wavefunction $\Psi(0,t)=0$. The position $L$ in Eq. (\ref{eq:OLPotential}) plays the role of the approximate localization of the lattice, chosen here to be repulsive, so the condensate is essentially confined between $0\leq x\lesssim L$. The lattice amplitude is gradually lowered from $V_0$ to $V_{\infty}$ as
\begin{equation}\label{eq:TDAmplitude}
    V(t)=\left\{ \begin{array}{cc}
V_0, & t\leq 0,\\
V_{\infty}+(V_{0}-V_{\infty})e^{-t/\tau}, & t>0 ,
\end{array}\right.
\end{equation}
where $\tau$ is the characteristic timescale of the process. This causes the condensate to outcouple through the optical lattice from the initial reservoir.

Quantitatively, the problem is described by the time-dependent GP equation
\begin{equation}\label{eq:TDGPConfined}
\left[-\frac{\partial_{x}^{2}}{2}+V(x,t)+|\Psi(x,t)|^{2}\right]\Psi(x,t) = i\partial_t\Psi(x,t),
\end{equation}
where the initial condition $\Psi(x,0)=\Psi_0(x)$ is the wavefunction describing the equilibrium condensate, which is solution of the time-independent GP equation
\begin{equation}\label{eq:TIGPConfined}
\left[-\frac{\partial_{x}^{2}}{2}+V(x,0)+|\Psi_0(x)|^{2}-\mu_{0}\right]\Psi_0(x) = 0,~\Psi_0(0)=0,
\end{equation}
with the chemical potential $\mu_0$ determined by the normalization condition (\ref{eq:Normalization}). This chemical potential also defines some relevant physical scales: density $n_0\equiv \mu_{0}$, length $\xi_0\equiv 1/\sqrt{\mu_{0}}$, velocity $c_{0}\equiv \sqrt{\mu_{0}}$, time $t_0\equiv 1/\mu_{0}$, and temperature $T_0\equiv \mu_0$. Typical orders of magnitude for $^{87}$Rb are $\xi_0\sim 0.1-1\,\mu\textrm{m}$, $t_0\sim 10^{-4}-10^{-3}\,\textrm{s}$, $c_0\sim 0.1-1 \,\textrm{mm/s}$, and $T_0\sim 1-10 \,\textrm{nK}$. The lattice period satisfies $d>\lambda/2\gtrsim \xi_0$ while the lattice amplitude can essentially take any value; we choose $V_0\gg \mu_0$ and $V_\infty\gtrsim \mu_0$, so the lattice goes from providing a tight confinement to allow some leakage. On the other hand, we require $\tau\gg t_0$, so the barrier lowering is adiabatic and does not introduce further distortions in the condensate flow, and $L\gg \xi_0$, so we are in the Thomas-Fermi regime where $\Psi_0(x)\simeq \sqrt{n_0}$ is the bulk value (for $0\lesssim x \lesssim L$) of the condensate amplitude. Therefore, we can regard $L$ as the approximate size of the reservoir, containing $N\sim n_0 L$ particles.

The time evolution of $\Psi(x,t)$ is displayed in Fig. \ref{fig:OLMF}, computed from numerical integration of Eq. (\ref{eq:TDGPConfined}). In the first row, we consider an {\it ideal} finite optical lattice, determined by an envelope
\begin{equation}\label{eq:idealOL}
f(x)=\chi\left(\frac{x+\frac{d}{2}}{L_{\rm lat}}\right),
\end{equation}
$\chi(x)$ being the characteristic function of the interval $[0,1]$. Thus, a lattice with instantaneous uniform amplitude $V(t)$ extends from $x_0=L-d/2$ to $x_1=x_0+L_{\rm lat}$, where the lattice length is chosen such that it contains an integer number of periods $n_{\rm osc}\sim 10-50$, $L_{\rm lat}\equiv n_{\rm osc}d$ and $V(x_0)=V(x_1)=0$.

In  Fig. \ref{fig:OLMF}a, we represent the time evolution of the sound speed profile $c(x,t)$, proportional to the square root of the local density, $c(x,t)=|\Psi(x,t)|=\sqrt{n(x,t)}$ (this choice improves the visibility of the condensate outside the reservoir as compared to using the density itself). After some transient times $t\sim 10^4 t_0$, the condensate achieves a quasi-stationary regime in which it flows through the lattice and eventually leaks outside. The sound and flow velocity profiles in the quasi-stationary regime are shown in Fig. \ref{fig:OLMF}b, where we observe that a BH configuration is achieved, with the downstream supersonic region located outside the lattice. 

We can quantify the degree of quasi-stationarity by defining a \textit{local} chemical potential as
\begin{equation}\label{eq:LocalChemicalPotential}
\mu(x,t)\equiv -\frac{1}{2}\frac{\partial_x^{2}\Psi(x,t)}{\Psi(x,t)}+V(x,t)+|\Psi(x,t)|^{2}.
\end{equation}
For a stationary solution, $\mu(x,t)=\mu$ is real and constant. The current is also constant and uniform for a 1D stationary solution. However, this latter condition is impossible to fulfill strictly, since the current is zero at $x=0$ due to the hard-wall boundary condition, while the leaked downstream flow carries a non-zero flux. Hence, there must be a current gradient, which, via the continuity equation, implies a time-dependent density. This also implies a non-homogeneous time-dependent complex chemical potential as
\begin{equation}
    \partial_t \ln n=2\,\text{Im}\,\mu .
\end{equation}
Nevertheless, in practice, such dependence can become so weak that one can neglect it, effectively achieving a quasi-stationary regime. This regime should be characterized by a sufficiently uniform local chemical potential $\mu(x,t)$, with small relative spatial fluctuations $\sigma(t)$ around its instantaneous average value $\bar{\mu}(t)$,  
\begin{eqnarray}
\bar{\mu}(t) & \equiv  & \frac{\int_{0}^{L_{g}}\mathrm{d}x~|\Psi(x,t)|^2\mu(x,t)}{\int_{0}^{L_{g}}\mathrm{d}x~|\Psi(x,t)|^2},\\ \nonumber 
\sigma(t) & \equiv  & \frac{1}{\bar{\mu}(t)}
\left[
\frac {\int_{0}^{L_{g}}\mathrm{d}x~|\Psi(x,t)|^2|\mu(x,t)-\bar{\mu}
(t)|^{2}} {\int_{0}^{L_{g}}\mathrm{d}x~|\Psi(x,t)|^2}
\right]^{\frac{1}{2}},
\label{eq:AverageChemicalPotential}
\end{eqnarray}
where $L_g$ is the total length considered for the average. Typically, $L_g$ is chosen well inside the downstream region, and the results are quite insensitive to its specific value.
%\text{Re}\,

The time evolution of the real part of the average chemical potential $\bar{\mu}(t)$ and its relative fluctuations $\sigma(t)$ is depicted in Fig. \ref{fig:OLMF}c (imaginary values can be neglected as $\textrm{Im}\,\bar{\mu}\sim 10^{-6}-10^{-7}\mu_0$). In order to understand their relation with $V(t)$, we also represent the instantaneous band structure of the lattice, computed using the linear Schr\"odinger equation since the non-linear interacting term is negligible within the lattice due to the smallness of the density. Specifically, the lowest conduction band is placed between $E_{0}(t)$ and $E_{1}(t)$, with a width $\Delta_c(t)=E_{1}(t)-E_{0}(t)$, where dimensional arguments show that
% \begin{equation}\label{eq:BandStructure}
%     E_{0,1}(t)=E_R\,  f_{\rm{min},\rm{max}}[\zeta(t)],~E_{\rm{min},\rm{max}}(t)=E_R\,  f_{\rm{min},\rm{max}}[\zeta(t)],~\zeta(t)\equiv \frac{V(t)}{16 E_R},
% \end{equation}
\begin{equation}\label{eq:BandStructure}
    E_{0,1}(t)=E_R\,  F_{0,1}[\zeta(t)],~\zeta(t)\equiv \frac{V(t)}{16 E_R},
\end{equation}
with $F_{0,1}$ increasing functions of the dimensionless parameter $\zeta$ that can be computed perturbatively, 
\begin{eqnarray}
    F_0(\zeta)&=& 8\zeta-8\zeta^2+O(\zeta^4),\\
    \nonumber F_1(\zeta)&=&1+4\zeta-2\zeta^2+O(\zeta^4),
\end{eqnarray}
and $E_R\equiv k^2_L/2$ the recoil energy of the lattice. In general, the solutions to the 1D Schr\"odinger equation with a sinusoidal potential can be obtained in terms of Mathieu functions. In practice, the functions $F_{0,1}$ can be easily evaluated numerically \cite{deNova2014a}. The resulting time-dependent conduction band (energy gap) is depicted as a white (gray) band. As we can see, the condensate smoothly approaches the bottom of the asymptotic conduction band, since tunneling is exponentially suppressed for $\mu_0<E_{0}$. In this regime of small leaking, the fluctuations of the chemical potential can become extremely small, $\sigma(t)\sim 10^{-4}$, ensuring a high-degree of quasi-stationarity. 

%In order to avoid agitated  transients that can reduce the quasi-stationarity, $\mu_0$ should be also chosen close to the asymptotic value of $E_{\rm{min}}$ to guarantee a smooth landing of the chemical potential on the bottom of the conduction band.

%$E_{\rm{min},\rm{max}}(t)\equiv E_{\rm{min},\rm{max}}[V(t)]$.

The formation of a quasi-stationary BH can be then easily understood from energetic arguments: in the upstream region, where the reservoir is placed, the flow velocity is negligible and the chemical potential is merely due to interactions, i.e., $\bar{\mu}\simeq n_u$ and $v_u\simeq 0$. Due to the conservation of the chemical potential as well as the small density there, in the downstream region the condensate flows with a high velocity $v_d\sim \sqrt{2\bar{\mu}}\gg c_d$, becoming supersonic. By continuity, this implies that there must be an acoustic horizon somewhere within the lattice. In the bulk of the lattice, the wavefunction is a Bloch wave, which is preferred to be subsonic due to its energetic stability \cite{Wu2003}. Thus, the acoustic horizon must be placed at the right edge of the lattice, as seen in Fig. \ref{fig:OLMF}b.

%$with $\tilde{w}=w/\cos(\theta/2)$ 

In the second row of Fig. \ref{fig:OLMF}, we analyze a more realistic Gaussian envelope
\begin{equation}\label{eq:actualpotential}
f(x)=e^{-2\tfrac{x^2}{\tilde{w}^2}},
\end{equation}
with $\tilde{w}$ the effective beam waist, which plays a similar role to $L_{\rm lat}$ for the ideal optical lattice. We require the length hierarchy 
\begin{equation}\label{eq:LengthHierarchy}
    d\ll \tilde{w}\ll L,
\end{equation}
where the second condition is imposed in order to have a sufficiently large and homogeneous condensate reservoir. The first condition is satisfied for typical waists, and implies that the overall Gaussian amplitude behaves as a spatially adiabatic envelope, so the potential can be regarded locally as an {\it ideal} optical lattice with an amplitude $V_{A}(x,t)$ \cite{Santos1998a,Santos1999,Carusotto2000}:
\begin{eqnarray}
V(x,t)&=&V_{A}(x,t)\cos^{2}\left[k_L(x-L)\right],\\
\nonumber V_{A}(x,t)&\equiv& V(t)\exp\left[-2\left(\frac{x-L}{\tilde{w}}\right)^2\right].
\end{eqnarray}
We observe in Figs. \ref{fig:OLMF}d,e the same trends as for the ideal lattice case, namely, a quasi-stationary BH solution is achieved for sufficiently long times. Due to the hierarchy (\ref{eq:LengthHierarchy}), we only depict the vicinity of the lattice peak $L-2.5\tilde{w}\leq x\leq L+2.5\tilde{w}$ to better observe the structure of the horizon; in turn, at this scale, the lattice structure cannot be resolved and the oscillations of the flow and sound velocities appear as broadened lines. In Fig. \ref{fig:OLMF}f, the average chemical potential also descends towards the bottom of the conduction band, achieving a highly quasi-stationary regime where the relative fluctuations $\sigma(t)$ are also insignificant, $\sigma(t)\sim 10^{-4}$. The band structure is now evaluated at the lattice peak $x=L$, where the local conduction band is determined by the energies $E_{0,1}(x,t)$ obtained by taking $\zeta(x,t)=V_A(x,t)/16E_R$ in Eq. (\ref{eq:BandStructure}). The fact that the lattice maximum is the transmission bottleneck can be understood from the increasing character of the asymptotic energies $E_{0,1}(x)\equiv E_{0,1}(x,\infty)$ with respect to the lattice envelope $V_A(x)\equiv V_A(x,\infty)$. Thus, in order to place the chemical potential within the local conduction band across the whole lattice, it must be satisfied
\begin{equation}
    E_{0}(x)\leq  E_{0}(L)< \mu_0 < E_R<E_{1}(x),
\end{equation}
where we have used that $F_1(0)=1$, so $E_1(0)=E_R$. This implies that the asymptotic lattice amplitude $V_\infty$ must be below a certain critical value $V_c$ so the condition $E_{0}(L)<E_R$ is met, which can be derived from
\begin{equation}
  F_0\left(\frac{V_\infty}{16E_R}\right)<1\Longrightarrow  V_\infty<V_c,~V_c\approx 2.33 E_R.
\end{equation}

Another remarkable feature is that the acoustic horizon is now placed exactly at the lattice peak (vertical dashed black line in Fig. \ref{fig:OLMF}e). This is no coincidence and a detailed local lattice calculation explicitly proves that the acoustic horizon must be placed at the extremes of the lattice envelope \cite{deNova2014a}. This is in agreement with another result derived for a smooth potential within the hydrodynamic approximation, stating that the acoustic horizon must be located at the potential maximum \cite{Giovanazzi2004}.

We summarize now the computation of the scattering matrix for the above quasi-stationary BH solutions, where the interested reader can consult Ref. \cite{deNova2017b} for more details. From Eq. (\ref{eq:LocalChemicalPotential}), and by invoking the time-dependent GP equation (\ref{eq:TDGPConfined}), it is easily shown that $i\partial_t \ln \Psi(x,t)=\mu(x,t)$. Hence,  
\begin{equation}\label{eq:quasistationarywavefunction}
\Psi(x,t)=\Psi(x,t_s)e^{-i\int_{t_s}^{t}\mathrm{d}t'\mu(x,t')} \equiv\Psi_{\infty}(x,t)e^{-i{\rm Re}\,\bar{\mu}(t-t_s)},
\end{equation}
where the rightmost term provides a definition for $\Psi_{\infty}(x,t)$. If one chooses $t_s$ well inside the quasi-stationary regime, when $\mu(x,t)\simeq \bar{\mu}\simeq {\rm Re}\,\bar{\mu}$, then one can approximate $\Psi_{\infty}(x,t)\simeq \Psi(x,t_s)=\Psi_{\infty}(x,t_s)\equiv \Psi_{\infty}(x)$. Thus, we can work with a fully stationary BH solution whose chemical potential is $\mu={\rm Re}\,\bar{\mu}$, and compute the Andreev and Hawking spectra from the $S$-matrix by solving the associated BdG scattering problem, where the external potential is the asymptotically stationary optical lattice, $V(x)=V_\infty(x)\equiv V(x,t=\infty)$. Specifically, the time-independent BdG equations for given $\omega$ are numerically integrated and eventually matched with the corresponding scattering channels at the asymptotic homogeneous subsonic (the upstream bulk of the condensate reservoir) and supersonic (the downstream leaking flux) regions. 

However, a major difficulty arises for the computation in the Gaussian lattice, since its large size makes that exponentially growing modes, corresponding to local Bloch waves with complex wavevector, explode above the propagating modes, making the matching equations singular within computer accuracy. This is known in general as the $\Omega d$ problem \cite{Lowe1995}, emerging in a wide range of scenarios, ranging from the propagation of ultrasonic and electromagnetic waves in multilayered media \cite{Lowe1995,Pernas2014} to Anderson localization \cite{Slevin2004}. Possible methods to deal with this specific $\Omega d$ problem in the BdG context are the Global Matrix method \cite{Lowe1995} or QR decomposition \cite{Slevin2004}.

As expected, the resulting Hawking and Andreev spectra display a highly non-thermal structure. In particular, since the energy contribution from the speed of sound is negligible as compared to that from the lattice potential, one can analyze the spectrum in terms of the underlying Schr\"odinger problem, characterized by the band structure shown in Figs. \ref{fig:OLMF}c,f. Thus, two main qualitative regimes arise: $\omega_{\rm{max}}<\Delta_c$, where the spectrum is cut at its natural cutoff frequency, and $\omega_{\rm{max}}>\Delta_c$, where the spectrum is abruptly cut by the upper end of the conduction band. Hence, an optical lattice can behave as a low-pass filter of Andreev-Hawking radiation, which may have potential applications in quantum transport and atomtronics \cite{Amico2021}.

\section{Quantum Andreev-Hawking radiation}\label{sec:QuantumHR}

Although resonant configurations are highly non-thermal as shown in the previous sections, this does not automatically imply that the observation of a non-thermal Hawking spectrum is a signature of the Hawking effect. This is because the BdG equations describe at the same time the linear dynamics of both perturbations of the GP wavefunction and quantum fluctuations of the field operator around the mean-field expectation value. Moreover, the quantum state of the system may be a highly thermal state. Thus, the observed non-resonant spectrum can result from the coherent or thermal stimulation of Hawking radiation by a classical source, instead of arising from a zero-point quantum origin. The same applies to the Andreev effect. In this section, we discuss how to unambiguously signal the genuine quantum character of the Andreev and Hawking effects, as opposed to classical stimulation, using different types of quantum correlations.

\subsection{Lessons from quantum optics}\label{subsec:QuantumOptics}

A major front of the quantum-classical frontier is present in the field of quantum optics, from where we can borrow a number of concepts and techniques; the interested reader is referred to Ref. \cite{Walls2008} for a pedagogical introduction to quantum optics and a thorough discussion of a number of fundamental quantum topics in that context. 

For simplicity, we begin by considering a single bosonic mode (as can be that of a photon) whose amplitude is given by an annihilation operator $\hat{a}$. The corresponding Hilbert space is the Fock space spanned by the number states $\ket{n}$, $\hat{a}^\dagger\hat{a}\ket{n}=n\ket{n}$, $n=0,1,2\ldots$ A coherent state is an eigenstate of the annihilation operator, $\hat{a}\ket{\alpha}=\alpha\ket{\alpha},~\alpha\in\mathbb{C}$, and can be expressed as
\begin{equation}
\ket{\alpha}=e^{-\frac{|\alpha|^2}{2}}\sum^{\infty}_{n=0}\frac{\alpha^n}{\sqrt{n!}}\ket{n}=D(\alpha)\ket{0},~D(\alpha)\equiv e^{\alpha \hat{a}^\dagger-\alpha^*\hat{a}},
\end{equation}
where $D(\alpha)$ is the displacement operator, which is unitary, $D^\dagger(\alpha)=D^{-1}(\alpha)=D(-\alpha)$. The coherent states form an overcomplete basis of the Hilbert space since
\begin{equation}\label{eq:Overcomplete}
    |\braket{\alpha|\beta}|^2=e^{-|\alpha-\beta|^2}\neq 0,\,\int\frac{\mathrm{d}^2\alpha}{\pi}\ket{\alpha}\bra{\alpha}=\sum^{\infty}_{n=0}\ket{n}\bra{n}=1,
\end{equation}
with $\mathrm{d}^2\alpha\equiv\mathrm{d}\alpha_x\mathrm{d}\alpha_y,~\alpha=\alpha_x+i\alpha_y$. Another relevant class of quantum states are the squeezed states
\begin{equation}\label{eq:Squeezing}
    \ket{\varepsilon}=S(\varepsilon)\ket{0},~S(\varepsilon)\equiv e^{\frac{\varepsilon^*\hat{a}^2-\varepsilon(\hat{a}^\dagger)^2}{2}},~\varepsilon\equiv r e^{i2\theta},
\end{equation}
where $S(\varepsilon)$ is the squeezing operator, also unitary as $S^\dagger(\varepsilon)=S^{-1}(\varepsilon)=S(-\varepsilon)$. These squeezed states are the vacuum of the annihilation operator $\hat{b}$, $\hat{b}\ket{\varepsilon}=0$, arising from the Bogoliubov transformation 
\begin{equation}  \hat{b}=S(\varepsilon)\hat{a}S^\dagger(\varepsilon)=\hat{a}\cosh r+\hat{a}^\dagger e^{i2\theta}\sinh r.
\end{equation}
By noticing that $\hat{a}^2,(\hat{a}^\dagger)^2,\hat{a}^\dagger\hat{a}+\hat{a}\hat{a}^\dagger$ form a closed Lie algebra [actually, they form a representation of $\mathfrak{su}(1,1)$], one can rewrite the squeezing operator in normal order as
\begin{equation}\label{eq:SqueezingOperatorNormal}
    S(\varepsilon)=\frac{1}{\sqrt{\cosh r}}e^{-\frac{g}{2}(\hat{a}^\dagger)^2}e^{f\hat{a}^\dagger\hat{a}}e^{\frac{g^*}{2}\hat{a}^2},
\end{equation}
with $g=e^{i2\theta}\tanh r$ and $f=-\ln \cosh r$. This allows to readily express the squeezed states as
\begin{equation}\label{eq:SqueezedFock}
    \ket{\varepsilon}=\frac{1}{\sqrt{\cosh r}}\sum^\infty_{n=0}(-1)^ne^{i2n\theta}\frac{\sqrt{2n!}\tanh^n r}{2^n\cdot n!}\ket{2n}.
\end{equation}
The most general quantum state in this Hilbert space is described by a density matrix $\hat{\rho}$ of the form
\begin{equation}\label{eq:GeneralQuantumStateFock}
    \hat{\rho}=\sum^\infty_{n,m=0} \rho_{nm}\ket{n}\bra{m},
\end{equation}
which is completely determined by its characteristic function
\begin{equation}
    \chi(\eta)\equiv \braket{e^{\eta\hat{a}^\dagger-\eta^*\hat{a}}}=\textrm{Tr}[e^{\eta\hat{a}^\dagger-\eta^*\hat{a}}\hat{\rho}].
\end{equation}
One can also work with its normal and anti-normal versions
\begin{eqnarray}\label{eq:CharacteristicParanormal}
      \nonumber \chi_N(\eta)&\equiv& \braket{e^{\eta\hat{a}^\dagger}e^{-\eta^*\hat{a}}},\\
      \chi_A(\eta)&\equiv& \braket{e^{-\eta^*\hat{a}}e^{\eta\hat{a}^\dagger}}.
\end{eqnarray}

Alternative representations to the Fock expansion (\ref{eq:GeneralQuantumStateFock}) are provided by the distributions resulting from the Fourier transform of the characteristic functions:
\begin{eqnarray}\label{eq:ProbabilitiesQuantumOptics}
      \nonumber P(\alpha)&\equiv &\int\frac{\mathrm{d}^2\eta}{\pi^2}e^{(\eta^*\alpha-\eta\alpha^*)}\chi_N(\eta),\\
      Q(\alpha)&\equiv &\int\frac{\mathrm{d}^2\eta}{\pi^2}e^{(\eta^*\alpha-\eta\alpha^*)}\chi_A(\eta),\\
      \nonumber W(\alpha)&\equiv &\int\frac{\mathrm{d}^2\eta}{\pi^2}~e^{(\eta^*\alpha-\eta\alpha^*)}\chi(\eta).
\end{eqnarray}
All of them are quasi-probability distributions, properly normalized,
\begin{equation}
\int\mathrm{d}^2\alpha~P(\alpha)=\int\mathrm{d}^2\alpha~W(\alpha)=\int\mathrm{d}^2\alpha~Q(\alpha)=1,
\end{equation}
but do not describe disjoint events since coherent states are not orthogonal, and can even take negative values, opening the door to genuine non-classical behavior. 

The Glauber-Sudarshan $P$ function is equivalent to a diagonal representation in the coherent basis,
\begin{equation}
\hat{\rho}=\int\mathrm{d}^2\alpha~P(\alpha)\ket{\alpha}\bra{\alpha},
\end{equation}
since its momenta provides the normal-ordered expectation values
\begin{equation}
\braket{(\hat{a}^\dagger)^n\hat{a}^m}=\int\mathrm{d}^2\alpha~(\alpha^*)^n\alpha^m P(\alpha),
\end{equation}
which are those typically characterizing correlation functions. This is where the crucial role of the $P$ function in the understanding of the classical-quantum frontier emerges: if it is non-negative, $P(\alpha)\geq 0$, we can understand these correlations as statistical averages over a continuous classical variable $\alpha$ with a probability distribution given precisely by $P(\alpha)$. Thus, quantum states with a non-negative $P$ function can be regarded as classical, admitting a conventional probabilistic description in terms of a stochastic complex amplitude. Examples of classical states are coherent states, chaotic states, as well as quantum thermal states. On the other hand, number states and squeezed states are intrinsically non-classical as they do not even have a well-defined $P$-representation.

%the observables of actual measurements

The $Q$-function yields the anti-normal expectation values
\begin{equation}
\braket{\hat{a}^m(\hat{a}^\dagger)^n}=\int\mathrm{d}^2\alpha~(\alpha^*)^n\alpha^mQ(\alpha),
\end{equation}
and it is easily evaluated by inserting the identity representation (\ref{eq:Overcomplete}) in Eq. (\ref{eq:CharacteristicParanormal}),
\begin{equation}    Q(\alpha)=\frac{\braket{\alpha|\hat{\rho}|\alpha}}{\pi},
\end{equation}
so it is non-negative and bounded, $0\leq Q(\alpha)\leq 1/\pi$.

Finally, the Wigner function $W(\alpha)$ characterizes symmetric expectation values such as
\begin{equation}
\frac{1}{2}\braket{\hat{a}^\dagger\hat{a}+\hat{a}\hat{a}^\dagger}=\int\mathrm{d}^2\alpha~|\alpha|^2\,W(\alpha).
\end{equation}
Further insight on the physical meaning of the Wigner function is obtained when working with the usual coordinate-momentum representation
\begin{equation}
    \hat{a}=\frac{\hat{q}+i\hat{p}}{\sqrt{2}},~\alpha=\frac{q+ip}{\sqrt{2}},
\end{equation}
whose eigenstates are labeled as $\ket{q},\ket{p}$, respectively. After proper normalization, the Wigner distribution in phase space reads
\begin{equation}\label{eq:WignerPosition}
W(q,p)=\frac{1}{2\pi}\int\mathrm{d}q'~\braket{q-\frac{q'}{2}|\hat{\rho}|q+\frac{q'}{2}}e^{ipq'}.
\end{equation}
Its marginal distributions
\begin{eqnarray} W(q)&=&\int\mathrm{d}p~W(q,p)=\braket{q|\hat{\rho}|q}\geq 0,\\
\nonumber W(p)&=&\int\mathrm{d}q~W(q,p)=\braket{p|\hat{\rho}|p}\geq 0.
\end{eqnarray}
are the spatial and momentum distributions of the quantum state. As a result, the Wigner function represents a quantum version of the classical Boltzmann distribution function. However, it is not necessarily positive, and the presence of negative values $W(q,p)<0$ is hence another quantum signature. Indeed, the negativity of the Wigner function is a stronger condition than that of the Glauber-Sudarshan function as they are related through the convolution
\begin{equation}
    W(\alpha)=\frac{2}{\pi}\int\mathrm{d}^2\beta~e^{-2|\alpha-\beta|^2}P(\beta),
\end{equation}
derived by noting that $\chi(\eta)=e^{-\frac{|\eta|^2}{2}}\chi_N(\eta)$. For instance, squeezed states do have a positive Wigner representation, while number states do not. An interesting approach to quantum optics from phase space can be found in Ref. \cite{Schleich2001}. 

All the above concepts can be straightforwardly extended to multipartite Hilbert spaces describing an ensemble of bosonic modes. Of particular interest is the case of bipartite systems composed by two modes, labeled as $i,j$, whose corresponding annihilation operators are $\hat{a}_{i,j}$. Their Hilbert space is spanned by the Fock product states $\ket{n_in_j}\equiv \ket{n_i}\otimes \ket{n_j}$, and accordingly coherent states and quasi-probability distributions now have two complex arguments $(\alpha_i,\alpha_j)$. For example, the $P$-representation reads
\begin{equation}\label{eq:PRepresentationBi}
\hat{\rho}=\int\mathrm{d}^2\alpha_i\mathrm{d}^2\alpha_j~P(\alpha_i,\alpha_j)\ket{\alpha_i\alpha_j}\bra{\alpha_i\alpha_j}.
\end{equation}

Nevertheless, there are genuine bipartite quantum states which cannot be expressed as a product of monomode states. One example is the two-mode squeezed state
\begin{equation}\label{eq:TwoModeSqueezed}
     \ket{\varepsilon}_{ij}\equiv U(\varepsilon)\ket{00},~U(\varepsilon)=e^{\varepsilon^*\hat{a}_i\hat{a}_j-\varepsilon\hat{a}^\dagger_i\hat{a}^\dagger_j}.
\end{equation}
In the same fashion of Eqs. (\ref{eq:SqueezingOperatorNormal}), (\ref{eq:SqueezedFock}), it is shown that
\begin{equation}
    \ket{\varepsilon}_{ij}=\frac{1}{\cosh r}\sum^\infty_{n=0}(-1)^ne^{i2n\theta}\tanh^n r\ket{nn}.
\end{equation}
Remarkably, the reduced state
\begin{equation}\label{eq:ThermalSqueezed}
\rho_i=\textrm{Tr}_j(\ket{\varepsilon}_{ij} \leftindex_{ij}{\bra{\varepsilon}})=\frac{1}{\cosh^2r}\sum^\infty_{n_i=0}(\tanh r)^{2n_i}\ket{n_i}  {\bra{n_i}}, %\sideset{_i}{\bra{n}}
\end{equation}
is then a thermal state, whose equivalent temperature for a mode with energy $\epsilon$ is obtained by $e^{-\beta \epsilon}=\tanh^2r$. The two-mode squeezed state is a non-classical state and it is of paramount importance in quantum optics because it describes the time-evolution of the so-called non-degenerate parametric amplifier, where one of the modes is designed as the signal and the other as the idler \cite{Walls2008}.

%with $\ket{n_i}$ a number state in the Fock space of the mode $i$, 

\subsection{Cauchy-Schwarz violation}
Experimentally, the quantumness of a bipartite state, such as the two-mode squeezed state (\ref{eq:TwoModeSqueezed}), can be characterized through the measurement of the first 
\begin{eqnarray}\label{eq:gdef}
g_{ij}\equiv \braket{\hat{a}_{i}^{\dagger}\hat{a}_{j}},~c_{ij}\equiv \braket{\hat{a}_{i}\hat{a}_{j}} ,
\end{eqnarray}
and second-order correlation functions
\begin{equation}\label{eq:GammaDef}
\Gamma_{ij}\equiv\braket{\hat{a}_{i}^{\dagger}
\hat{a}_{j}^{\dagger}\hat{a}_{j}\hat{a}_{i}}\geq 0\,.
\end{equation}
Since they are normal-ordered expectation values, they can be computed from the $P$-representation (\ref{eq:PRepresentationBi}). Interestingly, for classical states, $P(\alpha_i,\alpha_j)\geq 0$, and we can write the averages as a scalar product
\begin{equation}
g_{ij}=\int\mathrm{d}^2\alpha~P(\alpha_i,\alpha_j)\alpha^*_i\alpha_j\equiv (\alpha_i,\alpha_j)_C,
\end{equation}
with $c_{ij}=(\alpha^*_i,\alpha_j)_C$ and $\Gamma_{ij}=(|\alpha_i|^2,|\alpha_j|^2)_C$. By invoking the Cauchy-Schwarz (CS) inequality
\begin{equation}
    |(\alpha_i,\alpha_j)_C|\leq \sqrt{(\alpha_i,\alpha_i)_C(\alpha_j,\alpha_j)_C},
\end{equation}
one can prove that classical states obey the inequalities 
\begin{eqnarray}\label{eq:ClassicalIneq}
\nonumber |g_{ij}|^2&\leq& g_{ii}g_{jj},\\
    |c_{ij}|^2&\leq& g_{ii}g_{jj},\\
    \nonumber \Gamma_{ij}&\leq& \sqrt{\Gamma_{ii}\Gamma_{jj}}.
\end{eqnarray}
% \begin{equation}\label{eq:ClassicalIneq}
%     |c_{ij}|^2\leq g_{ii}g_{jj}\leq \Gamma_{ij}\leq \sqrt{\Gamma_{ii}\Gamma_{jj}}.
% \end{equation}
The violation of any of the above \textit{classical} CS inequalities requires a negative-valued $P$ function, a genuine signature of quantumness. The first violation of a CS inequality in photons was observed in 1974 \cite{Clauser1974}. In condensates, CS violation has been also observed in 2012 \cite{Kheruntsyan2012}.

Actual mathematical CS inequalities that are never violated can be proven for quantum operators, which we now review along the lines of the enlightening discussion from Adamek, Busch and Parentani \cite{Adamek2013}. For two operators $\hat{A},\hat{B}$, one can associate a scalar product to a quantum state $\hat{\rho}$ as
\begin{equation}\label{eq:ScalarProductOperator}
(\hat{A},\hat{B})_Q\equiv\braket{\hat{A}^{\dagger}\hat{B}}=\text{Tr}[\hat{A}^{\dagger}\hat{B}\hat{\rho}],
\end{equation}
which satisfies the usual properties of a scalar product, including the \textit{quantum} CS inequality
\begin{equation}\label{eq:CSmathematical}
|\braket{\hat{A}^{\dagger}\hat{B}}|^2\leq \braket{\hat{A}^{\dagger}\hat{A}}\braket{\hat{B}^{\dagger}\hat{B}}.
\end{equation}
Notice that the only assumption here is that $\hat{\rho}$ is a \textit{physical} quantum state, specifically, a non-negative operator, and this is satisfied by definition. We can now derive the quantum versions of the CS inequalities (\ref{eq:ClassicalIneq}), which are then strict mathematical inequalities. For instance, by substituting $\hat{A}=\hat{a}_{i}$ and $\hat{B}=\hat{a}_{j}$ in Eq. (\ref{eq:CSmathematical}), we find
\begin{equation}\label{eq:CSgijimpossible}
|g_{ij}|^2=|\braket{\hat{a}^{\dagger}_{i}\hat{a}_{j}}|^2
\leq\braket{\hat{a}^{\dagger}_{i}\hat{a}_{i}}\braket{\hat{a}^{\dagger}_{j}\hat{a}_{j}}=g_{ii}g_{jj}.
\end{equation}
This is the same CS inequality as in the classical case, so it is always verified. However, taking $\hat{A}=\hat{a}^{\dagger}_{i}$ and $\hat{B}=\hat{a}_{j}$ yields
\begin{equation}\label{eq:CScijpos}
|c_{ij}|^2=|\braket{\hat{a}_{i}\hat{a}_{j}}|^2
\leq\braket{\hat{a}_{i}\hat{a}^{\dagger}_{i}}
\braket{\hat{a}^{\dagger}_{j}\hat{a}_{j}}=(g_{ii}+1)g_{jj} \, .
\end{equation}
Interestingly, the \textit{quantum} CS inequality leaves the possibility of violating the \textit{classical} CS inequality $|c_{ij}|^2\leq g_{ii}g_{jj}$. In order to quantify this violation, we define the CS witness
\begin{eqnarray}\label{eq:CSviolation2}
\Delta_{ij}\equiv |c_{ij}|^2-g_{ii}g_{jj},
\end{eqnarray}
and denote the condition $\Delta_{ij}>0$ as {\it quadratic} CS violation. For the second-order correlation function, the associated quantum CS inequality is obtained by setting $\hat{A}=\hat{a}^{\dagger}_{i}\hat{a}_{i}$ and $\hat{B}=\hat{a}^{\dagger}_{j}\hat{a}_{j}$,
\begin{eqnarray}\label{eq:CSGammaijpos}
\nonumber |\Gamma_{ij}|^2&=&|\braket{\hat{a}^{\dagger}_{i}\hat{a}_{i}\hat{a}^{\dagger}_{j}\hat{a}_{j}}|^2
\leq\braket{\hat{a}^{\dagger}_{i}\hat{a}_{i}\hat{a}^{\dagger}_{i}\hat{a}_{i}}\braket{\hat{a}^{\dagger}_{j}\hat{a}_{j}\hat{a}^{\dagger}_{j}\hat{a}_{j}}\\
&=&(\Gamma_{ii}+g_{ii})(\Gamma_{jj}+g_{jj}),
\end{eqnarray}
which also leaves the possibility of violating the classical CS inequality $|\Gamma_{ij}|^2\leq\Gamma_{ii}\Gamma_{jj}$, as quantified by the CS witness
\begin{equation}\label{eq:CSviolation4}
\Theta_{ij}\equiv \Gamma_{ij}-\sqrt{\Gamma_{ii}\Gamma_{jj}}.
\end{equation}
In analogy to the quadratic CS violation, we refer to the condition $\Theta_{ij}>0$ as {\it quartic} CS violation. Remarkably, the origin of the violation of classical CS inequalities can be pin-pointed to the non-commutativity of quantum operators, a property present at the very core of quantum mechanics.

\subsection{Entanglement}

Entanglement is perhaps the most genuine quantum feature. It has been observed in a wide variety of systems as different as photons \cite{Aspect1982}, neutrinos \cite{Formaggio2016}, quarks \cite{ATLAS2024,CMS2024}, mesons \cite{Go2007}, atoms \cite{Hagley1997}, molecules \cite{Bao2023,Holland2023}, superconductors \cite{Steffen2006}, nitrogen-vacancy centers in diamond \cite{Pfaff2013}, and even macroscopic diamond itself \cite{Lee2011}. In general, entanglement is defined as the non-separability of the quantum state of a system \cite{Werner1989}. In turn, a quantum state in a bipartite Hilbert space is said to be separable \textit{iff} it can be written as a convex sum of product states,
\begin{equation}\label{eq:separability}
\hat{\rho}=\sum_n p_n\hat{\rho}^{(i)}_n\otimes\hat{\rho}^{(j)}_n,~ \sum_n p_n=1,~p_n\geq 0,
\end{equation}
where $\hat{\rho}^{(i),(j)}_n$ are states within the Hilbert subspaces associated to the $i,j$ modes, respectively. Classical states are separable, as directly seen from Eq. (\ref{eq:PRepresentationBi}).

In order to characterize entanglement, we make use of the generalized Peres-Horodecki (GPH) criterion \cite{Simon2000}, which extends the celebrated Peres-Horodecki (PH) criterion \cite{Peres1996,Horodecki1997} to continuous systems. The PH criterion results from the fact that, if $\hat{\rho}$ is separable, its partial transpose $\hat{\rho}_{t}$ with respect to one of the subsystems is also a physical density matrix and, in particular, a non-negative operator. Thus, the PH criterion states that, if $\hat{\rho}_{t}$ is not non-negative, then $\hat{\rho}$ is necessarily entangled. For $2\times 2$ and $2\times 3$ systems, the PH criterion is a necessary and sufficient condition for entanglement; in general, it is only a sufficient condition.

In order to obtain the partial transpose $\hat{\rho}_t$ of a density matrix $\hat{\rho}$, we make use of its Wigner function in phase space, $W(X)$, computed through the analogous version for bipartite systems of Eq. (\ref{eq:WignerPosition}), where we gather the phase-space variables in a single vector $X\equiv [q_i,p_i,q_j,p_j]^{T}$. Without loss of generality, we take the partial transpose with respect to the subsystem $j$, which amounts to transpose the matrix elements of $\hat{\rho}$ with respect to the Hilbert subspace of the mode $j$. It is straightforward to show then that the Wigner distribution $W_t(X)$ associated to $\hat{\rho}_t$ is simply given by 
\begin{equation}\label{eq:Wignertransposed}
W_t(X)=W(\Lambda X),~\Lambda=\rm{diag}[1,1,1,-1].
\end{equation}
Another way to put it is that the transposition operation amounts to a time reversal transformation in the Wigner function. 
%$W_t(q_i,p_i,q_j,p_j)=W(q_i,p_i,q_j,-p_j)$, or, in compact vector notation,

The effects of this seemingly innocuous transformation can become critical, as revealed when evaluating the uncertainties of the phase-space operators
\begin{equation}
\hat{Y}=\sum_{\alpha=1}^4u^\alpha\Delta\hat{X}_\alpha,
\end{equation}
where 
$\hat{X}\equiv [\hat{q}_i,\hat{p}_i,\hat{q}_j,\hat{p}_j]^{T}$ is the quantum version of the phase-space vector $X$, $\Delta \hat{X}\equiv \hat{X}-\braket{\hat{X}}$, and $u^\alpha$ are the components of an arbitrary four-dimensional complex vector $u$. Due to the positiveness of the scalar product (\ref{eq:ScalarProductOperator}),  $\braket{\hat{Y}^{\dagger}\hat{Y}}\geq 0$, which in compact vector notation reads
\begin{equation}\label{eq:Uncer}
u^{\dagger}Mu\geq0,~M_{\alpha\beta}=\braket{\Delta\hat{X}_{\alpha}\Delta\hat{X}_{\beta}}.
\end{equation}
This is an alternative expression of the uncertainty principle, holding for any complex vector $u$, which implies that $M$ must be a non-negative matrix, $M\geq 0$. 

For its computation, we separate the matrix $M$ into its symmetric and antisymmetric part as
\begin{eqnarray}\label{eq:Uncer}
\nonumber M&=&V+i\frac{L}{2},\\
V_{\alpha\beta}&=&\frac{\braket{\{\Delta\hat{X}_{\alpha},\Delta\hat{X}_{\beta}\}}}{2},\\
\nonumber  iL_{\alpha\beta}&=&\braket{[\Delta\hat{X}_{\alpha},\Delta\hat{X}_{\beta}]}=\braket{[\hat{X}_{\alpha},\hat{X}_{\beta}]}.
\end{eqnarray}
where $\{\ldots\}$ is the anticommutator. The matrix $V$ is the symmetric covariance matrix and, since it contains symmetric expectation values, it is readily evaluated with the help of the Wigner distribution,
\begin{eqnarray}\label{eq:SymmetricCovariances}
\braket{X_\alpha}&=&\int\mathrm{d}^4X~X_\alpha W(X),\\
\nonumber \frac{\braket{\{\Delta\hat{X}_{\alpha},\Delta\hat{X}_{\beta}\}}}{2}&=&\int\mathrm{d}^4X~\Delta X_\alpha \Delta X_\beta W(X).
\end{eqnarray}
On the other hand, the commutators between phase-space operators are proportional to the identity, and thus $L$ is a $4\times 4$ matrix independent of the state $\hat{\rho}$, 
\begin{eqnarray}\label{eq:XCommutators}
L=\left[\begin{array}{cc} J & 0\\
0 & J \end{array}\right],~J=\left[\begin{array}{cc}
0 & 1\\
-1 & 0
\end{array}\right],
\end{eqnarray}
$J$ being the symplectic matrix in two dimensions.

The above decomposition not only allows to evaluate $M$, but it also provides a straightforward way to derive the corresponding uncertainty principle for $\hat{\rho}_t$. Indeed, since its Wigner function $W_t(X)$ satisfies Eq. (\ref{eq:Wignertransposed}), it is immediate to see that the condition $\braket{\hat{Y}^\dagger\hat{Y}}_t\equiv \textrm{Tr}[\hat{Y}^\dagger\hat{Y}\hat{\rho}_t]\geq 0$ is equivalent to $M_t\geq 0$, with
\begin{eqnarray}\label{eq:Uncertrans}
M_t&\equiv&V_t+i\frac{L}{2},~V_t=\Lambda V \Lambda.
\end{eqnarray}
We can finally formulate quantitatively the GPH criterion: if $\hat{\rho}$ is separable, $\hat{\rho}_t$ must be a physical state satisfying the uncertainty principle, which implies $M_t\geq 0$. Therefore, if $M_t$ is not non-negative, the state is entangled. Notice that, since $M$ is always non-negative as the original $\hat{\rho}$ is a physical density matrix, by the Sylvester-Jacobi criterion, $M_t$ is non-negative \textit{iff} $\det M_t\geq 0$. The conditions $\det M,\det M_t\geq 0$ are respectively equivalent to $\mathcal{P}^{\pm}_{ij}\geq0$, where
\begin{eqnarray}\label{eq:GPHpm}
\mathcal{P}^{\pm}_{ij}&\equiv&\det A_i\det A_j+\left(\frac{1}{4}\mp \det C_{ij}\right)^2\\
\nonumber&-&\text{tr}(JA_iJC_{ij}JA_jJC_{ij}^{T})-\frac{1}{4}(\det A_i+\det A_j)
\end{eqnarray}
and the matrices $A_i,A_j,C_{ij}$ are the $2\times 2$ blocks forming the covariance matrix $V$, 
\begin{equation}\label{eq:WBlocks}
V=\left[\begin{array}{cc} A_{i} & C_{ij}\\
C^{T}_{ij} & A_{j} \end{array}\right].
\end{equation}
We can put together both conditions by defining the GPH function $\mathcal{P}_{ij}$ as
\begin{eqnarray}\label{eq:GPH}
\mathcal{P}_{ij}&\equiv&\det A_i\det A_j+\left(\frac{1}{4}-|\det C_{ij}|\right)^2\\
\nonumber&-&\text{tr}(JA_iJC_{ij}JA_jJC_{ij}^{T})-\frac{1}{4}(\det A_i+\det A_j),
\end{eqnarray}
where $\mathcal{P}_{ij}<0$ is a sufficient condition for entanglement. This is the entanglement witness that results from the GPH criterion. Notice that, whenever $\det C_{ij}\geq 0$, $\hat{\rho}$ is separable, because then $\mathcal{P}_{ij}=\mathcal{P}^{+}_{ij}\geq0$, so only states with $\det C_{ij}< 0$ can be entangled. 

%The condition $\mathcal{P}^{+}_{ij}\geq0$ is always satisfied since $\hat{\rho}$ is a physical density matrix, $M\geq 0$. However, when $\mathcal{P}^{-}_{ij}<0$, the state $\hat{\rho}_{t}$ is not physical and thus $\hat{\rho}$ is entangled.

In the usual case where the expectation values of the operators $\braket{\hat{X}_\alpha}=0$ vanish, the matrices $A_{k},C_{ij}$ are expressed in terms of the first-order correlation functions as
\begin{eqnarray}\label{eq:WBlocks}
\nonumber A_{k}&=&\left(g_{kk}+\frac{1}{2}\right)\mathbb{I}_2+\left[\begin{array}{cc}
\text{Re}~c_{kk} & \text{Im}~c_{kk}\\
\text{Im}~c_{kk} & -\text{Re}~c_{kk}
\end{array}\right],~k=i,j,\\
C_{ij}&=&
\left[\begin{array}{cc}
\text{Re}(g_{ij}+c_{ij}) & \text{Im}(g_{ij}+c_{ij})\\
\text{Im}(-g_{ij}+c_{ij}) & \text{Re}(g_{ij}-c_{ij})
\end{array}\right].
\end{eqnarray} 
Alternatively, we can work directly with the operators $\hat{X}_\alpha$ instead of their fluctuations $\Delta \hat{X}_\alpha$. This allows to prove that quadratic CS violation is a sufficient condition for the fulfillment of the GPH criterion. Indeed, suppose that $M_t\geq 0$. Then, we can define an associated scalar product as $(u,v)_t\equiv u^{\dagger}M_tv$, satisfying the CS inequality $|(u,v)_t|^2\leq (u,u)_t(v,v)_t$. By choosing \begin{equation}\label{eq:CSvectors4}
u=\frac{1}{\sqrt{2}}\left[\begin{array}{c}
0\\ 0\\ 1\\ i
\end{array}\right],~v=\frac{1}{\sqrt{2}}\left[\begin{array}{c}
1\\ i\\ 0\\ 0
\end{array}\right],
\end{equation}
we obtain the quadratic CS inequality $|c_{ij}|^2\leq g_{ii}g_{jj}$. Thus, quadratic CS violation implies that the matrix $M_t$ is not non-negative, and hence the GPH criterion is satisfied. 

% Notice that, if the expectation value of the annihilation and destruction operators is non-zero, we can simply modify the definition of the first-order correlations by substituting the operators $\hat{a}_k$ by their fluctuations $\Delta \hat{a}_k$.

More generally, from the definition of separability, Eq. (\ref{eq:separability}), we can apply the following chain of CS inequalities if the state is separable,  
\begin{align}\label{eq:separabilitycs}
\nonumber &|c_{ij}|=|\braket{\hat{a}_i\hat{a}_j}|=\left|\textrm{Tr}[\hat{a}_i\hat{a}_j\hat{\rho}]\right|=\left|\sum_np_n\braket{\hat{a}_i}_n\braket{\hat{a}_j}_n\right|\\
 &\leq\sum_n p_n|\braket{\hat{a}_i}_n\braket{\hat{a}_j}_n|\leq\sum_n p_n\sqrt{\braket{\hat{a}^\dagger_i\hat{a}_i}_n\braket{\hat{a}^\dagger_j\hat{a}_j}_n}\\
\nonumber &\leq  \sqrt{\sum_n p_n \braket{\hat{a}^\dagger_i\hat{a}_i}_n}\sqrt{\sum_n p_n \braket{\hat{a}^\dagger_j\hat{a}_j}_n}=\sqrt{g_{ii}g_{jj}},
\end{align}
where $\braket{\hat{A}}_n\equiv \textrm{Tr}[\hat{A}(\hat{\rho}^{(i)}_n\otimes\hat{\rho}^{(j)}_n)]$. Thus, quadratic CS violation is a sufficient condition for entanglement. However, the above derivation does not work for quartic CS violation. As a counterexample, the product of two number states $\hat{\rho}=\ket{nm}\bra{nm}$, with $n,m>0$, is clearly a separable state that nevertheless violates the quartic CS inequality.

Technically, the GPH criterion here explained is only a sufficient condition for the negativity of $\hat{\rho}_t$; in general, an infinite set of sufficient and necessary conditions for the negativity of $\hat{\rho}_t$, based on higher-order CS violations, can be derived \cite{Shchukin2005}. For multipartite entanglement, the role of the CS inequality is played by the H\"older inequality \cite{Wolk2014}, which is a generalization of the CS inequality. However, in practice, the GPH criterion provides a very powerful and simple tool to signal entanglement. In particular, for bipartite Gaussian states, the GPH criterion is a sufficient and necessary condition for entanglement \cite{Simon2000}.

% \begin{widetext}
%     \begin{eqnarray}\label{eq:GPHcomplete}
% \nonumber\mathcal{P}_{ij}&=&[g_{ii}g_{jj}-|c_{ij}|^2][(g_{ii}+1)(g_{jj}+1)-|c_{ij}|^2]+4\left(g_{ii}+\frac{1}{2}\right)\text{Re}(g_{ij}c^{*}_{jj}c_{ij})+4\left(g_{jj}+\frac{1}{2}\right)\text{Re}(g_{ij}c_{ii}c^{*}_{ij})-2\text{Re}(c^{2}_{ij}c^{*}_{ii}c^{*}_{jj})\\
% \nonumber&-&2\text{Re}(g^{2}_{ij}c_{ii}c^{*}_{jj})+|c_{ii}|^2|c_{jj}|^2+|g_{ij}|^4-|g_{ij}|^2(g_{ii}+g_{jj}+2g_{ii}g_{jj}+2|c_{ij}|^2)-g_{ii}(g_{ii}+1)|c_{jj}|^2-g_{jj}(g_{jj}+1)|c_{ii}|^2.\\
% \end{eqnarray}
% \end{widetext}

\subsection{Cauchy-Schwarz violation and entanglement in Andreev-Hawking radiation}

We finally switch back to analogue gravity and, in particular, we apply the above techniques to the study of Andreev-Hawking radiation in condensates. We first note that the mean-field formalism can be alternatively described in the Schr\"odinger picture by a coherent ansatz of the form
\begin{equation}
\ket{\Psi}=e^{\int\mathrm{d}\mathbf{x}\left[\Psi(\mathbf{x},t)\hat{\Psi}^\dagger(\mathbf{x})-\Psi^*(\mathbf{x},t)\hat{\Psi}(\mathbf{x})\right]}\ket{0}_{MB},
\end{equation}
where $\ket{0}_{MB}$ is the many-body vacuum containing no bosons, $\hat{\Psi}(\mathbf{x})\ket{0}_{MB}=0$. When this ansatz is inserted into a variational principle, such as the Dirac-Frenkel one, the time-dependent GP equation (\ref{eq:TDGP}) is retrieved.

%The equivalency of all these quantum phenomena for incoherent Gaussian states (\ref{eq:incoherent}) was originally proven in Ref. \cite{deNova2015}.

The genuine quantum character of the Andreev and Hawking effects is revealed by reexamining the relation of Eq. (\ref{eq:inoutmodesrelation}). After some algebra \cite{deNova2015}, it is shown that any matrix $S\in U(2,1)$ can be written as $S=B^{\dagger}S_{PA}A^{\dagger}$, with
% \begin{align}\label{eq:param}
% \nonumber S_{\rm PA}=\left[\begin{array}{ccc} e^{i\phi} &0&0\\
% 0&e^{-i2\theta}\cosh r&\sinh r\\
% 0&\sinh r &e^{i2\theta} \cosh r\end{array}\right]&\\
%  B\equiv \frac{1}{\sinh r}\left[\begin{array}{ccc} -S_{d1d2}&S_{ud2}&0\\
% S^*_{ud2}&S^*_{d1d2}&0\\
% 0&0&\sinh r\end{array}\right]&\\
% \nonumber A\equiv \frac{1}{\sinh r}\left[\begin{array}{ccc} -S_{d2d1}&S^*_{d2u}&0\\
% S_{d2u}&S^*_{d2d1}&0\\
% 0&0&\sinh r\end{array}\right]&
% \end{align}
\begin{align}\label{eq:param}
S_{PA}=\left[\begin{array}{ccc} e^{i\phi} &0&0\\
0&e^{-i2\theta}\cosh r&\sinh r\\
0&\sinh r &e^{i2\theta} \cosh r\end{array}\right],&\\
\nonumber A=\begin{bNiceArray}{cc|c}
\Block{2-2}<\large>{U_A} && 0  \\
  && 0 \\
  \hline
    0 & 0 & 1 
\end{bNiceArray},~U_A\equiv \frac{1}{\sinh r}\left[\begin{array}{cc} -S_{d2d1}&S^*_{d2u}\\
S_{d2u}&S^*_{d2d1}\end{array}\right],\\
\nonumber B=\begin{bNiceArray}{cc|c}
\Block{2-2}<\large>{U_B} && 0  \\
  && 0 \\
  \hline
    0 & 0 & 1 
\end{bNiceArray},~U_B\equiv\frac{1}{\sinh r}\left[\begin{array}{cc} -S_{d1d2}&S_{ud2}\\
S^*_{ud2}&S^*_{d1d2}\end{array}\right],&
\end{align}

% \begin{align}\label{eq:param}
%  B\equiv \frac{1}{\sinh r}\left[\begin{array}{ccc} -S_{d1d2}&S_{ud2}&0\\
% S^*_{ud2}&S^*_{d1d2}&0\\
% 0&0&\sinh r\end{array}\right]&\\
% \nonumber U_A\equiv \frac{1}{\sinh r}\left[\begin{array}{cc} -S_{d2d1}&S^*_{d2u}\\
% S_{d2u}&S^*_{d2d1}\end{array}\right]&
% \end{align}

% \begin{equation}
%     A\equiv \frac{1}{\sinh r}\left[\begin{array}{c|r} \displaystyle U_A & \begin{matrix} 0\\
% 0\end{matrix}\\
% \hline
% \begin{matrix}
%     0 & 0
% \end{matrix} & \begin{matrix} 1\end{matrix}\end{array}\right]
% \end{equation}

% \begin{equation*}
% \begin{bmatrix}
% \begin{matrix} a_{11} & a_{12} \\ a_{21} & a_{22} \end{matrix} & \begin{matrix} b_{11} & b_{12}\\ b_{21} & b_{22} \end{matrix}\\
% \begin{matrix} c_{11} & c_{12} \\ c_{21} & c_{22} \end{matrix} & \begin{matrix} d_{11} & d_{12} \\ d_{21} & d_{22} \end{matrix}
% \end{bmatrix}
% \end{equation*}

where
\begin{eqnarray}
    \nonumber e^{i\phi}&=&\det S,\\
    e^{i2\theta}\cosh r&=&S_{d2d2},\\
    \nonumber \sinh r&=&\sqrt{|S_{ud2}|^2+|S_{d1d2}|^2}=\sqrt{|S_{d2u}|^2+|S_{d2d1}|^2}.
\end{eqnarray}
Since the matrices $U_A,U_B$ are unitary, the above transformation amounts to a change of basis in the normal $u-d1$ sector,
\begin{equation}\label{eq:AHTransform}
\left[\begin{array}{c}
\hat{a}_{s}\\
\hat{a}_{AH}
\end{array}\right]=U^\dagger_A 
\left[\begin{array}{c}
\hat{a}_{u}\\
\hat{a}_{d1}
\end{array}\right],~\left[\begin{array}{c}
\hat{b}_{s}\\
\hat{b}_{AH}
\end{array}\right]=U^\dagger_B 
\left[\begin{array}{c}
\hat{b}_{u}\\
\hat{b}_{d1}
\end{array}\right].
\end{equation}
In this new basis, the scattering of the spectator (s) channel is a fully normal, unitary process. On the other hand, the normal hybrid Andreev-Hawking (AH) channel couples to the anomalous $d2$ channel as
\begin{equation}\label{eq:ParametricInOut}
\left[\begin{array}{c}
\hat{b}_{AH}\\
\hat{b}^{\dagger}_{d2}
\end{array}\right] =\left[\begin{array}{cc}
e^{-i2\theta}\cosh r&\sinh r\\
\sinh r &e^{i2\theta} \cosh r\end{array}\right]\left[\begin{array}{c}
\hat{a}_{AH}\\
\hat{a}_{d2}^{\dagger}
\end{array}\right]\, .
\end{equation}
Apart from a trivial phase, this is precisely the same Bogoliubov relation arising from a two-mode squeezed operator $U(\varepsilon)$, Eq. (\ref{eq:TwoModeSqueezed}).
Moreover, the unitary transformation (\ref{eq:AHTransform}) does not change the incoming or outgoing vacua. As a result, the ``in'' vacuum can be regarded as a two-mode squeezed state for the ``out'' modes, and the spontaneous production of outgoing AH modes simultaneously describes both the Andreev and Hawking effects,
\begin{align}\label{eq:AHEffect}
\nonumber &\bra{0_{\rm{in}}}\hat{b}_{AH}^{\dagger}(\omega)\hat{b}_{AH}(\omega')\ket{0_{\rm{in}}}=\delta(\omega-\omega')\sinh^2r(\omega)\\
 &=\delta(\omega-\omega')\left[|S_{ud2}(\omega)|^2+|S_{d1d2}(\omega)|^2\right].
\end{align}
Thus, in the quantum optics jargon, the joint AH effect is nothing else than a non-degenerate parametric amplifier, where the outgoing AH mode is the signal and the outgoing partner $d2$ mode is the idler. Furthermore, as discussed in Eq. (\ref{eq:ThermalSqueezed}), the reduced state for the AH mode is a thermal state, in analogy with the original prediction of Hawking. Notice, however, that this transformation applies independently to each frequency, which results in a $\omega$-dependent effective temperature.

Although the AH mode is very appealing from the theoretical point of view, in practice, the channels $i,j=u,d1,d2$ are still more convenient since i) they are located in separated physical regions and ii) quantum states are typically defined in this basis. Specifically, a most relevant class of quantum states is the family of incoherent Gaussian states characterized by the following first and second-order momenta in the incoming basis:
\begin{eqnarray} \label{eq:incoherent}
\nonumber \braket{\hat{a}_{i}(\omega)\hat{a}_{j}(\omega')}&=& \braket{\hat{a}_{i}(\omega)}=0\\
\braket{\hat{a}_{i}^{\dagger}(\omega)\hat{a}_{j}(\omega')}
&=&n_i(\omega)\delta_{ij}\delta(\omega-\omega')\, .
\end{eqnarray}
Since the state is Gaussian, any higher-order momentum can be put in terms of these expectation values via Wick theorem. The above class of states includes thermal states of incoming modes. A typical choice \cite{Macher2009a,Busch2014} is that where the incoming modes have thermalized in the comoving frame of the condensate,
\begin{equation}\label{eq:thermalin}
    n_i(\omega)=\frac{1}{e^{\beta\Omega_{i-in}(\omega)}-1},
\end{equation}
with $\Omega_{i-in}(\omega)=\Omega(k_{i-in}(\omega))$ the comoving Bogoliubov frequency in the corresponding asymptotic region; see Eq. (\ref{eq:dispersionrelation}). In order to assess the quantumness of the AH effect for incoherent Gaussian states (\ref{eq:incoherent}), we compute the first- and second-order correlation functions of Eqs. (\ref{eq:gdef}) and (\ref{eq:GammaDef}) for the ``out'' modes at given $\omega$,
\begin{eqnarray}\label{eq:CSGamma}
\nonumber g_{ij}(\omega)&\equiv&\braket{\hat{b}_{i}^{\dagger}(\omega)\hat{b}_{j}(\omega)},\\
c_{ij}(\omega)&\equiv&\braket{\hat{b}_{i}(\omega)\hat{b}_{j}(\omega)},\\
\nonumber\Gamma_{ij}(\omega)&\equiv&\braket{\hat{b}_{i}^{\text{\ensuremath{\dagger}}}(\omega)\hat{b}_{j}^{\dagger}(\omega)\hat{b}_{j}(\omega)\hat{b}_{i}(\omega)},
\end{eqnarray}
where we obviate all Dirac delta factors; a discussion on how to regularize the infinities resulting from $\delta(\omega=0)$ is present at the beginning of Sec. \ref{subsec:ExperimentalCS}. %, for the moment,

\begin{figure*}[t]
\begin{tabular}{@{}ccc@{}}\stackinset{l}{0pt}{t}{0pt}{(a)}{\includegraphics[width=0.66\columnwidth]{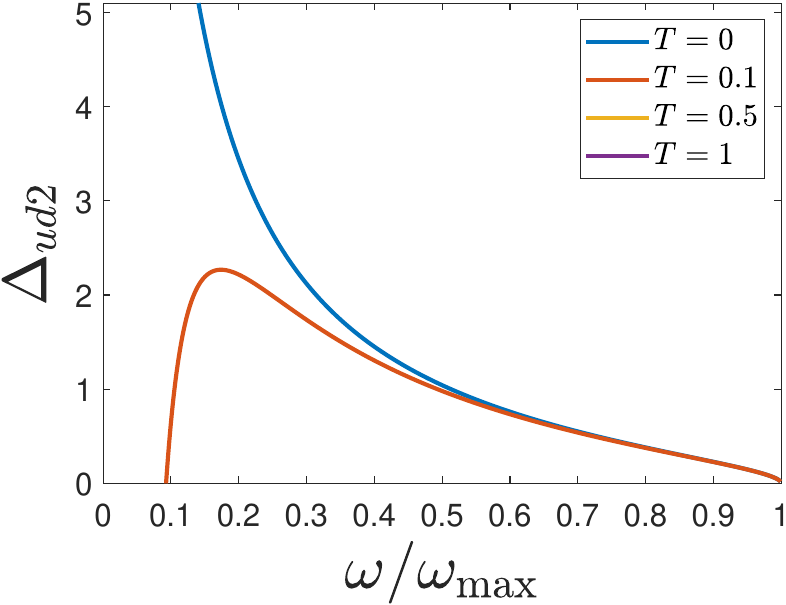}}~ &~
    \stackinset{l}{0pt}{t}{0pt}{(b)}
    {\includegraphics[width=0.66\columnwidth]{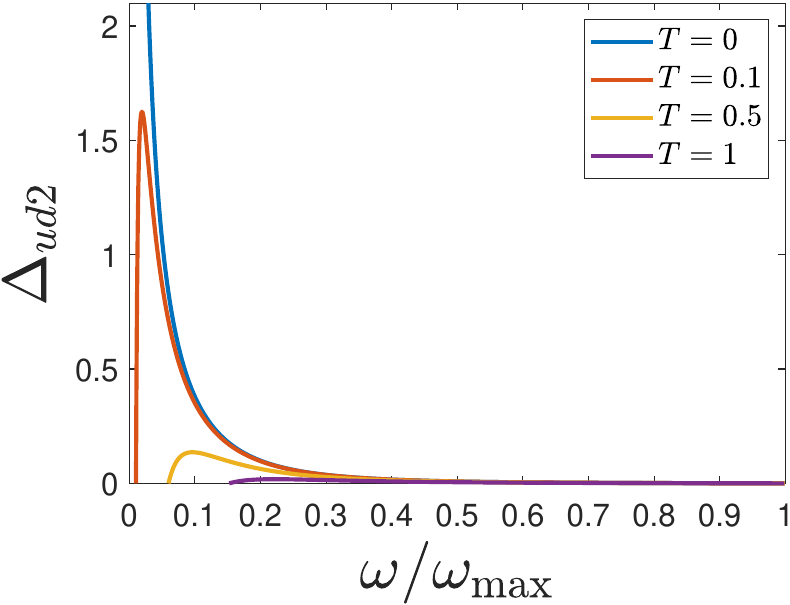}}~ &~ 
    \stackinset{l}{0pt}{t}{0pt}{(c)}{\includegraphics[width=0.66\columnwidth]{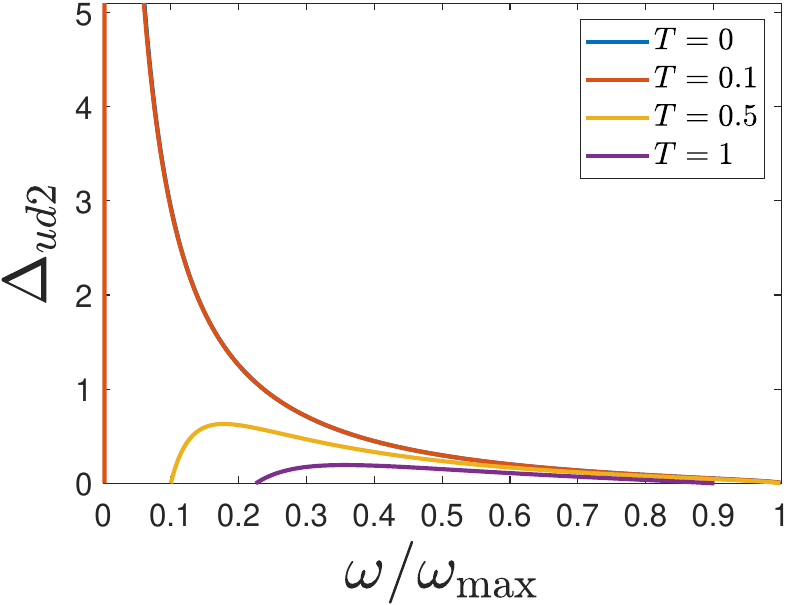}}  \\
    \stackinset{l}{0pt}{t}{0pt}{(d)}{\includegraphics[width=0.66\columnwidth]{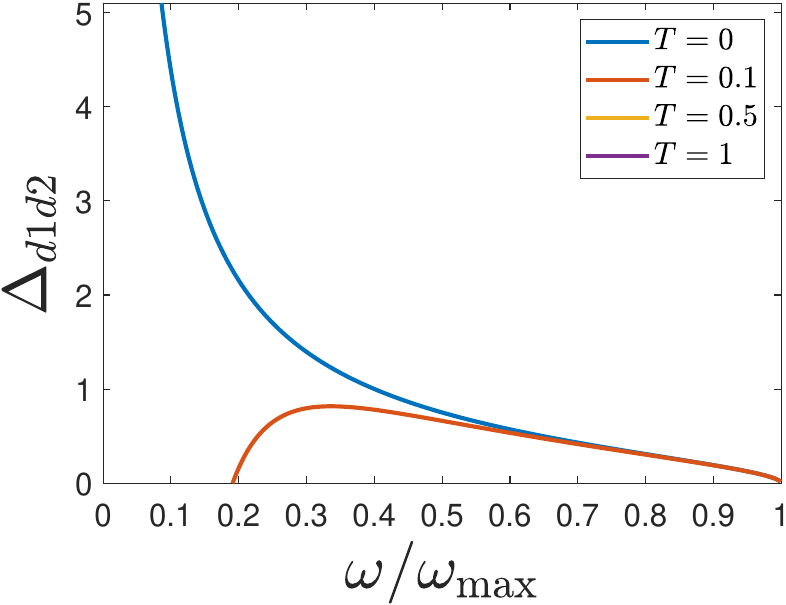}}~ &~
    \stackinset{l}{0pt}{t}{0pt}{(e)}
    {\includegraphics[width=0.66\columnwidth]{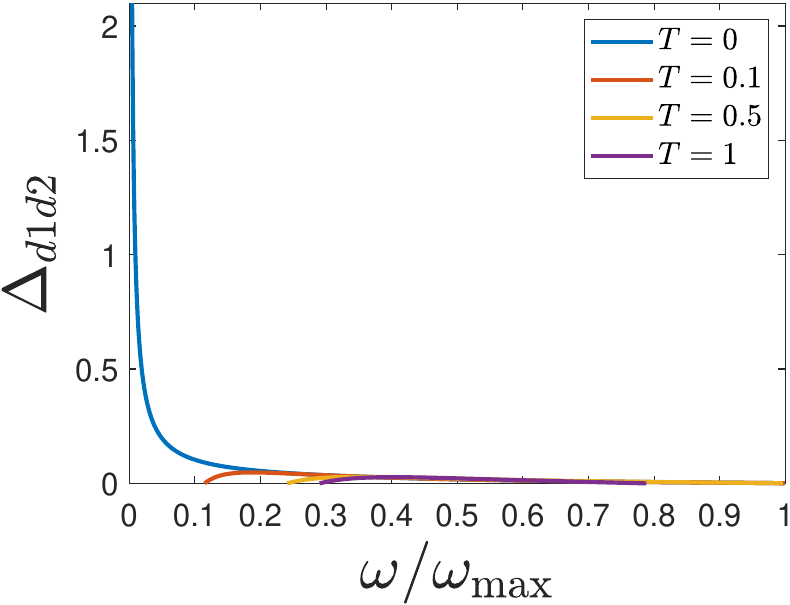}}~ &~ 
    \stackinset{l}{0pt}{t}{0pt}{(f)}{\includegraphics[width=0.66\columnwidth]{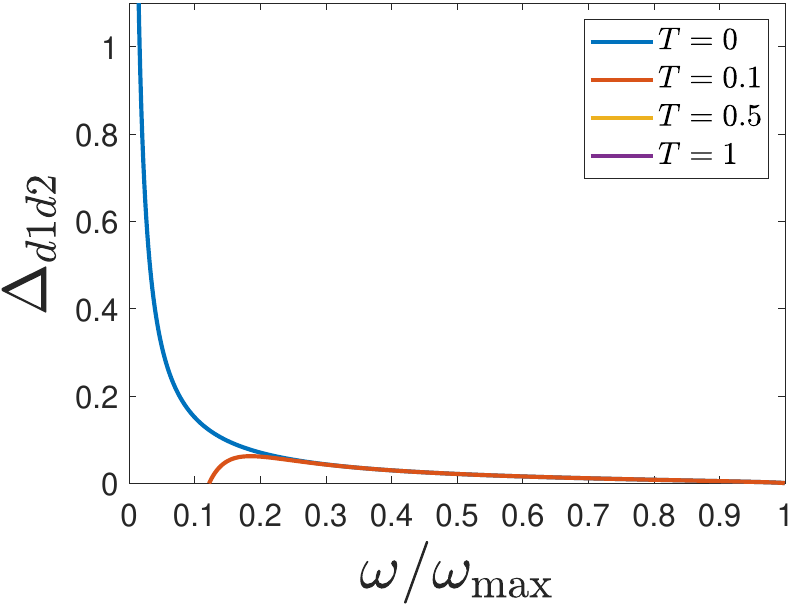}}
\end{tabular}
\caption{Entanglement witness $\Delta_{Id2}$ as function of $\omega$ for an incoherent thermal Gaussian state, Eqs. (\ref{eq:incoherent}), (\ref{eq:thermalin}). Different colors label different temperatures, as indicated in the legend of each panel. a)-c) $\Delta_{ud2}$ for the BH solutions of Fig. \ref{fig:BHModels}a-c. d)-f) Same as a)-c) but for $\Delta_{d1d2}$.}
\label{fig:Witness}
\end{figure*}

\begin{figure*}[t]
\begin{tabular}{@{}cc@{}}\stackinset{l}{0pt}{t}{0pt}{(a)}{\includegraphics[width=\columnwidth]{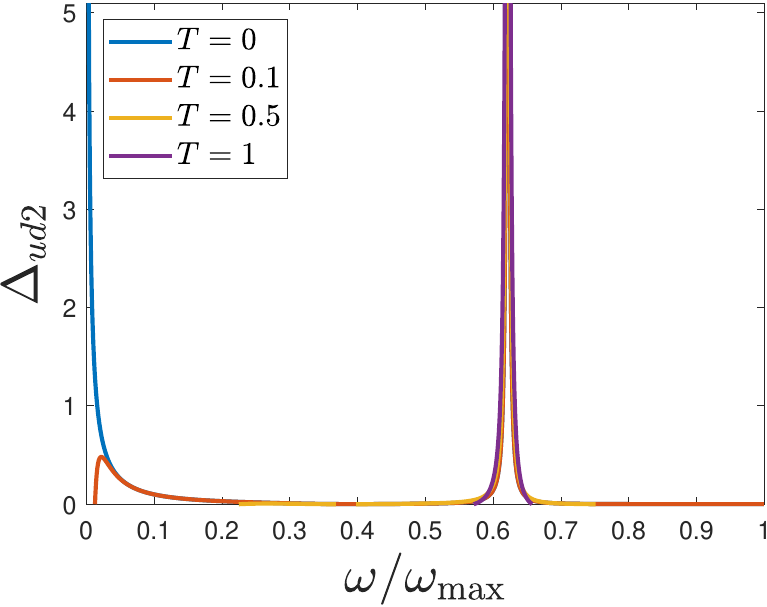}}~ &~
    \stackinset{l}{0pt}{t}{0pt}{(b)}
    {\includegraphics[width=0.975\columnwidth]{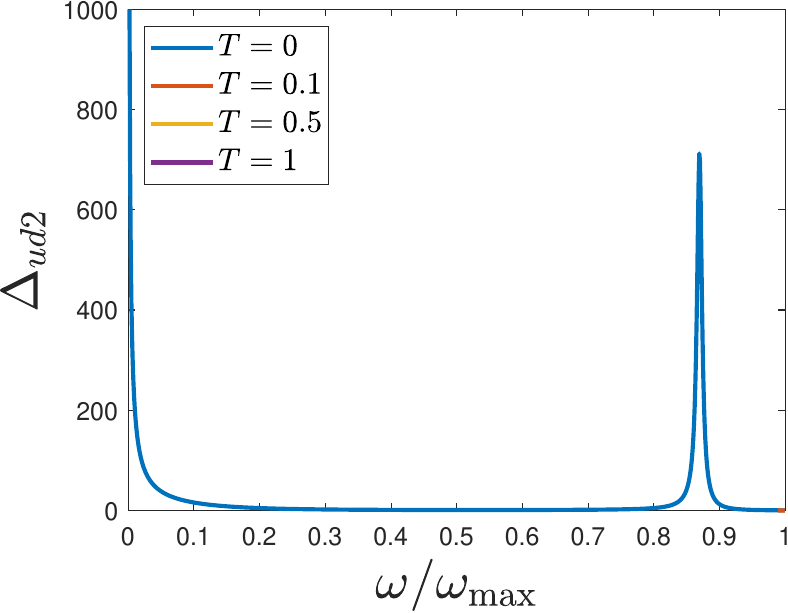}}\\
    \stackinset{l}{0pt}{t}{0pt}{(c)}{\includegraphics[width=\columnwidth]{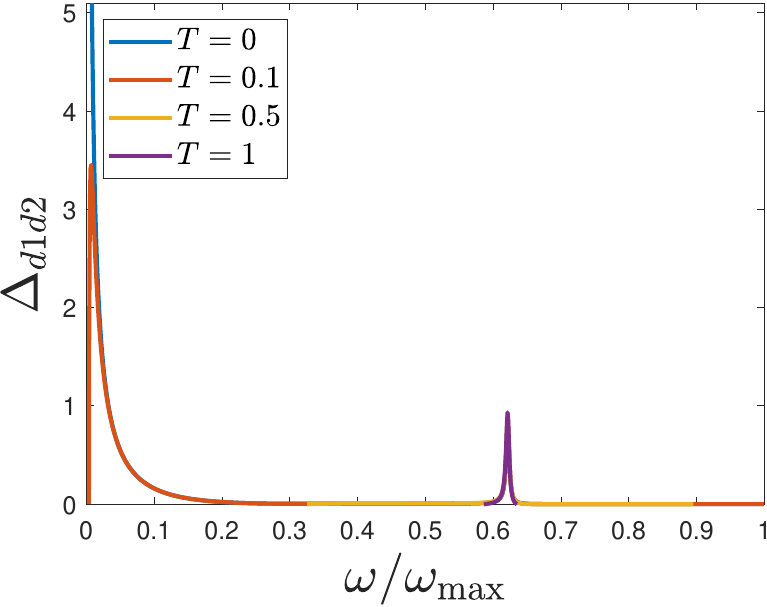}}~ &~
    \stackinset{l}{0pt}{t}{0pt}{(d)}
    {\includegraphics[width=0.975\columnwidth]{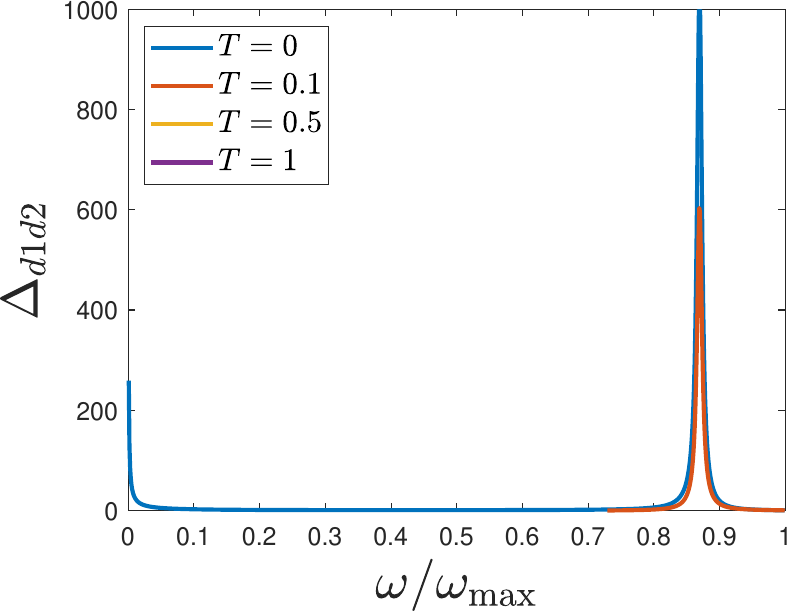}}
\end{tabular}
\caption{Entanglement witness $\Delta_{Id2}$ as function of $\omega$ for an incoherent thermal Gaussian state, Eqs. (\ref{eq:incoherent}), (\ref{eq:thermalin}). Different colors label different temperatures, as indicated in the legend of each panel. a)-b) $\Delta_{ud2}$ for the resonant BH solutions of Fig. \ref{fig:Resonant}a-b. c)-d) Same as a)-b) but for $\Delta_{d1d2}$.}
\label{fig:ResonantWitness}
\end{figure*}

From the previous definitions, it is easy to show that the only non-zero first-order correlations are
\begin{eqnarray}\label{eq:quadraticorrs}
\nonumber g_{IJ}&=& \alpha_{I}^{\dagger}\cdot\alpha_{J}, \\
g_{d2d2}&=&|\alpha_{d2}|^{2}-1,\\
\nonumber c_{Id2}&=&\alpha_{d2}^{\dagger}\cdot\alpha_{I}.
\end{eqnarray}
On the other hand, since we are working with Gaussian states, the second-order correlation functions are expressed in terms of the first-order ones as
\begin{eqnarray}\label{eq:CSWick}
\nonumber \Gamma_{IJ} & = & g_{II}g_{JJ}+|g_{IJ}|^2=|\alpha_{I}|^{2}|\alpha_{J}|^{2}+|\alpha_{I}^{\dagger}\cdot\alpha_{J}|^2,\\
\nonumber \Gamma_{d2d2} & = & 2g^2_{d2d2}=2(|\alpha_{d2}|^{2}-1)^{2},\\
 \Gamma_{Id2} & = & |c_{Id2}|^2+g_{II}g_{d2d2}=\\
\nonumber &=&|\alpha_{d2}^{\dagger}\cdot\alpha_{I}|^{2}+|\alpha_{I}|^{2}(|\alpha_{d2}|^{2}-1)\, .
\end{eqnarray}
In the above equations, we have compacted the notation by defining the complex vector
\begin{eqnarray}\label{eq:defcorrs}
\alpha_i(\omega)&\equiv&\left[\begin{array}{c}
S_{iu}(\omega)\sqrt{n_u(\omega)}\\
S_{id1}(\omega)\sqrt{n_{d1}(\omega)}\\
S_{id2}(\omega)\sqrt{n_{d2}(\omega)+1}
\end{array}\right]\, .
\end{eqnarray}
This is equivalent to working with complex vectors $\alpha_i$ with components $(\alpha_i)_j=S_{ij}$, and whose scalar product is given by the metric $g=\textrm{diag}[n_u,n_{d1},n_{d2}+1]$.

With the help of the above results, we quantify the quartic and quadratic CS violations using $\Delta_{ij}(\omega)$ and $\Theta_{ij}(\omega)$, Eqs. (\ref{eq:CSviolation2}), (\ref{eq:CSviolation4}), respectively. Notice that, as explained after Eq. (\ref{eq:CSgijimpossible}), no CS violation is possible for $g_{ud1}$. Moreover, the condition for quartic CS violation in the normal $u-d1$ correlations is equivalent to that for quadratic CS violation, $\Theta_{ud1}=\Delta_{ud1}=|g_{ud1}|^2-g_{uu}g_{d1d1}\leq 0$. Thus, only the anomalous correlations characterizing the Andreev and Hawking effects can give rise to genuine quantum correlations, quantified by
\begin{eqnarray}\label{eq:CSnonseparable}
\Delta_{Id2}&=&|\alpha_{d2}^{\dagger}\cdot\alpha_{I}|^{2}-|\alpha_{I}|^{2}(|\alpha_{d2}|^{2}-1),\\
\nonumber \Theta_{Id2}&=&\Delta_{Id2}.
\end{eqnarray}
Similarly, only anomalous processes can be entangled. Since our state is Gaussian, entanglement is equivalent to the condition $\mathcal{P}_{Id2}(\omega)<0$, Eq. (\ref{eq:GPH}), where the GPH function reads
\begin{equation}\label{eq:GPHanomalous1}
\mathcal{P}_{Id2}=-\Delta_{Id2}[(g_{II}+1)(g_{d2d2}+1)-|c_{Id2}|^2].
\end{equation}
Since $(g_{II}+1)(g_{d2d2}+1)\geq |c_{Id2}|^2$, Eq. (\ref{eq:CScijpos}), we observe that, for the class of states (\ref{eq:incoherent}), entanglement, quadratic and quartic CS violations are all equivalent conditions, $\Delta_{Id2}(\omega)>0$. Therefore, we can use $\Delta_{Id2}(\omega)$ simultaneously as both entanglement and CS witness. By invoking the pseudounitarity of $S$, Eq. (\ref{eq:pseudounitarity}), a simple explicit expression for $\Delta_{Id2}$ can be derived:
\begin{eqnarray}\label{eq:WitnessExplicit}
\nonumber \Delta_{Id2}&=&|S_{Id2}|^2(1+n_u+n_{d1}+n_{d2})\\
    &-&|S_{d2d1}|^2n_u-|S_{d2u}|^2n_{d1}\\
    \nonumber &-&|S_{I'u}|^2n_{d1}n_{d2}-|S_{I'd1}|^2n_{u}n_{d2}-|S_{I'd2}|^2n_{u}n_{d1},
\end{eqnarray}
where here $I'=d1,u$ is the complementary normal channel to $I=u,d1$.

We compute $\Delta_{Id2}(\omega)$ for different BH solutions assuming a thermal quantum state in the incoming channels, Eqs. (\ref{eq:incoherent}), (\ref{eq:thermalin}). In particular, at zero temperature, there is entanglement across the whole Andreev-Hawking spectrum,
\begin{equation}
    \Delta_{Id2}(\omega,T=0)=|S_{Id2}(\omega)|^2>0.
\end{equation}
At finite temperature, $n_i(\omega)\neq 0$; in particular, $n_{u}(\omega)$ is the only divergent occupation number since $\Omega_{u-\rm{in}}(\omega)\sim \omega$ at low frequencies, while the other comoving frequencies are finite at zero frequency since their wavevector is $k_{i-\rm{in}}(\omega=0)=k_{\rm{BCL}}$, $i=d1,d2$. Moreover, by invoking pseudounitarity and noting that $|S_{ij}(\omega)|^2\sim 1/\omega$ only for $j=d1,d2$, we have that $|S_{Id2}|^2<|S_{d2d1}|^2$ in this regime, which implies that there is no entanglement close to $\omega\simeq 0$ at finite temperature. On the other side of the spectrum, $\omega\simeq \omega_{\rm{max}}$, $|S_{Id2}(\omega)|^2\sim \sqrt{\omega_{\rm{max}}-\omega}$, and entanglement is lost again.

The results for the non-resonant BH solutions of Fig. \ref{fig:BHModels} are depicted in Fig. \ref{fig:Witness}. We observe that, for the flat-profile model (left column), entanglement only survives at low temperatures and high frequencies for both Andreev and Hawking radiation. However, for the delta and waterfall models (center and right columns, respectively), entanglement is present even at relatively high temperatures of the order of the chemical potential for the Hawking effect (upper row), with a significant reduction of the entanglement signal for the Andreev effect (lower row). These features can be understood from i) the  smallness of $\omega_{\rm max}$ for the flat-profile model results in large occupation numbers $n_i(\omega)$ even at low temperatures, and ii) for the delta and waterfall models, $|S_{d1d2}(\omega)|^2\ll |S_{ud2}(\omega)|^2$ in the relevant part of the spectrum, explaining the reduction of the Andreev entanglement.

The above picture is modified for the resonant structures of Fig. \ref{fig:Resonant}, as shown in Fig. \ref{fig:ResonantWitness}. For the double-delta configuration (left column), there is a strong entanglement signal near the resonant peak, close to the zero-temperature value even at high temperatures for both the Andreev and Hawking effects. For the resonant flat-profile configuration (right column), although entanglement is lost for low temperatures because of the smallness of $\omega_{\rm max}$, the Andreev entanglement signal is now larger than the Hawking one close to the resonant peak since there $|S_{d1d2}(\omega)|^2> |S_{ud2}(\omega)|^2$. These results clearly demonstrate the potential of resonant structures for studying the quantumness of the AH effect, in particular that of the Andreev effect, greatly attenuated in non-resonant structures.

To conclude the discussion, we examine the physical implications of both incoherent and Gaussianity conditions. Gaussianity results from the BdG approximation, see Eq. (\ref{eq:GrandCanonicalEnergy}), and it is only expected to be broken in a strongly interacting regime beyond Bogoliubov. In 1D, this regime is only achieved at low densities [see discussion after Eq. (\ref{eq:RelativeCorrelationsBdG})]. On the other hand, incoherence results from a stationary description in which the incoming modes that eventually scatter at the horizon are populated in the asymptotic regions, for instance by a thermal bath. 

\begin{table}[!tb]
\centering
\begin{tabular}[c]{|c|c|c|r|c|}
\hline
Incoherent &  Gaussian & $\text{CS4}$ & $\text{CS2}$~~~ & GPH\\
\hline
\textcolor{blue}{\checkmark} & \textcolor{blue}{\checkmark}& $\Leftrightarrow$ GPH &$\Leftrightarrow$ GPH &$\Leftrightarrow$ Entanglement\\
\hline
$\textcolor{blue}{\checkmark}$ & $\textcolor{red}{\times}$ &~EI & $\Leftrightarrow$  GPH & $\Rightarrow$ Entanglement\\
\hline
$\textcolor{red}{\times}$& $\textcolor{blue}{\checkmark}$ &~EI & $\Rightarrow$  GPH &$\Leftrightarrow$ Entanglement\\
\hline
$\textcolor{red}{\times}$ &$\textcolor{red}{\times}$ &~EI & $\Rightarrow$ GPH &$\Rightarrow$ Entanglement\\
\hline
\end{tabular}
\caption{Logical relations between the different quantum criteria considered here, where CS2, CS4 stand for quadratic and quartic CS violations, and EI means ``Entanglement independent''.}
\label{TCS}
\end{table}

%in which the physics is driven by the asymptotic regions, where the incoming modes that eventually scatter at the horizon are populated. 

%In particular, incoherence is expected to emerge for an adiabatic formation of the horizon.

If we remove the incoherence in the incoming basis but not Gaussianity, the GPH criterion is still equivalent to entanglement. However, the quadratic CS violation becomes only a sufficient condition for the GPH criterion and the quartic CS violation is then independent of the GPH criterion [see Eq. (\ref{eq:CSvectors4}) and subsequent discussion]. On the other hand, if we remove Gaussianity but not incoherence, the GPH criterion, now only a sufficient entanglement condition, is independent from the quartic CS violation but equivalent to the quadratic CS violation. For a general state which is neither Gaussian nor incoherent in the incoming basis, the quadratic CS violation is only a sufficient condition for the GPH criterion, which in turn is a sufficient condition for the presence of entanglement; all of them are independent from quartic CS violation. The above logical relations are summarized in Table \ref{TCS}.

\subsection{Experimental considerations and conceptual exports}\label{subsec:ExperimentalCS}

Experimental proposals for the detection of CS violation and entanglement in an analogue context were performed using time-of-flight (TOF) techniques in Ref. \cite{deNova2014} (see also Ref. \cite{Boiron2015}), and density-density correlations in Ref. \cite{Steinhauer2015}. A regularization procedure for the infinities arising from the Dirac delta factors ignored in Eq. (\ref{eq:CSGamma}), based on the use of windowed Fourier transforms, was presented in Refs. \cite{deNova2014,deNova2015} for each experimental scheme, respectively. The first claimed observation of the Hawking effect \cite{Steinhauer2016} was indeed based on the detection of the quadratic CS violation $\Delta_{ud2}>0$ from the measurement of density-density correlations. Later observations of the Hawking effect \cite{deNova2019,Kolobov2021}, although exhibiting a more accurate agreement with the theoretical prediction for the Hawking correlations $c_{ud2}(\omega)$, did not address the question of entanglement or CS violation. Entanglement in the Andreev-Hawking effect still represents an active topic of research, and it can be of interest for quantum technologies, since then an analogue horizon behaves as a source of entangled phonons.

Remarkably, the concepts and techniques discussed here to signal quantum correlations can be exported to qudits through the $P$-representation developed in Ref. \cite{Giraud2008}. For simplicity, we consider the particular case of a qubit, i.e., a two-level quantum system consisting of two states $\ket{+},\ket{-}$. Furthermore, for the sake of definiteness, we identify these states as spin projections along the $z$-axis of a spin-1/2 particle, $\sigma_z\ket{\pm}=\pm\ket{\pm}$, although the discussion can be trivially adapted to any type of qubit by using the pseudospin formalism. 

The quantum state of a spin-1/2 particle is described by a $2\times 2$ density matrix of the form
\begin{equation}\label{eq:QubitState}
    \rho=\sum_{n,m=\pm} \rho_{nm}\ket{n}\bra{m}=\frac{I_2+\mathbf{B}\cdot \boldsymbol{\sigma}}{2}
\end{equation}
with $I_n$ the $n\times n$ identity matrix and $\boldsymbol{\sigma}$ a vector containing the Pauli matrices. The Bloch vector $\mathbf{B}$ fully determines the quantum state, representing the spin polarization of the particle, $\mathbf{B}=\braket{\boldsymbol{\sigma}}=\textrm{Tr}[\boldsymbol{\sigma}\rho]$, where we have invoked the trace orthogonality of the Pauli matrices, $\textrm{Tr}[\sigma_i]=0$, $\textrm{Tr}[\sigma_i\sigma_j]=2\delta_{ij}$. The above expression can be regarded as the qubit version of the Fock expansion (\ref{eq:GeneralQuantumStateFock}), where the spin states $\ket{\pm}$ play the role of the number states $\ket{n}$. Actually, the Fock space of a fermion mode is also a qubit, spanned by the two number states $\ket{n}$, $n=0,1$.

The analogue of the coherent states $\ket{\alpha}$ are the spin-coherent states $\ket{\mathbf{\hat{n}}}$, 
\begin{equation}\label{eq:CoherentState1/2}
    \ket{\mathbf{\hat{n}}}=\cos\frac{\theta}{2}e^{-i\frac{\phi}{2}}\ket{+}+\sin\frac{\theta}{2}e^{i\frac{\phi}{2}}\ket{-},
\end{equation}
which are spin eigenstates with maximum projection along the direction of the unit vector $\mathbf{\hat{n}}=[\sin \theta \cos\phi,\sin \theta \sin\phi,\cos \theta]$, $(\mathbf{\hat{n}}\cdot \boldsymbol{\sigma})\ket{\mathbf{\hat{n}}}=\ket{\mathbf{\hat{n}}}$. These states also form an overcomplete basis as
\begin{equation}\label{eq:Identity}
|\braket{\mathbf{\hat{n}}|\mathbf{\hat{n}}'}|^2=\frac{1+\mathbf{\hat{n}}\cdot \mathbf{\hat{n}}'}{2},\,\frac{1}{2\pi}\int \mathrm{d}\Omega\ket{\mathbf{\hat{n}}}\bra{\mathbf{\hat{n}}}=\sum_{n=\pm}\ket{n}\bra{n}=1,
\end{equation}
where $\Omega$ is the solid angle associated to the unit vector $\mathbf{\hat{n}}$. Moreover, there are spin-squeezed states for larger total spin \cite{Kitagawa1993}, which play a central role in metrology. 

The spin-coherent basis allows for a $P$-representation, 
\begin{equation}\label{eq:PRepresentationQubit}
    \rho=\int \mathrm{d}\Omega~P(\mathbf{\hat{n}})\ket{\mathbf{\hat{n}}}\bra{\mathbf{\hat{n}}},~\int \mathrm{d}\Omega~P(\mathbf{\hat{n}})=1.
\end{equation}
The function $P(\mathbf{\hat{n}})$ is also a quasi-probability distribution, representing the spin analogue of the Glauber-Sudarshan $P$-function. Since the density matrix (\ref{eq:QubitState}) is diagonalized in the $\ket{\pm \mathbf{\hat{n}}}$ basis, with $\mathbf{\hat{n}}\parallel \mathbf{B}$, $P(\mathbf{n})$ can be always chosen as non-negative for one qubit.

Genuine quantum signatures arise when considering two spin-1/2 particles, labeled as $i,j$, whose total quantum state is described by a $4\times 4$ density matrix 
\begin{equation}\label{eq:TwoQubitState}
    \rho=\frac{I_4+\mathbf{B}_i\cdot \boldsymbol{\sigma}_i+\mathbf{B}_j\cdot \boldsymbol{\sigma}_j+\boldsymbol{\sigma}_i\cdot \mathbf{C}\cdot \boldsymbol{\sigma}_j}{4},
\end{equation}
with $\boldsymbol{\sigma}_{i,j}$ a vector containing the Pauli matrices in each subspace, $\mathbf{B}_{i,j}$ the individual spin polarizations, and $\mathbf{C}$ the spin-correlation matrix. From the definition of separability, Eq. (\ref{eq:separability}), and by using that any one-qubit state is diagonalized in the spin-coherent basis, we find that separability for two-qubit states is equivalent to the existence of a non-negative $P$-representation, 
\begin{equation}\label{eq:PRepresentation2Qubit}   \rho=\int\mathrm{d}\Omega_i\mathrm{d}\Omega_j~P(\mathbf{n}_i,\mathbf{n}_j)\ket{\mathbf{n}_i\mathbf{n}_i}\bra{\mathbf{n}_i\mathbf{n}_j},~P(\mathbf{n}_i,\mathbf{n}_j)\geq 0.
\end{equation}
Hence, the absence of a non-negative $P$ representation is automatically a signature of entanglement. This implies that any CS violation is then a sufficient condition for entanglement. For instance, 
\begin{align}\label{eq:CSViolationSpin}
\nonumber |\textrm{Tr}[\mathbf{C}]|&=\left|\braket{\boldsymbol{\sigma}_i\cdot \boldsymbol{\sigma}_j}\right|=\left|\int\mathrm{d}\Omega_i\mathrm{d}\Omega_j~P(\mathbf{n}_i,\mathbf{n}_j)\mathbf{n}_i\cdot\mathbf{n}_j\right|\\
\nonumber &\leq \int\mathrm{d}\Omega_i\mathrm{d}\Omega_j~P(\mathbf{n}_i,\mathbf{n}_j)\left|\mathbf{n}_i\cdot\mathbf{n}_j\right|\\
&\leq \int\mathrm{d}\Omega_i\mathrm{d}\Omega_j~P(\mathbf{n}_i,\mathbf{n}_j)=1
\end{align}
is a classical CS inequality, based on the non-negativity of the $P$-function. Qualitatively, we can understand this CS inequality as the fact that the classical average of the scalar product of two unit vectors (such as the spin orientations $\mathbf{n}_i,\mathbf{n}_j$) is never larger than one. Thus, 
\begin{equation}\label{eq:EntanglementWitness}
    \Delta\equiv -\textrm{Tr}[\mathbf{C}]-1>0
\end{equation}
represents a CS violation that provides an entanglement witness.

Using the analogies above as a pipeline, and inspired by the techniques discussed here for the study of quantum Andreev-Hawking radiation, as well as by the fact that the Standard Model is based on a relativistic quantum field theory in a flat spacetime, it was recently shown that quantum correlations can be also studied at the LHC \cite{Afik2021}. In particular, it was proven that the spin quantum state of a pair of top-antitop quarks, the most massive fundamental particles known to exist, can be fully reconstructed from their decay products, implementing the so-called quantum tomography in quantum information jargon. This is possible because the large top mass is translated into a short lifetime that avoids any other process, including hadronization, to affect its spin before the decay. 

Another remarkable source of inspiration was the study of quantum steering in Hawking radiation by Robertson, Michel and Parentani \cite{Robertson2017}, which directly motivated the analysis of steering in top quarks \cite{Afik2023}. In general, the study of quantum information in high-energy physics is becoming an active topic of research (see for instance Refs. \cite{Fabbrichesi2021,Severi2022,Aguilar2022,Aoude2022,Barr2022,Bernal2023,Morales2023,Cheng2023,Sakurai2024,Dong2024,Afik2025}); the interested reader is referred to Ref. \cite{Afik2022} for a pedagogical introduction to the topic, aimed at an audience outside particle physics. Moreover, the experimental proposal of Ref. \cite{Afik2021} has been implemented by both the ATLAS and CMS collaborations \cite{ATLAS2024,CMS2024}, leading to the first observation of entanglement in quarks and to the highest-energy entanglement detection ever achieved. Specifically, the entanglement witness (\ref{eq:EntanglementWitness}) was directly measured from the angular distribution of the separation between the leptons arising from the top-antitop decay, obtaining $\Delta>0$ with more than $5\sigma$ (the standard candle for discovery in particle physics), which also represents the violation of a CS inequality.

\section{Black-hole lasers}\label{sec:BHL}

Another main topic of research involving resonant analogue configurations is the so-called black-hole laser (BHL) \cite{Corley1999}. The BHL emerges in a configuration similar to that of resonant BH solutions, but now the asymptotic downstream region is again subsonic. As a result, a BHL displays a pair of BH/WH horizons, and the resulting finite-size supersonic cavity becomes unstable due to the successive bouncing of Andreev-Hawking radiation between the horizons. 

%In condensates, the BHL effect arises due to the superluminal Bogoliubov dispersion relation, which allows the $d2$-in mode to travel back to the BH horizon. 

Qualitatively, we can understand the BHL instability as the partner $d2$ modes from the Andreev-Hawking effect being reflected at the WH horizon as $d2$-in modes that bounce back towards the BH, further stimulating the production of Andreev-Hawking radiation and thus leading to a process of self-amplification, similar to that occurring in a lasing cavity. Quantitatively, the BHL effect is characterized by a discrete BdG spectrum of dynamical instabilities, computed by extending the usual scattering problem [see Eq. (\ref{eq:solitonspinors}) and subsequent discussion] to complex frequencies and retaining only the asymptotically bounded modes outside the cavity. This procedure bears some resemblance to the computation of the discrete spectrum of bound states for an attractive potential in the Schr\"odinger equation, where the usual scattering problem for positive energies is extended to negative energies, keeping only the exponentially decaying solutions at infinity.

A systematic procedure for the quantization of the unstable lasing modes was provided by Finazzi and Parentani \cite{Finazzi2010}. For simplicity, we discuss the case of a single unstable BdG mode $z_I$ with complex frequency $\omega=\gamma+i\Gamma$, where $\gamma$ is the real, oscillatory part of the frequency, and $\Gamma$ is the imaginary part of the frequency, determining the growth rate of the instability. We also assume that the mode is non-degenerate, which means that $\gamma\neq0$ so $\omega\neq -\omega^*$. In general, any dynamically unstable mode $z_I$ has associated a stable mode $z_S$ with frequency $\omega^*$ \cite{Leonhardt2003}. Their eigenvalue equation reads 
\begin{equation}
M_0z_I=\omega z_I,~M_0z_S=\omega^*z_S,~
\end{equation}
with $M_0$ the BdG matrix operator, Eq. (\ref{eq:BdGEigenmode}). Because of their complex frequency, both modes have zero norm $(z_S|z_S)=(z_I|z_I)=0$; see Eq. (\ref{eq:EigenOrto}). However, we can choose their normalization such that
\begin{equation}\label{eq:NormalizationLasing}
    (z_S|z_I)=-(\bar{z}_I|\bar{z}_S)=1.
\end{equation}
Properly normalized states are defined through
\begin{equation}\label{eq:UnstableToNormalModes}
Z_{+}\equiv \frac{1}{\sqrt{2}}(z_I+z_S),~Z_{-}\equiv\frac{1}{\sqrt{2}}(\bar{z}_I-\bar{z}_S),
\end{equation}
% \begin{eqnarray}\label{eq:UnstableToNormalModes}
% Z_{+}&=&\frac{1}{\sqrt{2}}(z_I+z_S),\\
% \nonumber Z_{-}&=&\frac{1}{\sqrt{2}}(\bar{z}_I-\bar{z}_S)
% \end{eqnarray}
satisfying 
\begin{equation}\label{eq:BHLOrto}
    (Z_{+}|Z_{+})=(Z_{-}|Z_{-})=1,~(\bar{Z}_{-}|Z_{+})=(Z_{-}|Z_{+})=0.
\end{equation}

As a result, their quantum amplitudes 
\begin{equation}
\hat{a}_{\pm}=(Z_{\pm}|\hat{\Phi})
\end{equation}
do behave as proper annihilation operators [see Eq. (\ref{eq:Aniquilacion})]. Their time evolution is easily derived from the quantum amplitudes of the original complex eigenmodes,
\begin{equation}\label{eq:UnstableToNormalAmplitudes}
\hat{a}_{I}=(z_S|\hat{\Phi})=\frac{1}{\sqrt{2}}(\hat{a}_{+}+\hat{a}^{\dagger}_{-}),~\hat{a}_{S}=(z_I|\hat{\Phi})=\frac{1}{\sqrt{2}}(\hat{a}_{+}-\hat{a}^{\dagger}_{-}).
\end{equation}
which are not annihilation operators as $[\hat{a}_{I},\hat{a}^\dagger_{I}]=[\hat{a}_{S},\hat{a}^\dagger_{S}]=0$, $[\hat{a}_{I},\hat{a}^\dagger_{S}]=[\hat{a}_{S},\hat{a}^\dagger_{I}]=1$. Nevertheless, they evolve as expected from Eq. (\ref{eq:BogoliubovQuantum}),
\begin{eqnarray}
i\partial_t\hat{a}_{I}&=&(z_S|M_0\hat{\Phi})=\omega \hat{a}_{I}\Longrightarrow
\hat{a}_{I}(t)=\hat{a}_{I}e^{-i\omega t},\\
\nonumber  i\partial_t\hat{a}_{S}&=&(z_I|M_0\hat{\Phi})=\omega^*\hat{a}_{S}\Longrightarrow
\hat{a}_{S}(t)=\hat{a}_{S}e^{-i\omega^* t}.
\end{eqnarray}
To invert the relation, it is quite convenient to employ matrix notation. First, Eq. (\ref{eq:UnstableToNormalAmplitudes}) can be rewritten as 
\begin{equation}\label{eq:LasingNormalRelation}
   \left[\begin{array}{c} \hat{a}_{I}\\
\hat{a}_{S}\end{array}\right]=U \left[\begin{array}{c} \hat{a}_{+}\\
\hat{a}^\dagger_{-}\end{array}\right],~ U\equiv e^{-i\frac{\pi}{4}\sigma_y}\sigma_z=\frac{1}{\sqrt{2}}\left[\begin{array}{rr} 1 & 1\\
1 &-1\end{array}\right].
\end{equation}
The matrix $U=U^\dagger=U^{-1}$ describes a spin inversion in the $x$-$y$ plane plus a rotation of $\pi/2$ around the $y$-axis, satisfying $U\sigma_{x,z} U^\dagger=U^\dagger\sigma_{x,z} U=\sigma_{z,x}$ and $U\sigma_y U^\dagger=U^\dagger\sigma_y U=-\sigma_y$. In this notation, the time evolution of the  amplitudes of the complex modes simply reads
\begin{equation}
    \left[\begin{array}{c} \hat{a}_{I}(t)\\
\hat{a}_{S}(t)\end{array}\right]=\left[\begin{array}{c} e^{-i\omega t}\hat{a}_{I}\\
e^{-i\omega^* t}\hat{a}_{S}\end{array}\right]=e^{-i\gamma t}e^{\Gamma t \sigma_z}\left[\begin{array}{c} \hat{a}_{I}\\
\hat{a}_{S}\end{array}\right].
\end{equation}
As a result, we trivially find
\begin{align}
\left[\begin{array}{c} \hat{a}_{+}(t)\\
\hat{a}^\dagger_{-}(t)\end{array}\right]&=U^\dagger \left[\begin{array}{c} \hat{a}_{I}(t)\\
\hat{a}_{S}(t)\end{array}\right]=e^{-i\gamma t}e^{\Gamma t \sigma_x}\left[\begin{array}{c} \hat{a}_{+}\\
\hat{a}^\dagger_{-}\end{array}\right]\\
\nonumber &=e^{-i\gamma t}\left[\begin{array}{cc} \cosh \Gamma t & \sinh \Gamma t\\
\sinh \Gamma t &\cosh \Gamma t\end{array}\right]\left[\begin{array}{c} \hat{a}_{+}\\
\hat{a}^\dagger_{-}\end{array}\right].
\end{align}
% By inverting the relation, we find
% \begin{eqnarray}
% \nonumber \hat{a}_{+}(t)&=&\frac{\hat{a}_{I}(t)+\hat{a}_{S}(t)}{\sqrt{2}}=e^{-i\gamma t}\left[ \hat{a}_{+}\cosh\Gamma t+\hat{a}^{\dagger}_{-}\sinh\Gamma t\right],\\
% \nonumber \hat{a}^{\dagger}_{-}(t)&=&\frac{\hat{a}_{I}(t)-\hat{a}_{S}(t)}{\sqrt{2}}=e^{-i\gamma t}\left[ \hat{a}_{+}\sinh\Gamma t+\hat{a}^{\dagger}_{-}\cosh\Gamma t\right],\\
% \end{eqnarray}
This is a similar evolution to that of a non-degenerate parametric amplifier. This can be better seen by examining the contribution from these modes to the field spinor $\hat{\Phi}$ of Eq. (\ref{eq:QuantumFieldFluctuations}),
\begin{eqnarray}
\nonumber \hat{\Phi}_L&=&Z_{+}\hat{a}_{+}+Z_{-}\hat{a}_{-}+\bar{Z}_{+}\hat{a}^{\dagger}_{+}+\bar{Z}_{-}\hat{a}^{\dagger}_{-}\\
&=&z_{I}\hat{a}_{I}+z_{S}\hat{a}_{S}+\bar{z}_{I}\hat{a}^{\dagger}_{I}+\bar{z}_{S}\hat{a}^{\dagger}_{S}.
\end{eqnarray}
When inserted into the Bogoliubov expansion for the grand-canonical Hamiltonian (\ref{eq:GrandCanonicalEnergy}), which governs the dynamics, we obtain an orthogonal contribution to that of the regular Bogoliubov sector with real frequencies,
\begin{eqnarray}
   \nonumber \hat{K}_L&=&\frac{1}{2}(\hat{\Phi}_L|M_0\hat{\Phi}_L)=\left[\hat{a}^\dagger_{I}~\hat{a}^\dagger_{S}\right]\left[\begin{array}{cc}0& \Omega^*\\
\Omega & 0\end{array}\right]\left[\begin{array}{c} \hat{a}_I\\
\hat{a}_S\end{array}\right]\\&=&\left[\hat{a}^\dagger_{+}~\hat{a}_{-}\right]\left(\gamma \sigma_z-\Gamma \sigma_y\right)\left[\begin{array}{c} \hat{a}_{+}\\
\hat{a}^\dagger_{-}\end{array}\right]\\
\nonumber&=&\gamma(\hat{a}^{\dagger}_{+}\hat{a}_{+}-\hat{a}^{\dagger}_{-}\hat{a}_{-})
+i\Gamma(\hat{a}^{\dagger}_{+}\hat{a}^{\dagger}_{-}-\hat{a}_{+}\hat{a}_{-}),
\end{eqnarray}
where we neglect zero-point contributions. This is precisely the same Hamiltonian of a non-degenerate parametric amplifier [see Eq. (\ref{eq:TwoModeSqueezed}) and ensuing discussion]. 
%with $\varepsilon=-\Gamma t$ 
Another remarkable feature of dynamical instability is that there is no well-defined vacuum for the unstable modes \cite{Ribeiro2022}. This is immediately seen by noticing that any Bogoliubov transformation
\begin{equation}
\left[\begin{array}{c} \hat{b}_{+}\\
\hat{b}^\dagger_{-}\end{array}\right]=e^{u \sigma_x}\left[\begin{array}{c} \hat{a}_{+}\\
\hat{a}^\dagger_{-}\end{array}\right]=\left[\begin{array}{cc} \cosh u& \sinh u\\
\sinh u &\cosh u\end{array}\right]\left[\begin{array}{c} \hat{a}_{+}\\
\hat{a}^\dagger_{-}\end{array}\right],
\end{equation}
leaves invariant $\hat{K}_L$, with each $ \hat{b}_{\pm}$ giving rise to a different vacuum.

The above derivations can be easily adapted to the case of a degenerate unstable mode, $\gamma=0$. This implies that $\omega=-\omega^*$, so $z_I,z_S$ are proportional to $\bar{z}_{I},\bar{z}_{S}$. In particular, one can set $\bar{z}_{I}=z_I$ and $\bar{z}_{S}=-z_S$, and thus Eq. (\ref{eq:NormalizationLasing}) is again satisfied. However, now there is only one independent normalizable mode
\begin{equation}\label{eq:UnstableToNormalModesDegenerate}
Z\equiv \frac{1}{\sqrt{2}}(z_I+z_S),~\bar{Z}\equiv\frac{1}{\sqrt{2}}(z_I-z_S),
\end{equation}
whose amplitude $\hat{a}=(Z|\hat{\Phi})$ is a proper annihilation operator. Consequently, Eq. (\ref{eq:LasingNormalRelation}) now reads
\begin{equation}\label{eq:LasingNormalRelationDegenerate}
\left[\begin{array}{c} \hat{a}_{I}\\
\hat{a}_{S}\end{array}\right]=U \left[\begin{array}{c} \hat{a}\\
\hat{a}^\dagger\end{array}\right],
\end{equation}
giving rise to the time evolution
\begin{equation}
\left[\begin{array}{c} \hat{a}\\
\hat{a}^\dagger\end{array}\right]=e^{\Gamma t \sigma_x}\left[\begin{array}{c} \hat{a}\\
\hat{a}^\dagger\end{array}\right]=\left[\begin{array}{cc} \cosh \Gamma t & \sinh \Gamma t\\
\sinh \Gamma t &\cosh \Gamma t\end{array}\right]\left[\begin{array}{c} \hat{a}\\
\hat{a}^\dagger\end{array}\right],
\end{equation}
which results from the grand-canonical Hamiltonian
\begin{equation}
    \hat{K}=i\Gamma\frac{(\hat{a}^\dagger)^2-\hat{a}^2}{2}.
\end{equation}
This is indeed the Hamiltonian of a degenerate parametric amplifier, where the same mode is both the signal and the idler. Interestingly, this Hamiltonian is, after a trivial phase transformation, that of an unstable harmonic oscillator,
\begin{equation}\label{eq:UnstablePendulum}
    \hat{H}=\omega\frac{\hat{p}^2-\hat{q}^2}{2}=-\omega\frac{(\hat{a}^\dagger)^2+\hat{a}^2}{2}.
\end{equation}
The resulting time-evolution operator $e^{-i\hat{H} t}$ is just the squeezing operator (\ref{eq:Squeezing}) with a linearly increasing amplitude $\varepsilon=-i\omega t$. Therefore, dynamically unstable modes can be understood as unstable harmonic oscillators, in contrast to the regular Bogoliubov modes of purely real frequency, which behave as normal harmonic oscillators.

Returning to the BHL, the above quantization procedure is applied separately to each unstable lasing mode. Due to the exponential parametric amplification of the lasing modes, at some point the Bogoliubov approximation ceases to be valid and one needs to take into account higher-order interacting terms to describe the dynamics. Consequently, we separate our discussion of the BHL effect following the three different stages of its time evolution: i) short times, when the dynamics is still governed by the linear BdG equations; ii) intermediate times, when the evolution is driven solely by the dominant unstable mode up to the saturation regime, where the full interacting Hamiltonian is required again; and iii) long times, when the system reaches its final state after the collapse of the metastable state achieved in the saturation regime.

\subsection{Short times: Linear and non-linear spectra}

\begin{figure*}[t]
\begin{tabular}{@{}ccc@{}}\stackinset{l}{0pt}{t}{0pt}{(a)}{\includegraphics[width=0.66\columnwidth]{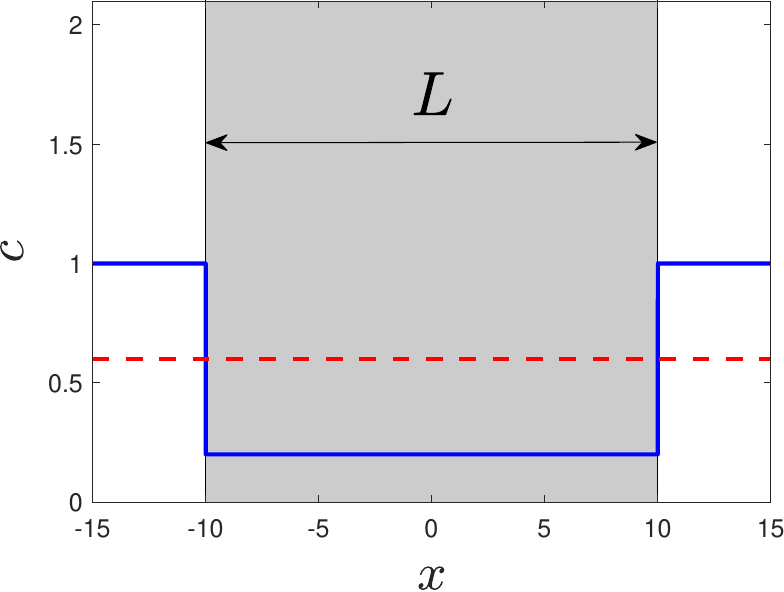}}~ &~
    \stackinset{l}{0pt}{t}{0pt}{(b)}
    {\includegraphics[width=0.66\columnwidth]{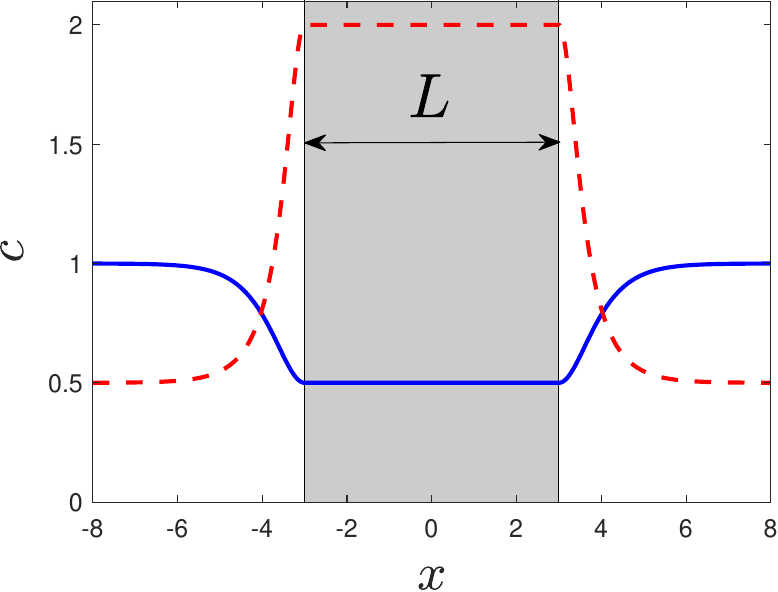}}~ &~ 
    \stackinset{l}{0pt}{t}{0pt}{(c)}{\includegraphics[width=0.66\columnwidth]{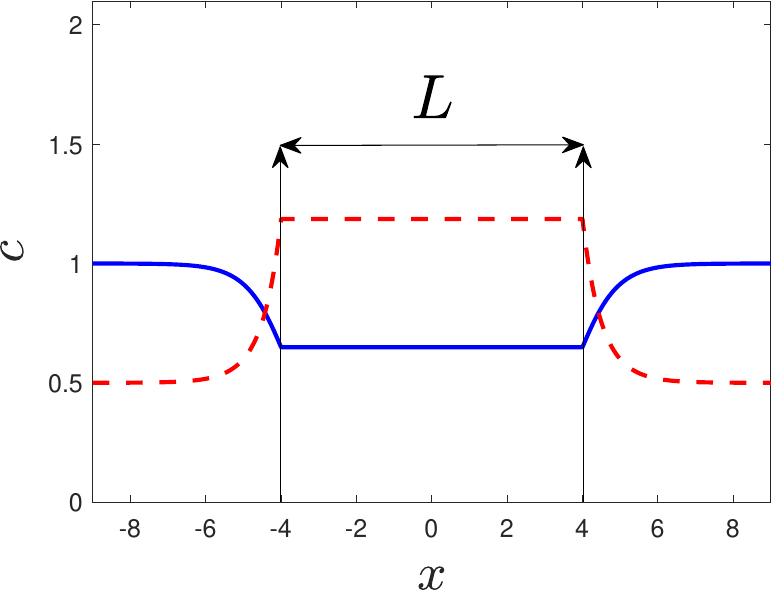}}  \\
    \stackinset{l}{0pt}{t}{0pt}{(d)}{\includegraphics[width=0.66\columnwidth]{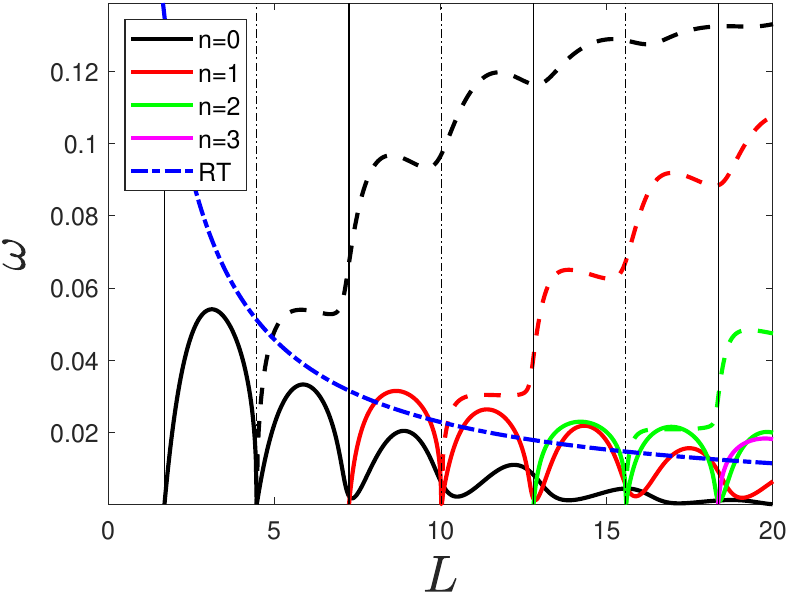}}~ &~
    \stackinset{l}{0pt}{t}{0pt}{(e)}
    {\includegraphics[width=0.66\columnwidth]{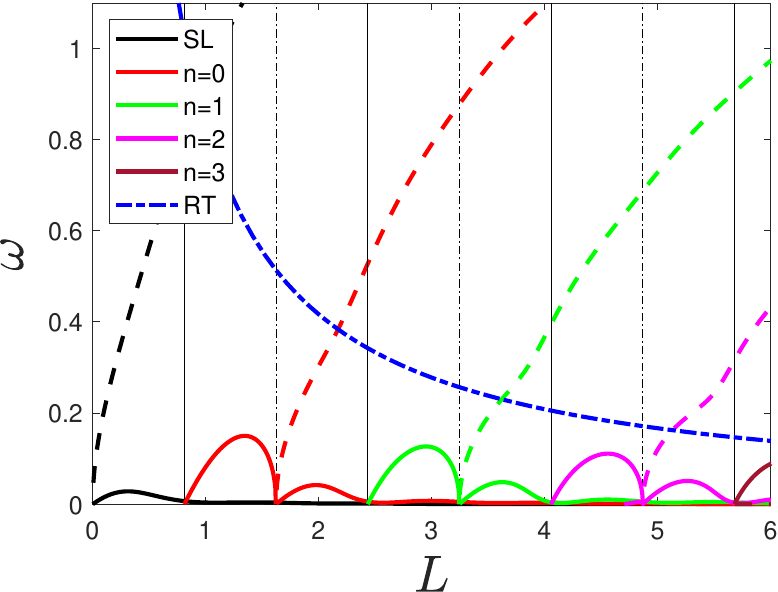}}~ &~ 
    \stackinset{l}{0pt}{t}{0pt}{(f)}{\includegraphics[width=0.66\columnwidth]{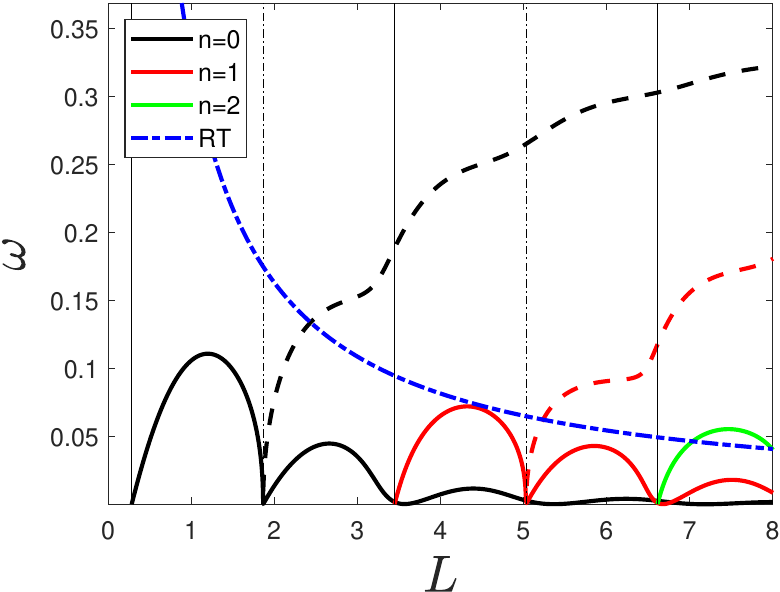}} \\
    \stackinset{l}{0pt}{t}{0pt}{(g)}{\includegraphics[width=0.66\columnwidth]{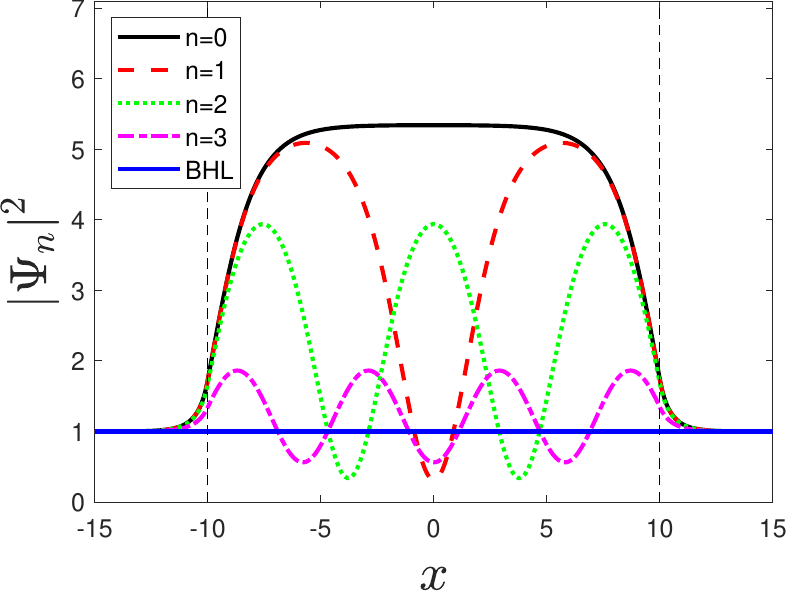}}~ &~
    \stackinset{l}{0pt}{t}{0pt}{(h)}
    {\includegraphics[width=0.66\columnwidth]{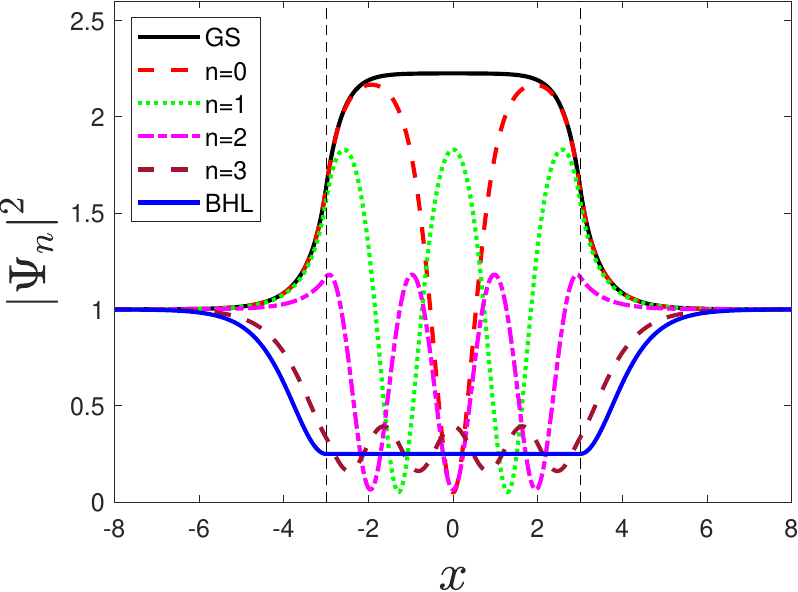}}~ &~ 
    \stackinset{l}{0pt}{t}{0pt}{(i)}{\includegraphics[width=0.66\columnwidth]{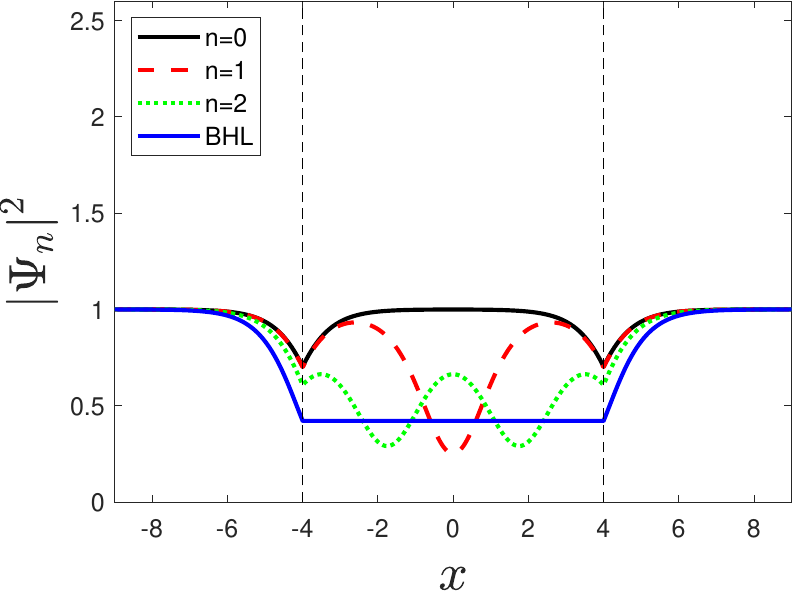}} 
\end{tabular}
\caption{a)-c) BHL solutions resulting from mirroring the BH solutions of Figs. \ref{fig:BHModels}a-c. d)-f) Linear BdG spectrum of dynamical instabilities for the BHL solutions of a)-c). Solid (dashed) lines represent the imaginary (real) part $\Gamma_n$ ($\gamma_n$) of the complex frequency $\omega_n$. The dash-dotted blue line is the inverse of the round-trip time. Solid (dashed) vertical lines mark the critical lengths $L_n$ ($L_{n+1/2}$). g)-i) Non-linear spectrum of stationary GP solutions $\Psi_n$ for the background configurations of a)-c), where the color code is chosen to match the associated lasing modes in d)-f). The density profile of the initial BHL solutions a)-c) is depicted as solid blue.}
\label{fig:BHLModels}
\end{figure*}

The microscopic derivation of Ref. \cite{Finazzi2010} was extended in another seminal work by Michel and Parentani \cite{Michel2013} using a simple analytical model based on the flat-profile configuration, Fig. \ref{fig:BHModels}a, where the flow velocity is homogeneous, $v(x)=q$, and the speed of sound is changed to $c(x)=c_2<q$ for $|x|<L/2$, Fig. \ref{fig:BHLModels}a (we still set the asymptotic subsonic sound speed to $c_u=1$). We label this stationary BHL solution as $\Psi_{\rm{BHL}}$. 

The discrete BdG spectrum of complex frequencies arising from $\Psi_{\rm{BHL}}$ is depicted in Fig. \ref{fig:BHLModels}d as a function of $L$. The critical lengths of the cavity $L=L_n$ at which the $n$-th dynamically unstable mode emerges (vertical solid lines) are given by
\begin{equation}\label{eq:CriticalLengths}
    L_n=\frac{\varphi_0+2\pi n}{k_{\rm{BCL}}}=L_0+n\lambda_{_{\rm{BCL}}},~n=0,1\ldots
\end{equation}
where
\begin{equation}\label{eq:CriticalFP}
    \varphi_0=2\arctan\sqrt{\dfrac{1-q^2}{q^2-c_2^2}},~k_{\rm{BCL}}=2\sqrt{q^2-c_2^2}.
\end{equation}
This equation can be simply understood as that, after some threshold length $L_0=\varphi_0/k_{\rm{BCL}}$ at which the first unstable lasing mode appears, the cavity gives birth to a new unstable mode each BCL wavelength $\lambda_{{\rm{BCL}}}=2\pi/k_{\rm{BCL}}$. The lasing modes are initially degenerate, i.e., they have purely imaginary frequency, $\omega_n=i\Gamma_n$. For lengths $L>L_{n+1/2}$, with $L_{n+1/2}$ obtained by inserting half-integer values $n+1/2$ in the above equation (vertical dashed lines), the $n$-th unstable mode becomes non-degenerate, developing a non-vanishing real part of the frequency $\gamma_n\neq 0$. 

 The dominant mode is that with the largest growth rate $\Gamma_n$, and determines the overall growth rate $\Gamma$ of the lasing instability, $\Gamma=\max_{n}\Gamma_n$. For short cavities, this is typically the mode with the largest $n$. However, as the cavity becomes longer and longer, the competition between the different unstable modes becomes stronger and stronger. We can compare these exact results with an estimation for the growth rate resulting from the qualitative picture of bouncing Hawking radiation, $\Gamma\sim 1/\tau_{\rm{RT}}$, with $\tau_{\rm{RT}}$ the roundtrip time for a zero-frequency $d2$ mode to travel back and forth between the horizons; the zero-frequency choice is motivated by the small value of $\gamma_n$ for the dominant mode observed in the plot. The result is depicted in dashed-dotted blue line, finding that it provides a decent estimation for long cavities. An elaborated WKB calculation shows a much more accurate agreement with the exact BdG results \cite{Michel2013}; however, it completely misses the existence of degenerate unstable modes, which are the dominant ones in short cavities. Thus, WKB prescriptions can only be used reliably in the long-cavity limit.

Interestingly, the work by Michel and Parentani \cite{Michel2013} further established a perfect correspondence between the emergence of dynamical instabilities in the BdG spectrum and the emergence of non-linear stationary solutions in the GP equation $\Psi_n(x),~n=0,1\ldots$. Specifically, these are stationary GP solutions for the same underlying Hamiltonian that are smoothly connected (as a function of $L$) to $\Psi_{\rm{BHL}}$, sharing the same conserved current $J$ and chemical potential $\mu$. The correspondence is shown in Fig. \ref{fig:BHLModels}g, where the spectrum of stationary non-linear GP solutions for the largest value of $L$ in Fig. \ref{fig:BHLModels}d is represented using the same color code of the associated lasing modes. The GP solution $\Psi_n(x)$ first emerges at $L=L_n$ as a sinusoidal oscillation around the supersonic cavity with the BCL wavelength [see for instance $\Psi_3(x)$ in dashed-dotted magenta], eventually becoming a non-linear cnoidal wave as the cavity enlarges. These solutions have lower grand-canonical energy (\ref{eq:energeticstability}) than the original BHL solution $\Psi_{\rm{BHL}}$, following the hierarchy
\begin{equation}
    K[\Psi_0]<K[\Psi_1]<K[\Psi_2]<\ldots<K[\Psi_{\rm{BHL}}].
\end{equation}
Michel and Parentani \cite{Michel2015} conjectured that all of them are also dynamically unstable except for the ground state solution $\Psi_0(x)$, which accumulates particles in the cavity in order to become fully subsonic and evaporate the horizons. It can be proven that the degeneracy breaking at $L=L_{n+1/2}$ of the $n$-th unstable mode can be also attributed to the emergence of a new non-linear GP solution, but this one is asymmetric and contains one soliton minimum outside the cavity, thus being energetically unfavored with respect to the symmetric solutions $\Psi_n(x)$ (i.e., their density $|\Psi_n(x)|^2$ has even parity with respect to the center of the lasing cavity). Remarkably, this non-linear spectrum of solutions is similar to that arising in a superconductor/normal/superconductor junction \cite{Sols1994} due to the GP-GL correspondence.

%Even though quite appealing from the theoretical point of view due to its simplicity, the flat-profile BHL configuration is very unrealistic from the experimental point of view. This motivated the search for m

More realistic BHL models were developed in Ref. \cite{deNova2017a}, where it was shown that any BH solution leading to a homogeneous supersonic region, as those of Fig. \ref{fig:BHModels}, can be mirrored to produce a WH solution by parity inversion of the Hamiltonian and time-reversal symmetry of the GP wavefunction. By matching the two BH/WH solutions in the homogeneous supersonic region, one obtains a symmetric BHL solution $\Psi_{\rm{BHL}}(x)$ with a homogeneous lasing cavity of arbitrary length $L$. Indeed, the flat-profile BHL solution of Fig. \ref{fig:BHLModels}a is a particular example of this general result. Two more examples are the attractive well and double-delta BHL solutions of Figs. \ref{fig:BHLModels}b,c, obtained from the corresponding waterfall and delta BH solutions of Figs. \ref{fig:BHModels}b,c. Their spectrum of dynamical instabilities is computed in Figs. \ref{fig:BHLModels}e,f, which exhibits the same trends as the flat-profile case. In particular, the critical lengths $L_n$ at which a new dynamical instability emerges are also given by Eq. (\ref{eq:CriticalLengths}), where
\begin{equation}
    \varphi_0=\pi,~k_{\rm{BCL}}=2\sqrt{v^2_d-c^2_d}=2\sqrt{\frac{1}{q^2}-q^2},
\end{equation}
for the attractive-well BHL solution, while for the double-delta BHL solution 
\begin{eqnarray}
    \varphi_0&=&4\arcsin\sqrt{\frac{1-r}{2}},~k_{\rm{BCL}}=2\sqrt{v^2_d-c^2_d},\\\nonumber r&=&c^2_d\sqrt{\frac{2(M^2_d-1)}{2Z^2c^2_d+\sqrt{q^4+8q^2}(1-c^2_d)}},
\end{eqnarray}
with $M_d=v_d/c_d=q/c^3_d$ the supersonic Mach number and $Z$ the amplitude of the delta barrier; see Eqs. (\ref{eq:CompactBHWF})-(\ref{eq:CompactDelta}) and ensuing discussion for the details of the waterfall and delta configurations. Each lasing mode has again associated a non-linear symmetric GP solution $\Psi_n(x)$ smoothly connected to $\Psi_{\rm{BHL}}(x)$ as a function of the cavity length, Figs. \ref{fig:BHLModels}h,i. Moreover, the $n$-th unstable mode also becomes non-degenerate at $L=L_{n+1/2}$, coinciding with the emergence of an asymmetric GP solution. The inverse of the roundtrip time still provides an estimation for the growth rate that improves for long cavities. In summary, all the trends predicted by Michel and Parentani in Ref. \cite{Michel2013} are further confirmed by these alternative BHL solutions. The only exception is the appearance of a dynamical instability in the attractive square well for $0<L<L_0$, labeled as the short-length (SL) mode in Fig. \ref{fig:BHLModels}e. This instability is not related to the BHL effect itself but its origin lies in the fact that  $\Psi_{\rm{BHL}}(x)$ is here smoothly connected to the soliton solution (\ref{eq:GraySoliton}) when no potential is present ($L=0$). However, the soliton has larger energy than the homogeneous plane wave $\Psi_0(x)=e^{iqx}$, which in turn is smoothly connected to the actual ground state, labeled as GS in Fig. \ref{fig:BHLModels}h. Thus, the SL mode is simply a consequence of the energetic instability of the BHL solution for any $L>0$. This provides further numerical evidence for the conjecture of Ref. \cite{Michel2015}: in flowing scattering configurations, energetic and dynamical instability are equivalent conditions.

\begin{figure*}[t]
\begin{tabular}{@{}ccc@{}}\stackinset{l}{2pt}{t}{3pt}{\large{(a)}}{\includegraphics[width=0.66\columnwidth]{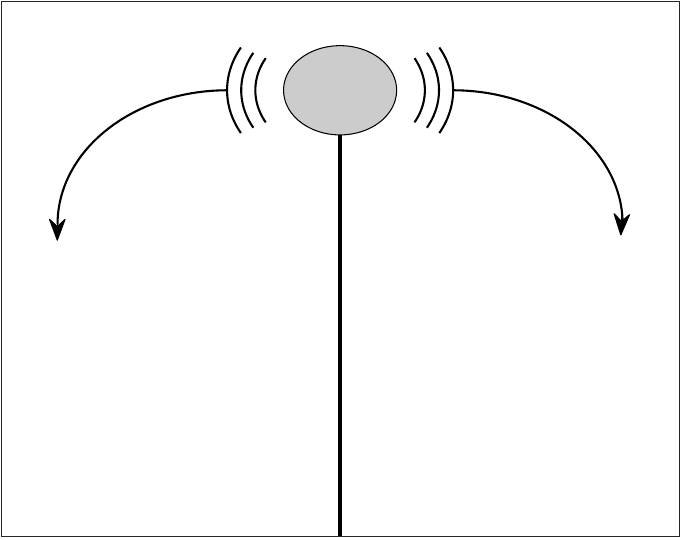}}~ &~
    \stackinset{l}{2pt}{t}{3pt}{\large{(b)}}
    {\includegraphics[width=0.66\columnwidth]{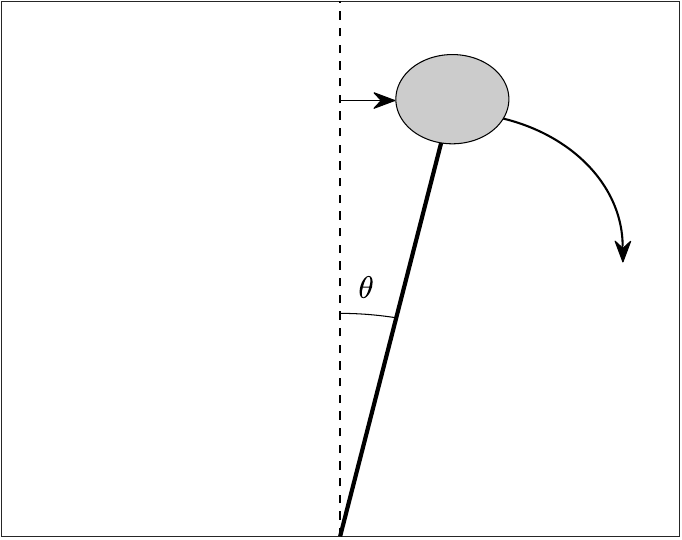}}~ &~ 
    \stackinset{l}{2pt}{t}{3pt}{\large{(c)}}{\includegraphics[width=0.66\columnwidth]{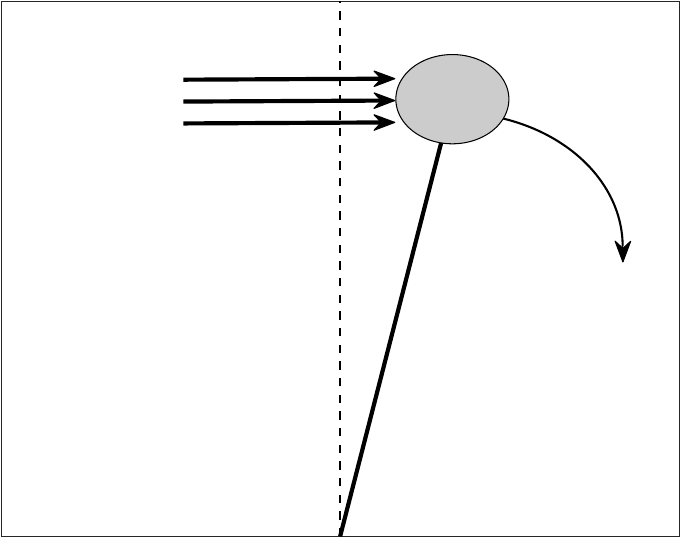}} 
\end{tabular}
\caption{Schematic depiction of the three different regimes of the BHL-BCL crossover using an analogy with an unstable pendulum. (a) Quantum BHL: Quantum fluctuations cause the unstable equilibrium position to collapse due to the Heisenberg uncertainty principle. (b) Classical BHL: A small kick on the pendulum displaces it some angle $\theta$ from the unstable equilibrium position, falling down with a well-defined classical trajectory as a result. (c) BCL: An external force (horizontal arrows) pushes the pendulum out of equilibrium, governing the dynamics instead of gravity.}
\label{fig:Pendula}
\end{figure*}

\subsection{Intermediate times: Quantum amplification in the BHL-BCL crossover}

The exponential amplification of the dominant lasing mode drives the linear Bogoliubov dynamics for times $\Gamma t\gtrsim 1$ until the saturation regime, when it typically reaches the non-linear stationary GP solution with the largest $n$. However, due to the energetic instability of the supersonic cavity, the exponential growth of the dominant lasing mode can be overshadowed by the coherent stimulation of the BCL wave resulting from the presence of an obstacle in the flow. For instance, as originally shown in Refs. \cite{Wang2016,Wang2017}, this can be the case of the WH horizon itself in highly time-dependent configurations, far from the fine-tuned stationary BHL solutions of Fig. \ref{fig:BHLModels}. Moreover, since the real part of the frequency of the dominant mode is very small, it contains wavevectors close to that of the BCL mode, something that greatly complicates their clear distinction in real setups \cite{Steinhauer2014,Tettamanti2016,Steinhauer2017,Wang2016,Wang2017,Llorente2019,Tettamanti2021,Kolobov2021,Steinhauer2022,deNova2023}. All of this leads to a strong competition between the BCL and BHL mechanisms.

% Moreover, since the dominant mode has a small real part of the frequency, it contains wavevectors close to that of the BCL mode, something that strongly complicates their clear distinction in real setups \cite{Steinhauer2014,Tettamanti2016,Steinhauer2017,Wang2016,Wang2017,Llorente2019,Tettamanti2021,Kolobov2021,Steinhauer2022,deNova2023}. All of this leads to a strong competition between the BCL and BHL mechanisms.

We can understand coherent BCL stimulation from a simple model based on linear response theory \cite{Carusotto2006}. We consider a general stationary condensate, solution of the time-independent GP equation (\ref{eq:TIGP}). At $t=0$, a small external perturbation described by a potential $W(x,t)$ is introduced. Expanding the GP wavefunction as
\begin{equation}\label{eq:ClassicalExpansion}
\Psi(\mathbf{x},t)=\left[\Psi_0(\mathbf{x})+\varphi(\mathbf{x},t)\right]e^{-i\mu t}
\end{equation}
in the time-dependent GP equation (\ref{eq:TDGP}) leads, at linear order in the external perturbation, to the classical BdG equations with a source,
\begin{eqnarray}\label{eq:TIBdGSource} 
[i\hbar\partial_t-M_0]\Phi(\mathbf{x},t)&=&iW(\mathbf{x},t)z_\theta(\mathbf{x}),\\
\nonumber \Phi(\mathbf{x},t)&=&\left[\begin{array}{r}\varphi(\mathbf{x},t)\\ \varphi^*(\mathbf{x},t)\end{array}\right],\,z_\theta(\mathbf{x})=\left[\begin{array}{r}-i\Psi_0(\mathbf{x})\\ i\Psi^*_0(\mathbf{x})\end{array}\right],
\end{eqnarray}
where $z_\theta$ is the zero-frequency Nambu-Goldstone mode associated to the spontaneous $U(1)$-symmetry breaking by the coherent GP wavefunction \cite{Lewenstein1996}; see Eq. (\ref{eq:NambuGoldstone}) and ensuing discussion. This equation can be solved by performing a classical expansion in terms of the complete set of BdG eigenmodes, analogous to that of Eq. (\ref{eq:QuantumFieldFluctuations}),
\begin{equation}
    \Phi(\mathbf{x},t)=\sum_{n}a_{n}(t)z_{n}(\mathbf{x})e^{-i\omega_nt}+a^*_{n}(t)\bar{z}_{n}(\mathbf{x})e^{i\omega_nt},
\end{equation}
where we subtract here the intrinsic time evolution of each mode. This leads to
\begin{eqnarray}\label{eq:BCLStimulation}
    \partial_t a_n&=&e^{i\omega_n t}(z_n|W(t) z_\theta)\Longrightarrow \\
    \nonumber a_n(t)&=& \int^t_0\mathrm{d}t'~e^{i\omega_n t'}(z_n|W(t') z_\theta).
\end{eqnarray}
Thus, for sufficiently long times, the amplitude of the collective modes, corresponding here to the BdG modes, is determined by the Fourier spectrum of the external perturbation, as expected from the usual theory of linear response. In the specific case of a supersonic condensate, since the BCL mode has zero-frequency, it is resonantly stimulated by any static obstacle in the supersonic flow: this is precisely the origin of Landau criterion for superfluidity. Notice that this stimulation is coherent, imprinted on the GP wavefunction, and thus it has a completely classical nature.

In order to study the BHL-BCL crossover, we first model the initial background condensate (before the BHL and/or BCL onsets) within the bulk of the lasing cavity by a supersonic plane wave of the form $\Psi_0(x)\simeq \sqrt{n_0}e^{iq_0x}$, with an associated healing length $\xi_0$. For the analysis, we focus on the expectation values of the density and its correlations, measurable in the laboratory through \textit{in situ} imaging after averaging over ensembles of repetitions of the experiment \cite{Shammass2012,Steinhauer2014,Steinhauer2016,deNova2019,Kolobov2021}. After expanding the density in terms of the quantum fluctuations of the background condensate, one obtains
\begin{align}\label{eq:Density expansion}
     \nonumber &\hat{n}(x,t)=\hat{\Psi}^\dagger(x,t)\hat{\Psi}(x,t)= n_0+\delta\hat{n}^{(1)}(x,t)+\delta\hat{n}^{(2)}(x,t),\\
    \nonumber &\delta\hat{n}^{(1)}(x,t)=\Psi_0^*(x)\hat{\varphi}(x,t)+\Psi_0(x)\hat{\varphi}^\dagger(x,t)=
    -iz^\dagger_\theta \sigma_z \hat{\Phi},\\
    &\delta\hat{n}^{(2)}(x,t)=\hat{\varphi}^\dagger(x,t)\hat{\varphi}(x,t),
\end{align}
where we separate the linear contribution in the field fluctuations $\delta\hat{n}^{(1)}(x,t)$ [the same of Eq. (\ref{eq:BdGfieldequationTIHydrodynamic2}), obtained within the BdG approximation], from the quadratic contribution $\delta\hat{n}^{(2)}(x,t)$. Since dimensional analysis implies that the  quantum fluctuations around $\Psi_0$ scale as $\hat{\varphi}\sim 1/\sqrt{\xi_0}$, we have the scalings 
\begin{equation}\label{eq:DensityScalings} %[see also Eq. (\ref{eq:dispersionrelation}) and related derivations]
    \delta\hat{n}^{(1)}\sim \sqrt{\frac{n_0}{\xi_0}},~\delta\hat{n}^{(2)}\sim \frac{1}{\xi_0}.
\end{equation}
Using these results, we characterize the BHL and BCL mechanisms through the first-order correlation function
\begin{equation}
     G^{(1)}(x,t)\equiv \frac{\braket{\hat{n}(x,t)}}{n_0}-1,
\end{equation}
which measures the amplitude of the developing density modulation above the background condensate, and the \textit{normalized} density-density correlation function 
\begin{equation}\label{eq:NormalizedRelativeCorrelationsBdG}
    G^{(2)}(x,x',t)\equiv n_0\xi_0 g^{(2)}(x,x',t),
\end{equation}
which in turn measures the quantum fluctuations around the density modulation, $g^{(2)}(x,x',t)$ being the \textit{relative} density-density correlation function 
\begin{eqnarray}\label{eq:RelativeCorrelationsBdG}
    \nonumber g^{(2)}(x,x',t)&\equiv& \frac{\braket{\hat{n}(x,t)\hat{n}(x',t)}-\braket{\hat{n}(x,t)}\braket{\hat{n}(x',t)}}{n^2_0}\\
    &\simeq& \frac{\braket{\delta\hat{n}^{(1)}(x,t)\delta\hat{n}^{(1)}(x',t)}}{n^2_0}\sim \frac{1}{n_0\xi_0},
\end{eqnarray}
where we take the leading contribution in the Bogoliubov approximation. The relative density-density correlation function $g^{(2)}$ provides the relative amplitude of the quantum fluctuations, so we can regard $(n_0\xi_0)^{-1}$ as the initial strength of the quantum fluctuations, which must be small for the Bogoliubov approximation to be valid, $n_0\xi_0\gg 1$. On the other hand, the normalization of $G^{(2)}$ ensures that it is a dimensionless function that does not depend explicitly on $n_0$ in the Bogoliubov approximation, only implicitly through the healing length $\xi_0$. Since both BHL and BCL mechanisms involve modes with well-defined wavevectors within the lasing cavity, we will use the Fourier transforms in the supersonic region of the above observables as figures of merit,
\begin{eqnarray}
     \nonumber G^{(1)}_{\rm{peak}}(t)&\equiv& \max_{k} |G^{(1)}(k,t)|,\\
     G^{(2)}_{\rm{peak}}(t)&\equiv& \max_{k,k'} |G^{(2)}(k,k',t)|.
\end{eqnarray}
In real space, this peaked Fourier structure is translated into a ripple in the ensemble-averaged density profile and into a checkerboard pattern in the density-density correlations, respectively.

By borrowing the analogy with an unstable pendulum from Eq. (\ref{eq:UnstablePendulum}), we can distinguish three main regimes in the BHL-BCL crossover, represented in Fig. \ref{fig:Pendula}, depending on the interplay between quantum fluctuations and classical BCL stimulation, where the former are controlled by the dimensionless amplitude 
\begin{equation}\label{eq:QFAmplitude}
     A_{\rm{QF}}\sim \frac{\hat{\varphi}}{\Psi_0}\sim \frac{\delta\hat{n}^{(1)}}{n_0} \sim \frac{1}{\sqrt{n_0\xi_0}}\ll 1,
\end{equation}
while the latter is controlled by the relative amplitude of the coherent BCL wave with respect to the background condensate,  
\begin{equation}
    A_{\rm{BCL}}\sim \frac{\varphi}{\Psi_0}.
\end{equation}

%, controlled by , and classical BCL stimulation, controlled by $A_{\rm{BCL}}$,

%where $a_{\rm{BCL}}$ is the BCL amplitude resulting from the external stimulation, see Eq. (\ref{eq:BCLStimulation}).

\subsubsection{Quantum BHL}

% By comparing  with the relative amplitude of the BCL wave with respect to the condensate, $A_{\rm{BCL}}$,

% The condition $n_0\xi_0\gg 1$ is required so that the relative density fluctuations are small, $g^{(2)}\sim (n_0\xi_0)^{-1}\ll 1$. Although, strictly speaking, $\xi_0=1$ in our units, we will write the complete dimensionless factor $n_0\xi_0$ to stress its physical meaning as the regulator of the strength of the quantum fluctuations. 

% From the previous considerations, one can anticipate three different regimes for the dynamics after the BHL formation depending on the amplitudes of the background Cherenkov wave, $A_{\rm{BCL}}$, and of the quantum fluctuations, $A_{\rm{QF}}$:

When $A_{\rm{BCL}}\ll A_{\rm{QF}}\ll 1$, the BHL instability is purely triggered by quantum fluctuations (e.g., there is no BCL stimulation, $A_{\rm{BCL}}=0$), and the dynamics is driven by the parametric amplification of the dominant lasing mode $z_I$, whose frequency and quantum amplitude are $\omega=\gamma+i\Gamma$ and $\hat{a}_I$, respectively. The contribution of the dominant mode to the quantum field fluctuations is
\begin{equation}\label{eq:QuantumDominant}
    \hat{\Phi}(x,t)\simeq e^{\Gamma t}\left[z_I(x)e^{-i\gamma t}\hat{a}_I+\bar{z}_I(x)e^{i\gamma t}\hat{a}^\dagger_I\right].
\end{equation}
In a quantum BHL, the phase of the amplitude of the dominant mode is expected to be random and hence we can take $\braket{\hat{a}_I\hat{a}_I}\simeq 0$. This assumption yields that $G^{(1)}_{\textrm{peak}},G^{(2)}_{\textrm{peak}}$ behave as
\begin{eqnarray}\label{eq:QBHLGrowth}
 \nonumber G^{(1)}_{\textrm{peak}}(t)&\sim& \frac{\braket{\delta\hat{n}^{(2)}}}{n_0}\sim \frac{e^{2\Gamma t}}{n_0\xi_0},\\
    G^{(2)}_{\textrm{peak}}(t)&\sim& n_0\xi_0 \frac{\braket{\delta\hat{n}^{(1)}\delta\hat{n}^{(1)}}}{n^2_0}\sim e^{2\Gamma t}.
\end{eqnarray}
Hence, for a quantum BHL, the correlation functions $G^{(1)}_{\textrm{peak}}(t)$, $G^{(2)}_{\textrm{peak}}(t)$ scale quadratically in the field fluctuations. In the case of $G^{(1)}_{\textrm{peak}}$, this is because the $\mathbb{Z}_2$ symmetry of a purely quantum BHL sets $\braket{\hat{\varphi}(x,t)}=0$ and thus, $\braket{\delta\hat{n}^{(1)}(x,t)}=0$, as originally discussed by Michel and Parentani \cite{Michel2015}. 

In the pendulum analogy, the unstable equilibrium position is the initial BHL solution and gravity is the lasing instability. Due to the Heisenberg uncertainty principle, the unstable equilibrium position collapses at the quantum level, and then the pendulum falls, Fig. \ref{fig:Pendula}a. This is akin to the parametric amplification of the lasing instability, where the $\mathbb{Z}_2$ symmetry can be understood as that of the unstable equilibrium position of the pendulum. 

The exponential growth of the dominant mode ceases when the system reaches the saturation regime, corresponding to one of the stationary GP solutions of the spectrum (typically, that with the largest $n$), where the density modulation becomes of the order of the background density itself, $G^{(1)}_{\textrm{peak}}\sim 1$; see lower row of Fig. \ref{fig:BHLModels}. Since this saturation stems from the amplification of the quantum fluctuations of the dominant mode, we also have that $g_{\rm{sat}}^{(2)}\sim 1$. As a result, the saturation values of both correlation functions are roughly
\begin{eqnarray}
 \nonumber G^{(1)}_{\textrm{sat}}&\sim& 1\sim \frac{e^{2\Gamma t_{\textrm{sat}}}}{n_0\xi_0 },\\
    G^{(2)}_{\textrm{sat}}&\sim& n_0\xi_0 \sim e^{2\Gamma t_{\textrm{sat}}},
\end{eqnarray}
where $t_{\textrm{sat}}$ is the time needed to reach saturation,
\begin{equation}
    t_{\textrm{sat}}\sim \frac{\ln n_0\xi_0}{2\Gamma}.
\end{equation}

\subsubsection{Classical BHL} 

%This regime is described by the linear response theory previously developed, Eq. (\ref{eq:ClassicalExpansion}) and subsequent results, where the perturbation of the flow gives a classical coherent amplitude to the dominant lasing mode through the stimulation of the BCL wave. As a result, the dynamics is driven by the exponential amplification of this classical amplitude,

Here, $A_{\rm{QF}}\ll A_{\rm{BCL}}\ll 1$, so BHL amplification still dominates the dynamics but the seed of the instability is now the classical amplitude of the BCL wave in the condensate, leading to a well-defined mean-field trajectory.
Specifically, the perturbation of the flow gives a classical coherent amplitude to the dominant lasing mode through the stimulation of the BCL wave [see Eq. (\ref{eq:ClassicalExpansion}) and subsequent results], which is then exponentially amplified as
\begin{equation}\label{eq:ClassicalDominant}
    \Phi(x,t)\simeq e^{\Gamma t}\left[z_I(x)e^{-i\gamma t}a_I+\bar{z}_I(x)e^{i\gamma t}a^*_I\right].
\end{equation}
Hence, $\braket{\delta\hat{n}^{(1)}}\neq 0$, and the $\mathbb{Z}_2$ symmetry is broken, 
\begin{equation}\label{eq:Z2BreakingLinear}
   G^{(1)}_{\textrm{peak}}(t)\sim \frac{\braket{\delta\hat{n}^{(1)}}}{n_0}\sim A_{\rm{BCL}}e^{\Gamma t}.
\end{equation}
Precisely because of its classical deterministic character, at the linear level the BCL amplitude does not show up in the density-density correlation function, and $G^{(2)}_{\textrm{peak}}(t)$ still follows Eq. (\ref{eq:QBHLGrowth}) in this regime. Therefore, the $\mathbb{Z}_2$ symmetry-breaking implies now
\begin{equation}\label{eq:Z2Breaking}
    G^{(1)}_{\textrm{peak}}(t)\sim e^{\Gamma t},~ G^{(2)}_{\textrm{peak}}(t)\sim e^{2\Gamma t}.
\end{equation}

In the pendulum analogy, a classical BHL is akin to separate the pendulum some small angle $\theta$ from its equilibrium position, which consequently falls following a well-defined classical trajectory, Fig. \ref{fig:Pendula}b. Here, the angle $\theta$ plays the role of the Cherenkov amplitude $A_{\rm{BCL}}$ that seeds the BHL instability, breaking the original $\mathbb{Z}_2$ symmetry of the problem. 

In the saturation regime, $G^{(1)}_{\textrm{peak}}\sim 1$, which now implies that
\begin{equation}\label{eq:SaturationTimeCBHL}
    G^{(1)}_{\textrm{sat}}\sim 1 \sim A_{\rm{BCL}}e^{\Gamma t_{\textrm{sat}}}.
\end{equation}
Therefore, the saturation time is predicted to behave as
\begin{equation}
    t_{\textrm{sat}}\sim - \frac{\ln A_{\rm{BCL}}}{\Gamma},
\end{equation} %according to linear response theory
and then
\begin{equation}\label{eq:SaturationAmplitudeCBHL}
    G^{(2)}_{\textrm{sat}}\sim e^{2\Gamma t_{\textrm{sat}}}\sim A^{-2}_{\rm{BCL}}.
\end{equation}

\begin{table*}[t]
\begin{tabular}[c]{|c|c|c|c|c|c|c|}
\hline
~ & $G^{(1)}_{\rm{peak}}(t)$ & $G^{(2)}_{\rm{peak}}(t)$ & $G^{(1)}_{\rm{sat}}$ & $G^{(2)}_{\rm{sat}}$ & $t_{\rm{sat}}$ & Monotonic \\
\hline
Quantum BHL & $\sim e^{2\Gamma t}/n_0\xi_0 $ & $\sim e^{2\Gamma t}$ & $\sim 1$ & $\sim n_0\xi_0 $ &  $\sim \ln n_0\xi_0 /2\Gamma$  & No\\
\hline
Classical BHL  & $\sim A_{\rm{BCL}}e^{\Gamma t}$ & $\sim e^{2\Gamma t}$  & $\sim 1$ & $\sim e^{2\Gamma t_{\rm{sat}}} \sim A_{\rm{BCL}}^{-2}$ &  $\sim -\ln A_{\rm{BCL}}/\Gamma$ & No\\
\hline
BCL  & ----- & ----- & $\sim A^2_{\rm{BCL}}$ & $F(A_{\rm{BCL}}) $ &  $\gtrsim \tau_{\rm{BCL}}$ & Yes\\
\hline
\end{tabular}
\caption{Summary of the different scalings for the three regimes of the BHL-BCL crossover: quantum BHL, classical BHL, and BCL. There is no analytical prediction for $G^{(1)}_{\rm{peak}}(t),G^{(2)}_{\rm{peak}}(t)$ in the BCL regime since they depend on the particular stimulation source. $F(A_{\rm{BCL}})$ is an increasing function of  $A_{\rm{BCL}}$ and $\tau_{\rm{BCL}}$ is the time that it takes the BCL wave to reach the BH horizon. The column ``Monotonic'' indicates a monotonic dependence on the background parameters of the flow. }
\label{TBCL}
\end{table*}

\subsubsection{BCL}  

When the BCL amplitude is highly non-linear, $A_{\rm{QF}}\ll A_{\rm{BCL}}\sim 1$, it dominates the mean-field dynamics towards the saturation regime, overshadowing the lasing mechanism. Since the BCL stimulation depends on the specific perturbation of the flow [see Eq. (\ref{eq:BCLStimulation})], in general, no analytical formula is available for the evolution of $G^{(1)}_{\textrm{peak}}(t),G^{(2)}_{\textrm{peak}}(t)$. In the pendulum analogy, the BCL regime is akin to applying a strong external force to the pendulum that overcomes gravity, Fig. \ref{fig:Pendula}c, so its evolution depends on the specific force.

%, especially in the non-linear regime where linear response theory fails

%which stems from the wavevectors of the lasing modes, whose quantum fluctuations are exponentially amplified

Regarding saturation, a large BCL amplitude is no longer described by linear response theory, but it rather requires the full GP equation. The saturation amplitude then scales as $G^{(1)}_{\textrm{sat}}\sim A^2_{\rm{BCL}}$ by definition of BCL amplitude. Regarding the density fluctuations $G^{(2)}$, the sharp peaked structure of the BCL wave now acts as a new mean-field background over which fluctuations evolve. This gives rise to a checkerboard pattern in the correlation function whose origin is completely different to that from a BHL, which there stems from the exponential amplification of the quantum fluctuations of the lasing modes, with wavevectors close to the BCL one. Therefore, we can expect $G^{(2)}_{\textrm{sat}}=F(A_{\rm{BCL}})$, where $F$ is in general a monotonically increasing function of $A_{\rm{BCL}}$ that also depends on the other parameters of the background flow. Due to the strong BCL stimulation, the saturation time is essentially limited by $t_{\textrm{sat}}\gtrsim \tau_{\rm{BCL}}$, where $\tau_{\rm{BCL}}$ is the time that it takes the BCL wave to expand along the whole lasing cavity.

% If a strong enough force is applied (), it will drive the pendulum motion outside its unstable equilibrium position instead of gravity, as in the case where BCL stimulation dominates the dynamics over the BHL mechanism.

\subsubsection{Upshot}

All the above scalings were originally derived in Ref. \cite{deNova2023} within a model based on a flat-profile BHL solution (Fig. \ref{fig:BHLModels}a) with a delta barrier placed exactly at the WH horizon to stimulate BCL radiation, which allows to isolate the contribution of both mechanisms to the dynamics, finding a good agreement with numerical results. Interestingly, it was also stressed there that each regime of the BHL-BCL crossover can be characterized according to its efficiency as a quantum amplifier. Specifically, we measure the quantum amplification using the relative density-density correlations, following Eq. (\ref{eq:RelativeCorrelationsBdG}). In the initial state,
\begin{equation}
    g_{\rm{peak}}^{(2)}(t=0)\sim \frac{1}{n_0\xi_0}\ll 1, %g^{(2)}_0\equiv
\end{equation}
which is the input of the quantum amplifier, while the output is the saturation value $g_{\rm{sat}}^{(2)}$. This implies that the gain of the quantum amplifier $\mathcal{G}$ is directly proportional to the saturation value $G_{\rm{sat}}^{(2)}$ since
\begin{equation}
    \mathcal{G}\equiv \frac{g_{\rm{sat}}^{(2)}}{g_{\rm{peak}}^{(2)}(t=0)}\propto n_0\xi_0 g_{\rm{sat}}^{(2)}= G_{\rm{sat}}^{(2)}. %\propto \frac{g_{\rm{sat}}^{(2)}}{g_0^{(2)}}
\end{equation}
For each regime, we find
\begin{eqnarray}
    \nonumber \mathcal{G}_{\rm{QBHL}}&\sim& n_0\xi_0,\\
        \mathcal{G}_{\rm{CBHL}}&\sim& e^{2\Gamma t_{\rm{sat}}},\\
    \nonumber \mathcal{G}_{\rm{BCL}}&\sim &F(A_{\rm{BCL}}).
\end{eqnarray}
This means that a quantum BHL behaves as a non-linear quantum amplifier, since it amplifies the initial quantum fluctuations up to the same saturation amplitude $g_{\rm{sat}}^{(2)}\sim 1$, so the gain depends on the input amplitude $1/n_0\xi_0$. On the other hand, classical BHL and BCL do behave as linear quantum amplifiers (i.e., their gain does not depend on the initial quantum strength $1/n_0\xi_0$). In a classical BHL, the gain is exponentially large in the saturation time $t_{\rm{sat}}$, which in turn decreases with the BCL amplitude $A_{\rm{BCL}}$. This is because $t_{\rm{sat}}$ is the lasing time during which the exponential amplification of quantum fluctuations takes place. Hence, for increasing $A_{\rm{BCL}}$, the system starts closer to the saturation regime and less amplification is needed to reach it. In the BCL regime, the gain is exponentially smaller as compared to that of a classical BHL since there is no microscopic mechanism of exponential amplification, and the enhancement of quantum fluctuations stems just from the large BCL modulation of the background mean-field density. This also implies that the function $F(A_{\rm{BCL}})$ determining the gain increases with $A_{\rm{BCL}}$, in stark contrast with the decrease expected for lasing amplification. Thus, the dependence of the gain with respect to the BCL amplitude can be used to distinguish between classical BHL and BCL in experiments. 

It is also remarkable that, in the quantum BHL and BCL regimes, the behavior of $G_{\rm{peak}}^{(1)},G_{\rm{peak}}^{(2)}$ is correlated since they are dominated by the same mechanism (either exponential amplification of quantum fluctuations or classical BCL stimulation), while classical BHL is a hybrid regime where $G_{\rm{peak}}^{(1)}$ has a classical nature while $G_{\rm{peak}}^{(2)}$ has a quantum one. Another qualitative criterion of distinguishability is the non-monotonic behavior of the growth rate of the density ripple and the checkerboard pattern of the density-density correlations with respect to the parameters determining the background flow (for instance, the cavity length $L$, as shown in central row of Fig. \ref{fig:BHLModels}), in contrast with the smooth behavior expected for BCL stimulation. This non-monotonicity is a typical feature of resonant structures, and it is also observed in the peak structure of the Andreev-Hawking spectrum discussed in Sec. \ref{sec:ResonantHawking} \cite{Zapata2011}. A summary of the main results for each regime is presented in Table \ref{TBCL}.

\subsection{Long times: Spontaneous Floquet states}

\begin{figure*}[t]
\begin{tabular}{@{}cc@{}}\stackinset{l}{0pt}{t}{0pt}{\large{(a)}}{\includegraphics[width=0.98\columnwidth]{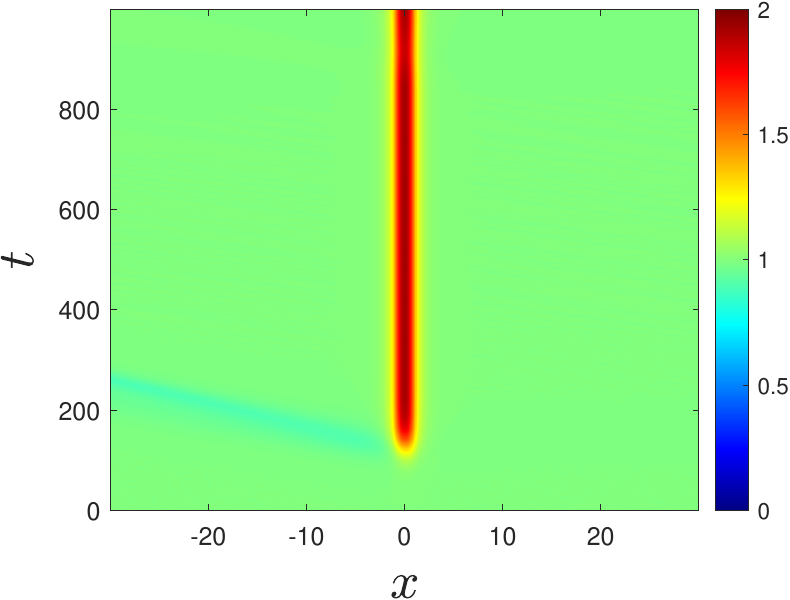}}~ &~
    \stackinset{l}{0pt}{t}{0pt}{\large{(b)}}
    {\includegraphics[width=\columnwidth]{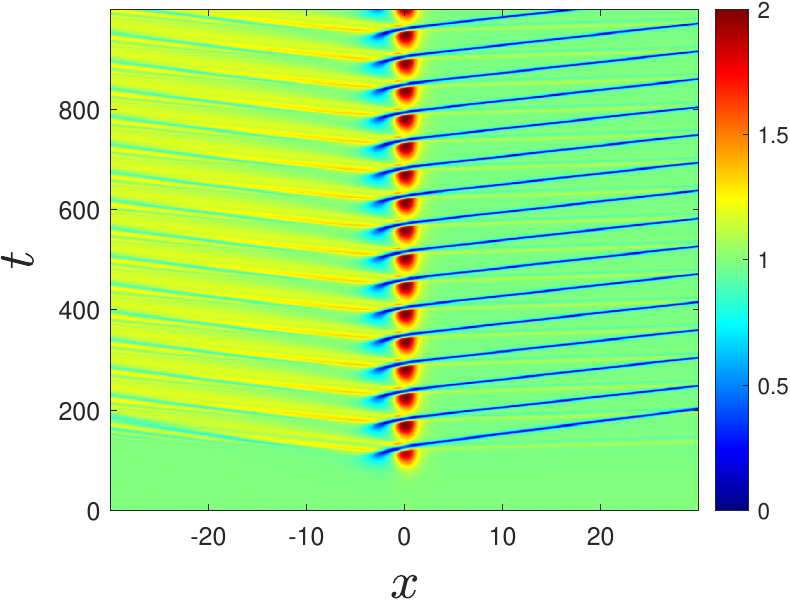}} \\
    \stackinset{l}{0pt}{t}{0pt}{\large{(c)}}{\includegraphics[width=0.98\columnwidth]{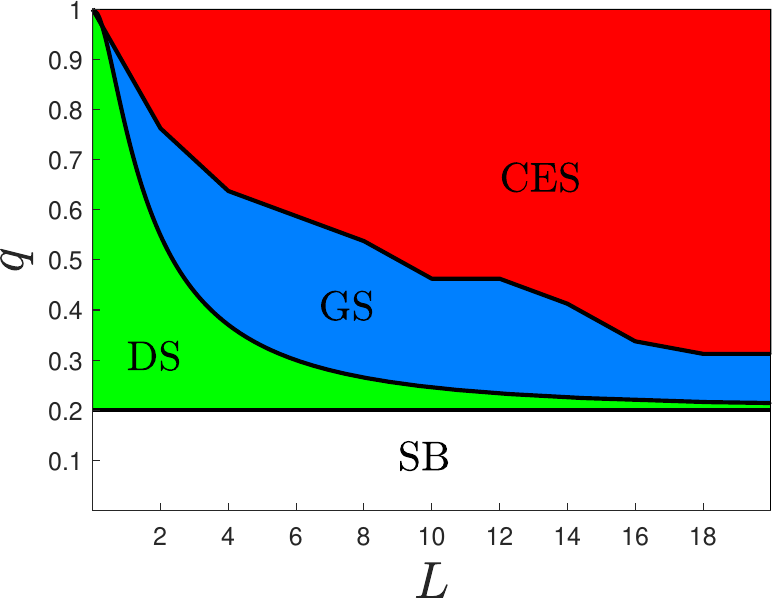}}~ &~
    \stackinset{l}{0pt}{t}{0pt}{\large{(d)}}
    {\includegraphics[width=\columnwidth]{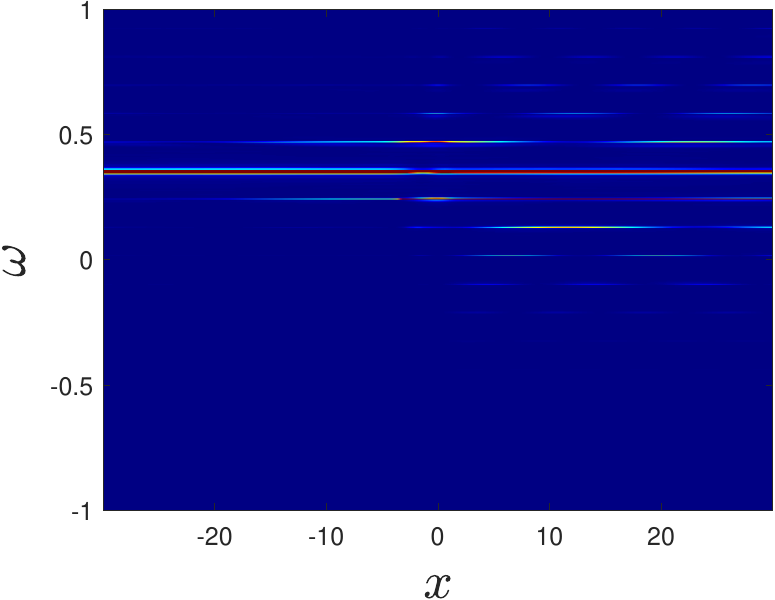}} 
\end{tabular}
\caption{a) Spatio-temporal density profile $|\Psi(x,t)|^2$ for an initial flat-profile BHL solution with $c_2=0.2$, $L=2$ and $q=0.7$. Some random noise is initially added to trigger the BHL instability. b) Same as a) but now $q=0.9$. c) Dynamical phase diagram as a function of $(L,q)$ with fixed $c_2=0.2$ for the final state of a flat-profile BHL, where SB denotes that the cavity is subsonic and DS denotes the dynamically stable region. d) Fourier spectrum $|\Psi(x,\omega)|^2$ of the simulation in b) once in the CES state.}%Dashed black line is a fit to a gray-body spectrum (\ref{eq:GrayBody}). Inset: Frequency-dependent Hawking temperature $T_H(\omega)$, Eq. (\ref{eq:OmegaTemperature}). Horizontal dashed green line marks the predicted Hawking temperature for a soliton, Eq. (\ref{eq:SolitonHawking}).
\label{fig:CES}
\end{figure*}

As explained, the lasing instability grows up to the saturation regime, where it reaches a certain stationary GP solution. However, this solution is also dynamically unstable, with a lifetime much longer than that of the initial BHL solution, and eventually collapses. After some (possibly very long) transient, where the system may intercept a number of intermediate metastable GP solutions, it was numerically observed  \cite{deNova2016} that the flat-profile BHL of Fig. \ref{fig:BHLModels}a only has two possible fates: it either reaches the true ground state (solid black line in Fig. \ref{fig:BHLModels}g) or the so-called Continuous Emission of Solitons (CES) state, where the system self-oscillates periodically while continuously emitting solitons into the downstream region. Due to its periodic nature, the CES state has been argued to be the \textit{bona-fide} black-hole laser \cite{deNova2016}.

Both trajectories are represented in Figs. \ref{fig:CES}a,b, where we show the time evolution of the density $|\Psi(x,t)|^2$ for two different flat-profile BHL configurations where some initial noise has been added to trigger the BHL instability. We take a short cavity $L=2$ in both cases since that implies a large growth rate for the lasing instability as well as only one stationary GP solution in the non-linear spectrum, the ground state $\Psi_0(x)$, considerably shortening the transient towards the final state. In Fig. \ref{fig:CES}a, for $q=0.7,c_2=0.2$, the system directly reaches the stable ground state $\Psi_0(x)$ in the saturation regime (vertical red stripe centered at $x=0$). After further increasing the initial flow velocity to $q=0.9$, Fig. \ref{fig:CES}b, the system also approaches the ground state by increasing the density in the cavity while expelling a gray soliton upstream to conserve particle number; however, now the flow velocity is high enough to drag the soliton back to the cavity (blue half-rings at the left of the vertical red stripe), which then passes to the downstream region and travels along the flow (diagonal blue lines downstream). The process is accompanied by the emission of waves (diagonal yellow lines upstream) to ensure conservation of particle number and energy. The passage of the dragged soliton through the cavity restarts the cycle, giving rise to a periodic behavior; this is the physical mechanism behind the CES state. 

The final state only depends on the background parameters of the flow $(L,q,c_2)$, being quite insensitive to the particular details of the transient or the initial noise. This gives rise to a dynamical phase diagram, shown in Fig. \ref{fig:CES}c as a function of $(L,q)$ for fixed $c_2=0.2$. Above the green region of dynamical stability (denoted as DS), whose upper boundary is given by the condition $L=L_0(q,c_2)$ [see Eqs. (\ref{eq:CriticalLengths}), (\ref{eq:CriticalFP})], the initial BHL solution first asymptotically reaches the ground state, denoted as GS (blue region). Above some critical flow velocity $q=q_c(L,c_2)$, numerically obtained, the final state is the CES state (red region), which can be then regarded as a non-linear extension of the Landau criterion. Later work \cite{deNova2021} numerically showed that the final fate of BHL solutions from an attractive well (Fig. \ref{fig:BHLModels}h) is also either the ground state or the CES state, suggesting the universality of the long-time behavior of a BHL. Moreover, it was observed that the same CES state can be reached even without starting from a BHL solution, indicating that the CES state is an intrinsic state of the system, and not some fine-tuned trajectory. 

This observation, along with the periodicity of the CES state, led to identify a novel type of quantum state \cite{deNova2022}: the so-called spontaneous Floquet state, which is a state of a time-independent Hamiltonian that oscillates like a Floquet state due to many-body interactions, without the need of external periodic driving. The emergence of a spontaneous Floquet state can be easily understood from the time-dependent GP equation for a time-independent Hamiltonian, which is a non-linear Schr\"odinger equation of the form
\begin{eqnarray}\label{eq:HamiltonianEffective}
   i\partial_t\Psi(x,t)&=&H_{\rm{GP}}(x,t)\Psi(x,t),\\
   \nonumber H_{\rm{GP}}(x,t)&=&-\frac{\partial_x^2}{2}+V(x)+g(x)|\Psi(x,t)|^2,
\end{eqnarray}
where we allow for a possible inhomogeneous coupling constant (as it is the case of a flat-profile BHL configuration). Notice that the only possible time dependence of $H_{\rm{GP}}(x,t)$ results from that of the GP wavefunction itself. A periodic density $|\Psi(x,t)|^2$, as that of the CES state, implies an effective periodic Hamiltonian, $H_{\rm{GP}}(x,t+T)=H_{\rm{GP}}(x,t)$, where $T$ is the oscillation period. Self-consistently, $\Psi(x,t)$ becomes a Floquet state of its own periodic Hamiltonian,
\begin{equation}\label{eq:FloquetWaveMu}
\Psi(x,t)=\Psi_0(x,t)e^{-i\tilde{\mu}t},~\Psi_0(x,t)=\sum^{\infty}_{n=-\infty}u_n(x)e^{-in\omega_0t},
\end{equation}
with $\Psi_0(x,t+T)=\Psi_0(x,t)$, $\omega_0=2\pi/T$, and $\tilde{\mu}$ the quasi-chemical potential, defined modulo $\omega_0$. By inserting this expansion into the GP equation, we get self-consistent equations for the Floquet components $u_n(x)$,
\begin{eqnarray}\label{eq:FloquetEquation}
    n\omega_0u_n(\mathbf{x})&=& \left[-\frac{\partial_x^2}{2}+V(x)-\tilde{\mu}\right]u_n(x)\\
    \nonumber&+&
    \sum^{\infty}_{m=-\infty}\sum^{\infty}_{k=-\infty} g(x)u^*_{k}(x)u_{k+n-m}(x)u_{m}(x).
\end{eqnarray}
The CES state provides a particular example of spontaneous Floquet state, as can be seen from its Fourier transform $\Psi(x,\omega)$, represented in Fig. \ref{fig:CES}d. We observe that the Fourier spectrum consists of a series of equispaced lines at frequencies $\omega=\tilde{\mu}+n\omega_0$, which can be identified as the Floquet components $u_n(x)$. The dominant Floquet component has a frequency of the order of the initial chemical potential $\omega\simeq q^2/2$ [notice that for the flat-profile BHL solution we subtract the interacting contribution to the chemical potential, as discussed after Eq. (\ref{eq:FlatProfileCondition})], which we can identify as a non-trivial quasi-chemical potential $\tilde{\mu}\neq 0$. 

%Therefore, the CES state is indeed a spontaneous Floquet state.

\begin{figure*}[t]
\begin{tabular}{@{}cc@{}}\stackinset{l}{0pt}{t}{0pt}{\large{(a)}}{\includegraphics[width=0.98\columnwidth]{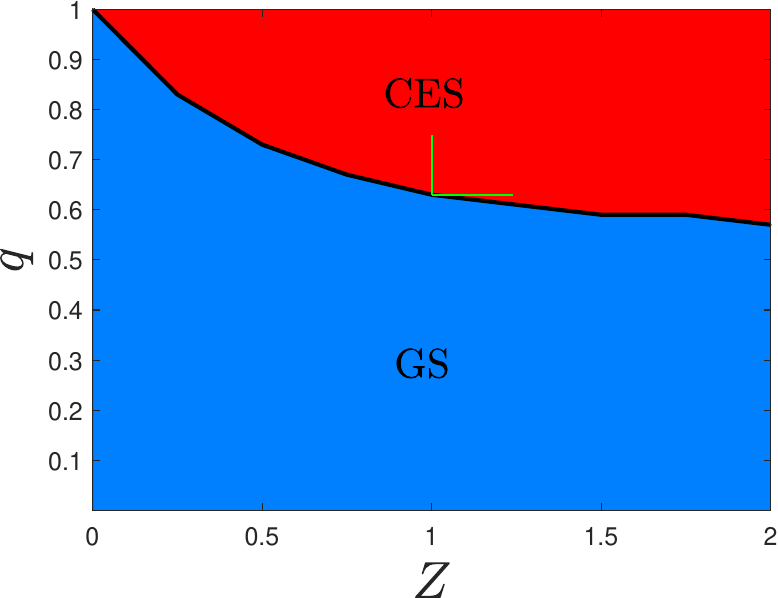}}~ &~
    \stackinset{l}{0pt}{t}{0pt}{\large{(b)}}
    {\includegraphics[width=\columnwidth]{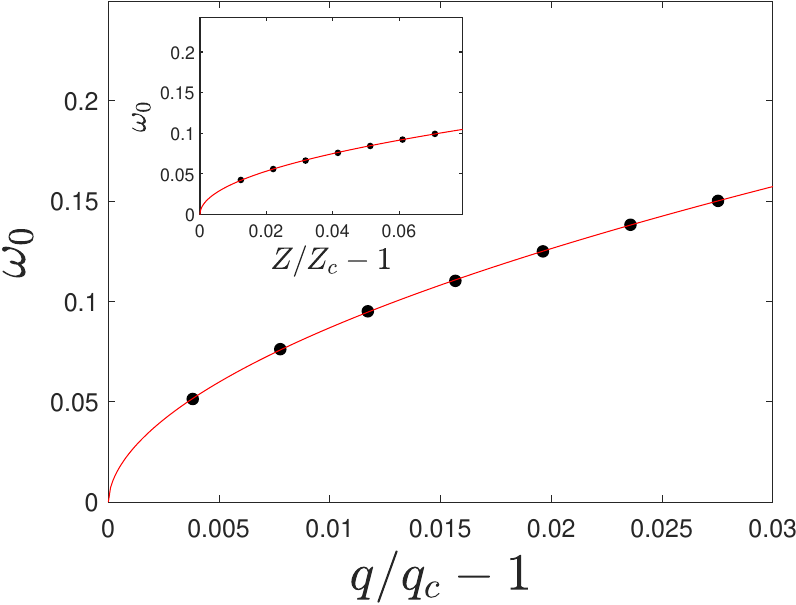}} 
\end{tabular}
\caption{a) Dynamical phase diagram as a function of $(Z,q)$ for the final state of an initially subsonic flowing condensate $\Psi(x,0)=e^{iqx}$ in which an attractive delta barrier $V(x)=-Z\delta(x)$ is quenched at $t=0$. (b) Critical behavior of the CES frequency $\omega_0$ close to the phase transition along the green lines in (a), where the red line represents a fit to a power law. Main panel: Velocity dependence. Inset: Delta-strength dependence.}
\label{fig:Critical}
\end{figure*}

Interestingly, the concept of spontaneous Floquet state can be extended to any many-body system whose dynamics can be described by a variational ansatz that leads to an effective self-consistent Hamiltonian like the GP equation, as proven in Ref. \cite{deNova2022}. This includes several canonical many-body descriptions such as the MultiConfiguration Time-Dependent Hartree method for bosons and fermions \cite{Caillat2005,Alon2008}, the Hartree-Fock equations for fermions \cite{Thouless2014}, or the Gutzwiller ansatz in Bose-Hubbard models \cite{Jaksch1998}.

Furthermore, since a spontaneous Floquet state breaks the time-translation symmetry of the underlying Hamiltonian, the CES state represents a realization of continuous time crystal \cite{Wilczek2012,Kongkhambut2022}. This in stark contrast with discrete time crystals \cite{Sacha2015,Else2016,Zhang2017,Choi2017}, arising in conventional, driven Floquet systems, where the periodicity is imposed by a subharmonic response to the external driving, and the resulting symmetry breaking is discrete, not continuous. Actually, the CES state was shown \cite{deNova2022} to satisfy typical time-crystal criteria of robustness (against the presence of time-dependent disorder or variations of the parameters of the Hamiltonian), independence from the initial state, and universality (i.e., it is a feature of a wide class of Hamiltonians). 

For instance, a simple realization of the CES state is achieved by quenching an attractive delta barrier $V(x)=-Z\delta(x)$ at $t=0$ in a homogeneous flowing condensate with velocity $q$, described by an initial GP wavefunction $\Psi(x,0)=e^{iqx}$. The resulting dynamics is deterministic, and the final state of the system is described by a similar dynamical phase diagram, solely function of $(Z,q)$, which only displays the ground state at low velocities and the CES state at high velocities, Fig. \ref{fig:Critical}a. The GS/CES phase diagram is an example of dynamical phase transition \cite{Moeckel2008,Sciolla2010,Lang2018}, where the ground state is the symmetry-unbroken phase, with continuous time translation symmetry, while the CES state is the time-crystalline phase, with discrete time translation symmetry. Indeed, as predicted by Renaud Parentani himself during our visit to Orsay in 2015, the oscillation frequency $\omega_0$ of a CES state exhibits a critical behavior close to the phase transition, where the critical exponents $\alpha_q,\alpha_Z$ for $q,Z$ (obtained from a fit in Fig. \ref{fig:Critical}b) are both approximately $\alpha_q\simeq \alpha_Z\simeq 0.50$, strongly suggesting a possible analytical derivation.

%This is an SMBF state since a) the effective Hamiltonian $H_{\rm{GP}}(,t)$ is periodic, and the wavefunction $\Psi(x,t)$ is a Floquet state; b) this periodicity is spontaneously induced by many-body interactions, and not tuned by some external field or symmetry transformation. 

% in terms of quasi-energy bands.

% which in turn take the form of a more usual linear Floquet system since $M(t)$ is periodic. 

Another remarkable feature of a spontaneous Floquet state is that it conserves energy due to the time-independence of the underlying Hamiltonian, unlike conventional Floquet systems. After developing the so-called $(t,\phi)$ formalism within the generalized Gibbs ensemble \cite{Cazalilla2006,Rigol2007,Langen2015}, Ref. \cite{deNova2025} showed that spontaneous Floquet states have a unique conserved magnitude, the Floquet charge $F$, whose conjugate thermodynamic variable is the frequency $\omega$. This allows to identify spontaneous Floquet states as isofloquetic, conserving the total energy, and driven Floquet states as isoperiodic, conserving the Floquet enthalpy, $I=E-\omega F$, in analogy with isochoric and isobaric systems, respectively. This characterization gives rise to the so-called Floquet thermodynamics, which describes Floquet systems with the same thermodynamical tools as in stationary states.

The quantum fluctuations of a spontaneous Floquet state in an atomic condensate are described by the BdG equations resulting from the expansion $\hat{\Psi}=[\Psi_0(x,t)+\hat{\varphi}(x,t)]e^{-i\tilde{\mu}t}$, namely
\begin{equation}\label{eq:BdGfieldequation}
i\partial_t\hat{\Phi}=M_0(t)\hat{\Phi},\,M_0(t)=\left[\begin{array}{cc} N_0(t) & A_0(t)\\
-A_0^*(t) &-N_0^*(t)\end{array}\right],
\end{equation}
where
\begin{eqnarray}
\nonumber N_0(t)&=&-\frac{\partial_x^2}{2}+V(x)+2g(x)|\Psi_0(x,t)|^2-\tilde{\mu},\\
A_0(t)&=&g(x)\Psi_0^{2}(x,t)
\end{eqnarray}
are periodic operators. Consequently, the BdG matrix $M_0(t)$ is a periodic linear operator, described by conventional Floquet theory, so its spectrum is given in terms of quasi-energy bands,
\begin{equation}\label{eq:BogoliubovFloquetSectorTime}
    [M_0(t)-i\partial_t]z_{\varepsilon,\nu}(t)=\varepsilon z_{\varepsilon,\nu}(t),
\end{equation}
with $z_{\varepsilon,\nu}(\mathbf{x},t+T)=z_{\varepsilon,\nu}(\mathbf{x},t)$, $\varepsilon$ the quasi-energy (defined again modulo $\omega_0$), and $\nu$ a discrete index labeling the solution. This BdG expansion also results from a conventional Floquet state, which takes the same form of Eq. (\ref{eq:FloquetWaveMu}) but the period $T$ there is imposed by the external driving, while for a spontaneous Floquet state $T$ is spontaneously chosen by the system.
 
Of particular interest is the presence of Nambu-Goldstone (NG) modes. In a stationary context, if the GP wavefunction spontaneously breaks one of the continuous symmetries of the Hamiltonian in such a way that, if $\Psi_0(x)$ is a stationary GP solution, then $\Psi_\alpha(x)=e^{-i\alpha G}\Psi_0(x)$ is another stationary GP solution, a zero-energy NG mode emerges in the BdG spectrum:
\begin{equation}\label{eq:NambuGoldstone}
    M_0z_\alpha=0,~z_\alpha=\left[\begin{array}{cc} \partial_\alpha \Psi_0 \\
\partial_\alpha \Psi^*_0\end{array}\right]=\left[\begin{array}{cc} -iG \Psi_0 \\
i (G\Psi_0)^*\end{array}\right].
\end{equation}
This can be proven by expanding the symmetry transformation to linear order in $\alpha$, where $G$ is the infinitesimal generator of the transformation. Since NG modes have zero norm, their amplitude does not behave as an annihilation operator but instead as a coordinate operator, with a conjugate momentum that describes the fluctuations of the conserved charge $Q$ associated to the broken continuous symmetry \cite{Lewenstein1996,Dziarmaga2004}. A major example is the Goldstone mode $z_\theta$ corresponding to the $U(1)$-symmetry breaking, see Eq. (\ref{eq:TIBdGSource}), resulting from the fact that $e^{-i\theta}\Psi_0$ is also a stationary GP solution for arbitrary $\theta$. When several symmetries are spontaneously broken, the quantization procedure is elegantly described by a geometric formalism involving the so-called Berry-Gibbs connection, which is the Berry connection associated to the GP wavefunction, the continuous parameters of the manifold being the conserved charges associated to the broken symmetries \cite{deNova2025}.

For Floquet states, both spontaneous and conventional, spontaneous symmetry breaking is translated into the emergence of Floquet-Nambu-Goldstone (FNG) modes with zero quasi-energy. In a condensate, this means $[M_0(t)-i\partial_t]z_\alpha(t)=0$, where $z_\alpha(t)$ is now periodic. In the specific case of a spontaneous Floquet state, a genuine temporal FNG mode arises from the spontaneous symmetry breaking of time-translational invariance \cite{deNova2025}. This is the characteristic hallmark of a \textit{bona-fide} spontaneous Floquet state, reflecting also its time-crystalline nature, which distinguishes it from trivial periodic behavior such as traveling soliton waves in a ring, which do not possess a proper temporal FNG mode. The CES state indeed has a genuine temporal FNG mode, stemming from the fact that $\Psi_0(x,t+t_0)e^{-i\tilde{\mu}t}$ is also a solution of the time-dependent GP equation (\ref{eq:HamiltonianEffective}) for arbitrary $t_0$. 

Interestingly, the quantum amplitude of the temporal FNG mode provides a unique realization of a time operator in quantum mechanics, which commutes with the linear fluctuations of the grand-canonical energy, no longer vanishing since $\Psi_0$ is not here an extreme of the grand-canonical energy as it is time-dependent \cite{deNova2025}. In general, quantum mechanics forbids the existence of a time operator $\hat{T}$ since the Hamiltonian should be its canonical conjugate, $[\hat{T},\hat{H}]=-i$. This implies that 
\begin{equation}
    e^{-i\hat{T}E_0}\hat{H}e^{i\hat{T}E_0}=\hat{H}-E_0.
\end{equation}
for arbitrary value of $E_0$. Hence, if $E$ is an eigenenergy, then $E-E_0$ is also an eigenergy, which contradicts the fact that Hamiltonians are bounded from below by the ground-state energy.

Nevertheless, this problem is similar to that of the phase operator \cite{Mora2003}. If a phase operator $\hat{\theta}$ exists, then it satisfies $[\hat{N},\hat{\theta}]=i$, which implies
\begin{equation}
    e^{-i\hat{\theta}N_0}\hat{N}e^{i\hat{\theta}N_0}=\hat{N}-N_0
\end{equation}
for arbitrary value of $N_0$. Hence, if $N$ is an eigenvalue of $\hat{N}$, then $N-N_0$ is also, violating two fundamental properties of the spectrum of $\hat{N}$: its positive definiteness and its discreteness. 
However, it is well known that one can define phase fluctuations in condensates involving large particle numbers, where one can neglect the discreteness of $N$ and states with low occupation number. In analogy, by identifying energy with particle number and time with phase, one can define time fluctuations for large energies well above the ground state. This correspondence indicates that spontaneous Floquet states must be highly excited states, shifted by a macroscopic energy from the ground state. This is indeed the case of the CES state, whose energy is shifted $\Delta E\simeq Nq^2/2$ above that of a condensate at rest. 

\section{Discussion}\label{sec:conclu}

Resonant configurations represent a rich paradigm in analogue gravity. In addition to the universal thermal behavior at low frequencies, they display a highly non-thermal peaked structure in the Andreev and Hawking spectra, since they act as a Fabry-Perot resonator for the negative-energy partners of the Andreev-Hawking effect. This contrasts with standard analogue configurations, whose spectra have a marked thermal character which is easy to misidentify with any other background thermal component, as in the real astrophysical scenario. Another interesting feature of resonant configurations is that they highly enhance the Andreev signal, which can even overcome the Hawking one, while in standard analogue configurations the former is typically highly suppressed.

A feasible experimental scheme to implement a resonant configuration using atomic condensates relies on outcoupling a large boson reservoir through an optical lattice, eventually achieving a quasi-stationary black-hole configuration. Remarkably, the optical lattice acts as a low-pass filter of Andreev-Hawking radiation \cite{deNova2017b}, something that could have potential applications in quantum transport and atomtronics \cite{Amico2021}, further motivating the interest for its experimental implementation. An alternative approach to achieve a non-thermal spectrum  is provided by dipolar condensates, where the departure from thermality is induced by the presence of a roton minimum in the dispersion relation \cite{Ribeiro2023}.

By applying concepts and techniques originally derived in quantum optics, we can understand the Andreev and Hawking effects as a joint phenomenon, resulting from the non-degenerate parametric amplification of the outgoing modes, where the hybrid Andreev-Hawking channel is the signal, and the anomalous channel is the idler. Regarding quantum correlations, we have analyzed the occurrence of violations of Cauchy-Schwarz inequalities and entanglement, which are equivalent conditions for a broad and relevant class of quantum states. We have observed that resonant configurations highly enhance entanglement near the resonant peaks of the spectrum for both the Andreev and Hawking effects, allowing for its detection even at high temperatures comparable to the chemical potential. Thus, they improve the performance of standard analogue configurations, where entanglement is highly attenuated with temperature, fading away at low temperatures in the Andreev case. 

The characterization of quantum correlations in the Andreev-Hawking effect, including tripartite entanglement \cite{Isoard2021} and Bell non-locality \cite{Ciliberto2024}, is still an active topic of research, which could lead to potential applications in quantum technologies, since an analogue horizon behaves as a spontaneous source of entangled phonons. 

An interesting spin-off of the study of quantum correlations in analogue gravity is the research on quantum information in high-energy colliders \cite{Afik2021}, which is rapidly becoming a whole topic of research by itself. It has already led to the first observation of entanglement in quarks, in turn the highest-energy entanglement detection ever, by the ATLAS and CMS collaborations \cite{ATLAS2024,CMS2024}. This observation paves the way to use high-energy colliders for the study of foundational quantum problems, something of great interest due to their genuine relativistic nature and fundamental character, operating at the current frontier of knowledge in Physics. In fact, a number of experimental analyses searching for genuine quantum signatures at the LHC are currently ongoing.  

A most important phenomenon arising in resonant configurations is the black-hole laser effect. For its discussion, we have separated the three main stages of its time evolution. At short times, the dynamics is governed by the spectrum of dynamical instabilities in the linear BdG equations. By analyzing several BHL models, we have observed the generality of the original predictions by Michel and Parentani \cite{Michel2013}, namely: i) the lasing modes emerge as degenerate at critical lengths equispaced by the BCL wavelength, becoming non-degenerate at halfway between the critical lengths; and ii) there is a perfect correspondence between the emergence and later degeneracy breaking of the lasing modes, and the emergence of stationary GP solutions in the non-linear spectrum. Our results also confirm the conjecture of Michel and Parentani \cite{Michel2015}: in flowing scattering configurations, energetic and dynamical instability are equivalent conditions, and the only stable solution is the ground state, which evaporates all the acoustic horizons to become fully subsonic.

At intermediate times, we have theoretically studied the BHL-BCL crossover, originally characterized in Ref. \cite{deNova2023}. By invoking the analogy with an unstable pendulum \cite{Leonhardt2003,Finazzi2010,Burkle2018,deNova2023}, three regimes can be identified: quantum BHL, classical BHL, and BCL. Their most characteristic trait is their efficiency as quantum amplifiers: a quantum BHL is a non-linear quantum amplifier, increasing quantum fluctuations up to the same saturation amplitude regardless of their initial strength, while classical BHL and BCL are linear quantum amplifiers, where the output is proportional to the input. In particular, quantum amplification in a classical BHL is exponentially large in the lasing time, and much larger than in the BCL regime, where there is no microscopic amplification mechanism, and the amplification just stems from the strong background modulation induced by the BCL wave. The characterization of each regime as a quantum amplifier can be complemented with a qualitative analysis of the monotonicity of their growth rate and their quantum gain, providing practical experimental criteria for the unambiguous detection of the BHL effect, a major remaining challenge in the analogue field. Furthermore, our analysis identifies the classical BHL regime as the most reachable target, where the BCL wave fuels the BHL effect by providing it with a classical seed, instead of undermining it.

Remarkably, the study of the BHL-BCL crossover in Ref. \cite{deNova2023} has also allowed to identify novel analogue phenomena such as Hawking-stimulated white-hole (HSWH) radiation at the start of the BHL process (when the partner modes of the Andreev-Hawking effect stimulate the continuous spectrum of white-hole radiation \cite{Mayoral2011}), or quantum BCL-stimulated Hawking radiation (the spontaneous resonant Hawking radiation above the non-linear saturated BCL wave). The analysis of the BHL-BCL crossover can be of interest for other analogue setups in which low-frequency undulations similar to the BCL wave hinder the BHL effect \cite{Coutant2012,Coutant2014,Bossard2023}. In general, a quantum/classical BHL provides an ideal testing ground for the study of quantum/classical backreaction \cite{Balbinot2005a,Patrick2021,Baak2022,Butera2023}. Apart from its intrinsic interest for the analogue field, a BHL behaves as a quantum amplifier, which could have potential applications in atomtronics.

At long times, a black-hole laser exhibits a dynamical phase diagram with two states: the ground state, with continuous time-translation symmetry, and the CES state, with discrete time-translation symmetry, resulting from its periodic nature. Indeed, the CES state is a universal feature of a flowing condensate, representing a particular example of the much more general concept of spontaneous Floquet state \cite{deNova2022}: a Floquet state arising from a time-independent Hamiltonian, whose periodicity is spontaneously set by many-body interactions.

Spontaneous Floquet states are by themselves a novel non-equilibrium paradigm. For instance, they conserve energy, in contrast to conventional Floquet states, which in turn conserve the so-called Floquet enthalpy as they arise from periodically driven Hamiltonians, operating at fixed frequency. These conserved magnitudes allow for a thermodynamic description of Floquet states completely analogous to that of stationary states, which has been labeled Floquet thermodynamics \cite{deNova2025}. 

In addition, spontaneous Floquet states spontaneously break continuous time-translation symmetry, representing a specific realization of a continuous time crystal. This results in the emergence of a genuine temporal Floquet-Nambu-Goldstone mode with zero quasi-frequency, whose amplitude provides a unique realization of a time operator in a tangible condensed-matter setup \cite{deNova2025}. We note that the construction of a time operator is a fundamental subject in quantum mechanics \cite{Aharonov1961,Susskind1964,Unruh1989,Honh2021}. Therefore, the identification of a time operator in an analogue gravity setup provides a rich scenario that could lead to fundamental research on the quantum foundations of spacetime.

Although our discussion is restricted to atomic condensates, the results of this work can be easily translated to optical and polaritonic analogues due to the similarity of the equations of motion. For instance, Andreev reflection has been studied in polaritonic condensates \cite{Septembre2021,Septembre2023}. The excitation of a quasi-normal mode from vacuum fluctuations has been recently numerically observed in a resonant polaritonic configuration \cite{Jacquet2023}. Quantum correlations in optical and polaritonic analogues have been also studied \cite{Busch2014a,Agullo2022,Brady2022,Delhom2024}. Similar periodic states to the CES state are predicted for polaritons \cite{Opala2018}.

As a global remark, it must be noted that the low-pass filter of Andreev-Hawking radiation provided by an optical lattice, the stationary source of entangled phonons provided by the Hawking effect \cite{Kolobov2021}, the behavior of a black-hole laser as a quantum amplifier, and the spontaneous Floquet state represented by the CES state, demonstrate the potential of interdisciplinary applications of analogue gravity concepts. This is further supported by the establishment of a line of research on quantum information in high-energy colliders, conceptually based on the study of quantum correlations in Andreev-Hawking radiation.

We would like to conclude by emphasizing the profound influence provided by the intellectual leadership of Renaud Parentani in the shaping of the field of analogue gravity. His ideas and inspiration permeate any coherent narration of the evolution of this research field, and in particular are ubiquitous in all the results discussed in the present article.

%The intellectual leadership of Renaud Parentani has had a profound influence in the shaping of the field of analogue gravity. His ideas and inspiration permeate any coherent narration of the evolution of this research field.

\acknowledgments
We devote this article to the memory of Renaud Parentani. This work has received funding from European Union's Horizon 2020 research and innovation programme under the Marie Sk\l{}odowska-Curie Grant Agreement No. 847635, from Spain's Agencia Estatal de Investigaci\'on through Grant No. PID2022-139288NB-I00, and from Universidad Complutense de Madrid through Grant No. FEI-EU-19-12.

\bibliography{Hawking}

\end{document}